\documentclass[fleqn,10pt]{wlscirep}
\usepackage{subcaption}
\usepackage[utf8]{inputenc}
\usepackage[T1]{fontenc}
\usepackage{multirow, booktabs, tabularx, longtable}
\usepackage{makecell}
\usepackage{url}
\usepackage{tikz}
\usepackage{xurl}
\usepackage{caption}
\usepackage{xfp}
\usepackage{color}
\usepackage{tabularray}
\usepackage{datetime}
\UseTblrLibrary{booktabs}

\DefTblrTemplate{caption}{supplementary}{
  \DefTblrTemplate{caption-tag}{supplementary}{Supplementary Table \thetable}%
  \DefTblrTemplate{caption-sep}{supplementary}{.}%
  \centering
  \UseTblrTemplate{caption-tag}{supplementary}%
  \UseTblrTemplate{caption-sep}{supplementary}
  \UseTblrTemplate{caption-text}{default}\par
}
\DefTblrTemplate{firsthead}{supplementary}{\UseTblrTemplate{caption}{supplementary}}
\DefTblrTemplate{middlehead}{supplementary}{\UseTblrTemplate{caption}{supplementary}}
\DefTblrTemplate{lasthead}{supplementary}{\UseTblrTemplate{caption}{supplementary}}
\NewTblrTheme{supplementary}{
  \SetTblrStyle{caption-tag}{font=\bfseries}
  \SetTblrTemplate{firsthead}{supplementary}
  \SetTblrTemplate{middlehead}{supplementary}
  \SetTblrTemplate{lasthead}{supplementary}
}
\DeclareCaptionFormat{bfdot}{\textbf{#1}.~#3}
\definecolor{light-gray}{HTML}{FFFFFF}
\definecolor{light-blue}{HTML}{F0F0F0} 

\DeclareMathAlphabet\mathbfcal{OMS}{cmsy}{b}{n}
\DeclareSymbolFont{operators}   {OT1}{cmr} {m}{n}
\DeclareSymbolFont{letters}     {OML}{cmm} {m}{it}
\DeclareSymbolFont{symbols}     {OMS}{cmsy}{m}{n}

\def\A{\mathcal A}
\def\X{\mathcal X}
\def\Y{\mathcal Y}
\def\V{\mathcal V}
\def\R{\mathbb R}
\def\N{\mathbb N}

\usepackage{soul}
\usepackage{color}

\usepackage[section]{placeins}

\RequirePackage[explicit]{titlesec}
\usepackage{hyperref}
\titleformat{name=\section,numberless}
  {\large\sffamily\bfseries}
  {}
  {0em}
  {\leavevmode\MakeLinkTarget[section]{}\ignorespaces#1}
  []  
\ExplSyntaxOn\makeatletter
\def\ttl@straight@i#1[#2]#3{%
  \tl_if_empty:nTF {#2}
   {\NR@gettitle{#3}}
   {\NR@gettitle{#2}}
    \gdef\ttl@savemark{\csname#1mark\endcsname{#3}}%
  \let\ttl@savewrite\@empty
  \def\ttl@savetitle{#3}%
  \gdef\thetitle{\csname the#1\endcsname}%
  \if@noskipsec \leavevmode \fi
  \par
  \ttl@labelling{#1}{#2}%
  \ttl@startargs\ttl@straight@ii{#1}{#3}}
\ExplSyntaxOff\makeatother

\usepackage{xr}
\makeatletter

\newcommand*{\addFileDependency}[1]{%
\typeout{(#1)}
\@addtofilelist{#1}
\IfFileExists{#1}{}{\typeout{No file #1.}}
}\makeatother

\newcommand*{\myexternaldocument}[1]{%
\externaldocument{#1}%
\addFileDependency{#1.tex}%
\addFileDependency{#1.aux}%
}
\myexternaldocument{SI_V1}

\title{A severe local flood and social events show a similar impact on human mobility}

\author[1,2,*]{Simone Loreti}
\author[3,4]{Margreth Keiler}
\author[1,2]{Andreas Paul Zischg}

\affil[1]{University of Bern, Institute of Geography, Bern, 3012, Switzerland}
\affil[2]{University of Bern, Oeschger Centre for Climate
 Change Research, Mobiliar Lab for Natural Risks, Bern, 3012, Switzerland}
\affil[3]{University of Innsbruck, Department of Geography, Innsbruck, 6020, Austria}
\affil[4]{Austrian Academy of Sciences, Institute of Interdisciplinary Mountain Research, Innsbruck, 6020, Austria}

\affil[*]{To whom correspondence should be addressed. E-mail: simone.loreti@unibe.ch 
\newline Date: \today
}

\begin{abstract}

\vspace{-0.5cm}

While a social event, such as a concert or a food festival, is a common experience to people, a natural disaster is experienced by a fewer individuals. The ordinary and common ground experience of social events could be therefore used to better understand the complex impacts of uncommon, but devastating natural events on society, such as floods. Based on this idea, we present a comparison - in terms of human mobility -, between an extreme local flood that occurred in 2017 in Switzerland, and social events which took place in the same region, in the weeks before and after the inundation. Using mobile phone location data, we show that the severe local flood and social events have a similar impact on human mobility, both at the national scale and at a local scale. At the national level, we found a small difference between the distributions of visitors and their travelled distances among the several weeks in which the events took place. At the local level, instead, we detected the anomalies (in time series) in the number of people travelling each road and railway, and we found that the distributions of anomalies, and of their clusters, are comparable between the flood and the social events. Hence, our findings suggest that the knowledge on ubiquitous social events can be employed to characterise the impacts of rare natural disasters on human mobility. The proposed methods at the local level can thus be used to analyse the disturbances in complex spatial networks and, in general, as complementary approaches for the analyses of complex systems.

\end{abstract}

\begin{document}

\flushbottom
\maketitle
\thispagestyle{empty}

\section*{Introduction}
\label{Introduction}

Being an innate need of human beings \cite{Colonna2012}, mobility represents one of the essential components of human societies and their economic activities \cite{Rodrigue2016}. Understanding the complex underlying mechanisms of human mobility is crucial for a wide range of applications, such as urban planning, traffic prediction, diseases spread, disaster evacuation and response \cite{González2008,WangTaylor2014,Barbosa2018,Wang2021_A_Bibliometric_Analysis,Wang2022_review_human_mobility}. 
The growth and proliferation of technological innovations, as Internet, mobile phones  \cite{Barth2014,Gupta2011,Smith2014}, Wi-Fi, Global Positioning System (GPS), which led to an unprecedent availability of geo-referenced data on human mobility, 
enhanced our observations, assessments and modelling of human mobility dynamics
\cite{Lepri2017,Meekan2017,Barbosa2018,Yabe2022_Mobile_phone}.
In particular in the last decade, human mobility was primarily investigated through mobile phone data, i.e. the Call Detail Records (CDR) and Global Positioning System (GPS), the latter including geo-located social media \cite{Wang2018_Applying_mobile_phone,Muniz-Rodriguez2020,Haraguchi2022,Yabe2022_Mobile_phone,Wang2022_review_human_mobility,Barbosa2018,Pappalardo2023}.
The need to evaluate policies and interventions during the COVID-19 period, and likely the recent concerns about a possible increase of climatic hazards
(related to a worldwide rapid urbanization and climate change), spurred further the usage of mobile phone data for human mobility studies \cite{Oliver2020,Yabe2022_Mobile_phone}.
Forecasts about climatic hazards (weather and climate extreme events) refer to a possible increase in both their probability of occurrence and impacts \cite{Diffenbaugh2017,IPCC2023,Lange2020,Kreibich2022}, and
can be justified by the sharp increase in the number of natural loss events
$-$ mostly climate-related disasters $-$, 
that occurred between the period from $1980$ to $1999$, and the following twenty years, i.e., from $2000$ to $2019$ \cite{CRED_UNDRR_2020,WMO_2021}.
A similar trend continued in $2020$ and $2021$, where
climate and weather-related loss events and disasters
represented the most frequent hazards (not considering the COVID-19 pandemics \cite{CRED_UNDRR_2021}), outstripping geological and technological disasters \cite{IFRC2023}. 
More recently, in $2022$,
floods and storms had the highest number of occurrences, causing the largest economic losses, and droughts affected the largest number of people \cite{CRED_2023}. 
Despite the abundance of studies that leveraged mobile phone data to analyse mobility patterns during disaster events, we still lack: (i) a universal framework for modeling the impacts of natural hazards related disasters,
crossing geographical regions and different natural hazards related 
disaster events \cite{Li2022,Yabe2022_Toward_data_driven}, (ii) a full understanding of the complexity, dynamics, and interdependencies among social, economic and technical systems, which could be explored by integrating advanced techniques (e.g. AI) \cite{Wang2021_A_Bibliometric_Analysis} and different data sources, such as CDR, social media and satellite imagery, (iii) an effective and simplified transfer of mobility insights to policy-makers and 
stakeholders
\cite{Yabe2022_Toward_data_driven}, (iv) real-time tools to predict mobility patterns during real-world disasters, (v) a comprehensive analysis of long-term recovery and resilience patterns, (vi) a transparent procedure for data collection and elaboration \cite{Yabe2022_Mobile_phone}, and (vii) a clear comprehension of the fluctuations of human activities \cite{Farahmand2022}.
Instead of comparing different disaster events among each other \cite{Li2022,Yabe2022_Toward_data_driven}, in this work, we compare one natural loss event,
i.e. a flood, with social events, both occurred in Switzerland, between June and July 2017. 
Indeed, we argue
that possible similarities or differences among  scheduled artificial events and unscheduled natural loss event
would help us in better understanding the transition between the laws of ordinary and unperturbed mobility \cite{Barbosa2018,Schlapfer2021} and the laws of perturbed mobility (e.g. that one proposed by Li et al. (2022) \cite{Li2022}). 
By using mobile phone data provided by a major telecommunications company of Switzerland (i.e. Swisscom, with a market share of around $60\%$, at the end of 2017 \cite{OFCOM_2023}), we show that a severe
local flood, causing more than $90$ million Swiss Francs of property damage \cite{FOEN_2018} (i.e. $\sim 93$ million USD, in July 2017), has an impact on human mobility similar to that one of major social events, as food festivals and concerts, both at a national and at a local scale.

\section*{Results}
\label{Results}

\subsection*{Similarities between social events and a local flood at a national scale}
\label{similarities_national}

\begin{figure}[htp]
    \setlength{\lineskip}{0pt}
    \centering
    
    \begin{tikzpicture}
        \node[anchor=north west,inner sep=0pt] at (0,0)
        {\includegraphics[width=13.6cm] {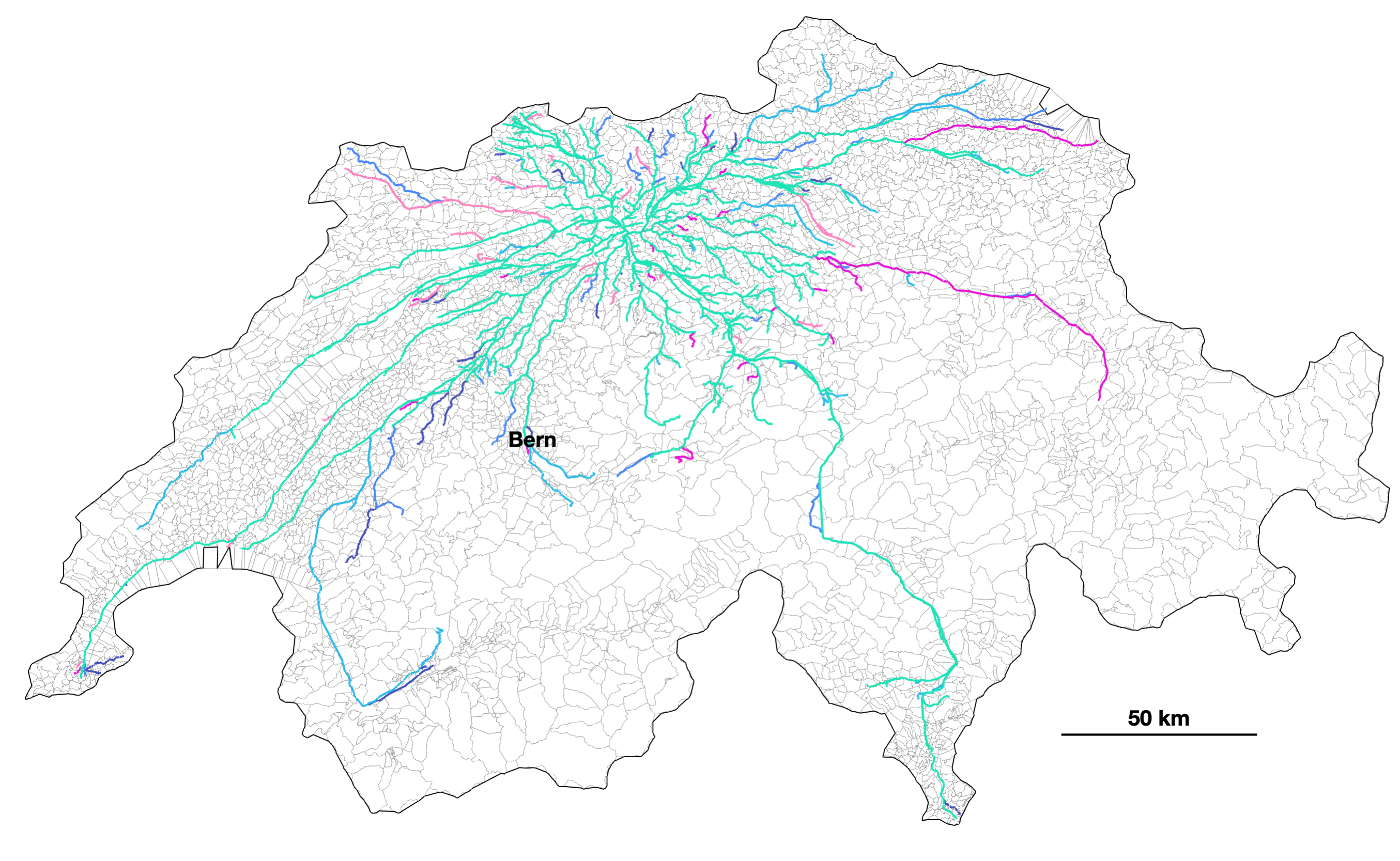}}; 
        \node[font=\sffamily\bfseries\large] at (12.7cm,-1.2cm) {a};
    \end{tikzpicture}\\
    \begin{tikzpicture}
        \node[anchor=north west,inner sep=0pt] at (0,0){\includegraphics[width=14.5cm]{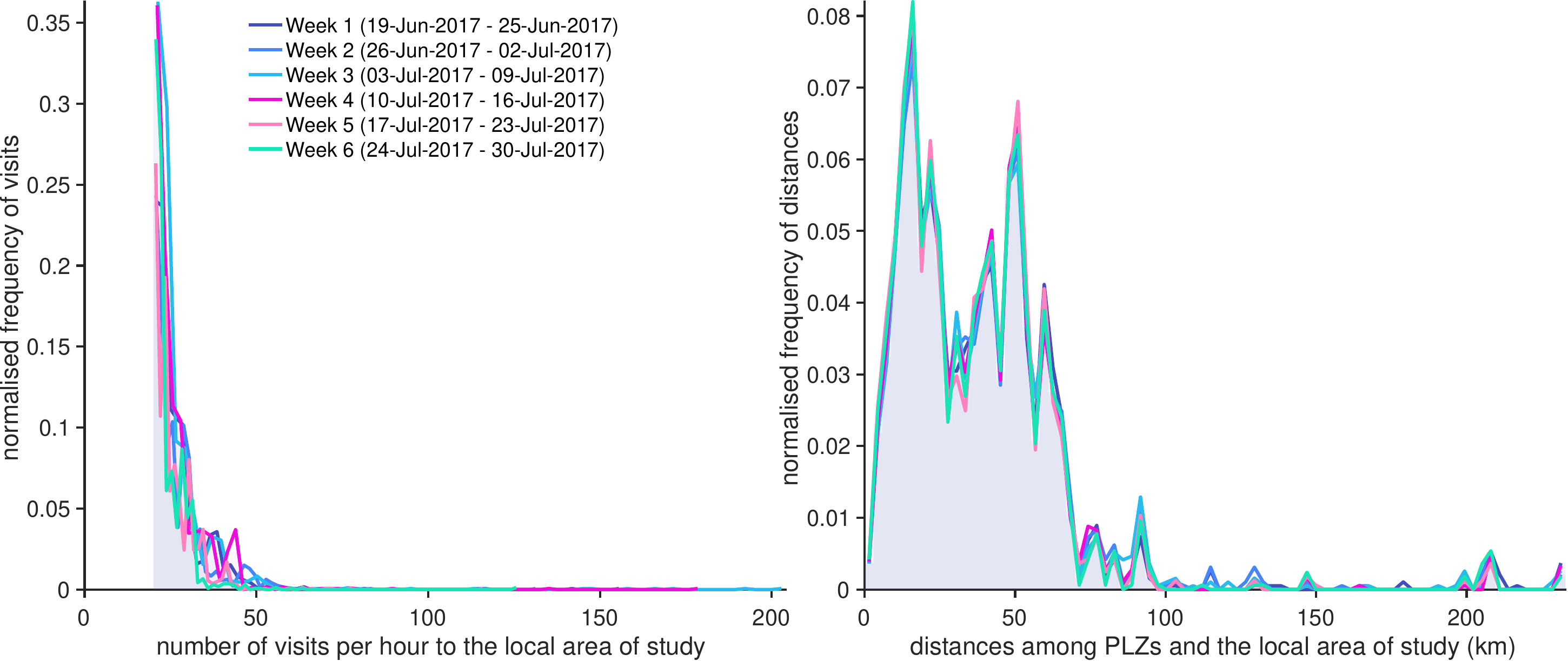}};
        \node[font=\sffamily\bfseries\large] at (6.2cm,-0.5cm) {b};
        \node[font=\sffamily\bfseries\large] at (12.7cm,-0.5cm) {c};
    \end{tikzpicture}\\
    \begin{tikzpicture}
        \node[anchor=north west,inner sep=0pt] at (0,0){\includegraphics[width=15cm]{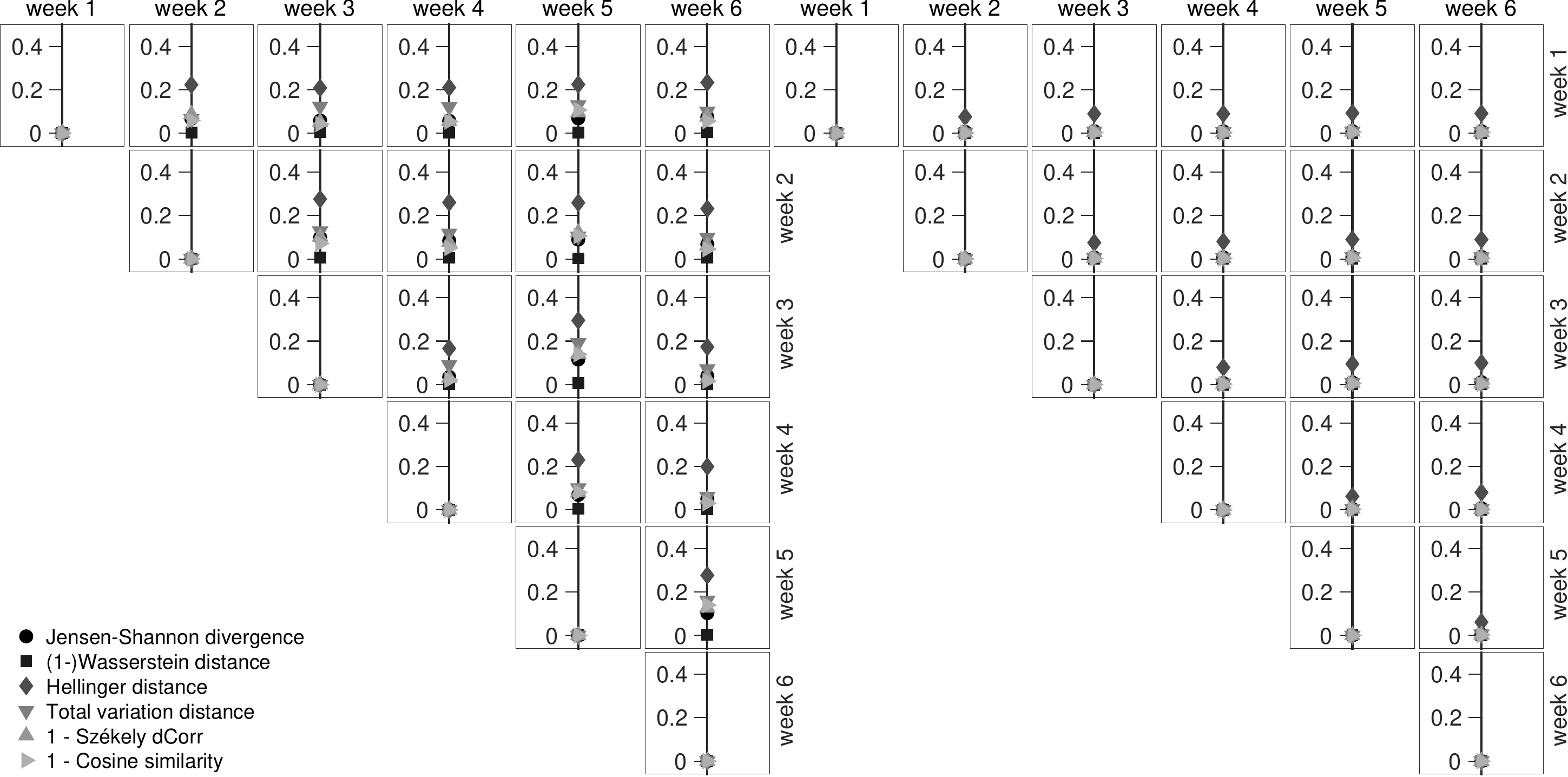}};
        \node[font=\sffamily\bfseries\large] at (0.5cm,-4cm) {d};
        \node[font=\sffamily\bfseries\large] at (8cm,-4cm) {e};
    \end{tikzpicture}
    
    \caption{\textbf{(a)} Graphical representation of the impacts of social and flood events at a national scale. The paths show the distances travelled by individuals who visited the local area of study, during $6$ weeks. \textbf{(b)} Relative frequency histograms representing the empirical distributions of visits to the local area of study, during $6$ weeks. \textbf{(c)} Relative frequency histograms representing the empirical distributions of distances travelled by people who visited the local area of study, during $6$ weeks. \textbf{(d)} Comparison of the $6$ histograms that represent the number of visits in each week. \textbf{(e)} Comparison of the $6$ histograms that represent the distances covered by visitors, in each week.
    }
    \label{fig:national_scale}
    
\end{figure}
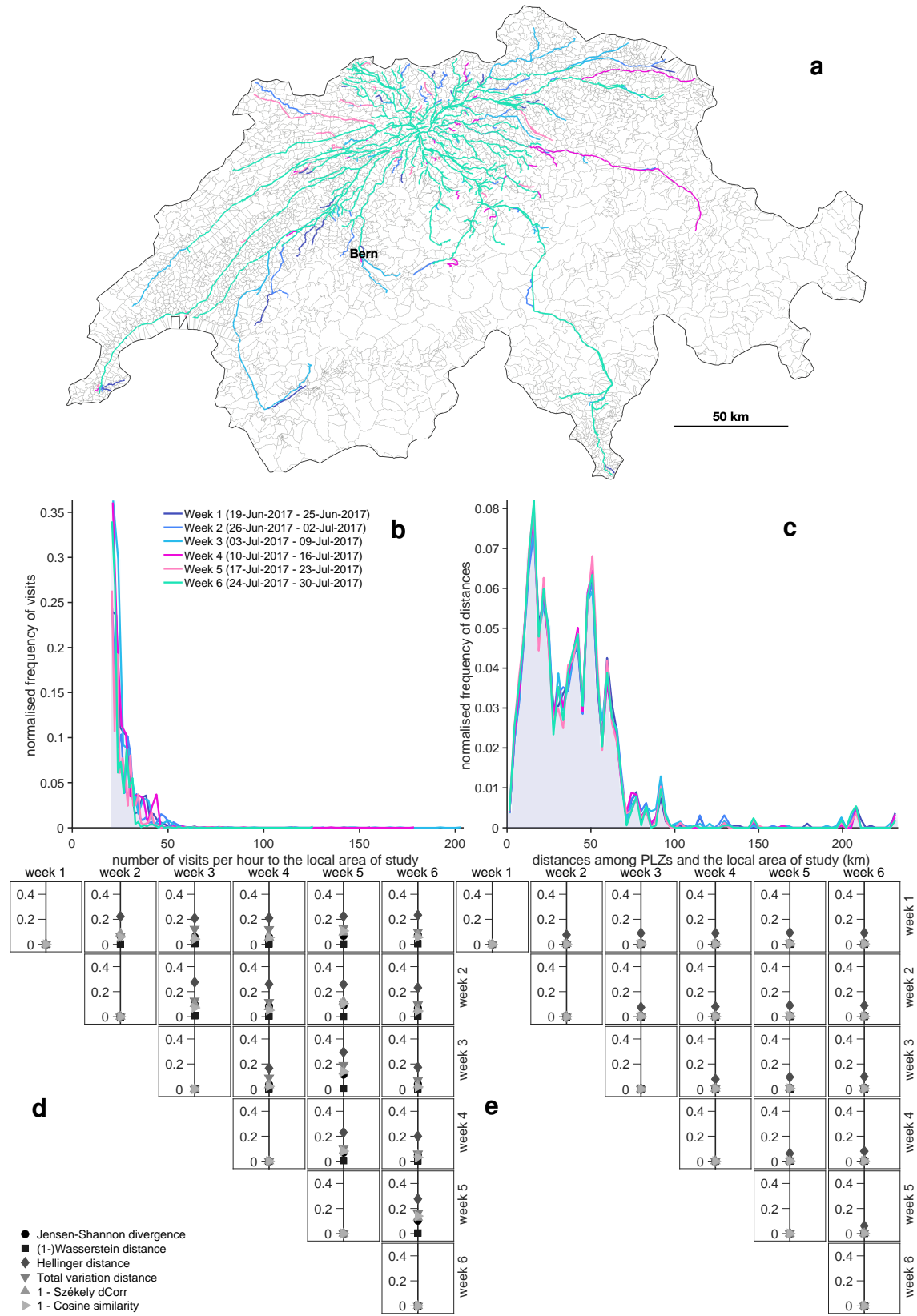

We consider the entire Swiss territory to describe the long-range impacts of medium-to-large social events (such as music and food festivals) and the impacts of a severe local flood, with a return period of $100$ years, on human mobility (see \cref{fig:national_scale}(a)). The investigated social events took place between the $17^{th}$ of June and the $30^{th}$ of July 2017, i.e. in a time window of $44$ days. The flood, instead, occurred on Saturday the $8^{th}$ of July $2017$, i.e. exactly in the middle of the temporal window under investigation 
(see \textcolor{blue}{Supplementary Text} and \textcolor{blue}{Supplementary Table 1}
for details on the flood and social events, respectively).
Both types of events occurred within a geographical region that measures around $20$ km times $15$ km, and that crosses three Swiss cantons, named, Solothurn, Aargau, and Lucerne. Throughout the entire manuscript, we will call that region as the \emph{local area of study}, to distinguish it from the \emph{national area of study}, the latter corresponding to the entire Switzerland. To capture potential differences of mobility among the diverse types of events, we could simply focus, for each event, on the exact days and hours in which a specific event occurred, and then measure the corresponding impacts on mobility. However, the examined events occurred both during weekends (most of them) and weekdays, i.e. during days that are usually characterised by different mobility patterns. Indeed, for example, people go to work and to schools during weekdays, while they are generally more involved in leisure activities during the weekends. In addition, the studied events lasted for different lengths of time, making the assessment of their impacts on mobility more challenging. One possibility to overcome potential biases could be to analyse the six weeks, composing the $44$-day period, each week lasting from Monday to Sunday, both days included. This approach would require an observation period that is multiple of $7$ (days), such as a $42$-day period. We therefore neglect the first two days, a Saturday and a Sunday, of the examined temporal window of $44$ days. For each of the six weeks, we then build two relative frequency distributions, displayed through histograms ($12$ histograms in total). One related to the number of visits (see \cref{fig:national_scale}(b), and \nameref{Methods} about the datasets details), and a second one related to the distances travelled by those individuals who visited the local area of study, during the $42$ days (see \cref{fig:national_scale}(c), and \nameref{Methods} about the datasets details). The paths in \cref{fig:national_scale}(a) show that, during the entire period of $42$ days, people arrived to the local area of study from all around Switzerland. Then, we compare the $6$ distributions of visits $-$ among themselves $-$ pairwise (see \cref{fig:national_scale}(b)) and, lastly, we compare the $6$ distributions of distances travelled, still pairwise (see \cref{fig:national_scale}(c)). 
Such comparisons represent,
at a national scale, our idea of assessment of the impacts of both social and flood events, on human mobility. 
Indeed, in one of the examined weeks, 
we would anticipate
a considerable perturbation of mobility, either in the number of visits or in the distances travelled by visitors, if the event responsible for such a perturbation were impactful enough.
For example, we might expect a strong perturbation of mobility due to a severe flood, as that one occurring at the end of the $3^{rd}$ week.
To measure such differences among weeks, both for the distributions related to the visits and the distributions related to the distances, we use a set of common statistical measures of distributional dissimilarity.
The values returned by both the Chi-square test and the one-sample Kolmogorov-Smirnov test show that the null hypothesis is rejected at the $5\%$ significance level, for each pair of estimated probability distributions $P_i$ and $P_j$ at different weeks, i.e. at $i-$week and $j-$week. This would indicate that the $6$ distributions containing the counts of visits - to the local area of study -, would be different. The same consideration would hold for the $6$ distributions representing the distances travelled by visitors to the local area of study. However, by using less strict measures of dissimilarity among distributions, the differences of mobility would appear less significant among the different weeks, implying a similar impact of social and flood events on human mobility. Indeed, the values returned by the Jensen-Shannon divergence, the Wasserstein distance, $W_1$, the Hellinger distance, the total variation distance, the $1-$Szekely's distance correlation and the $1-$Cosine similarity (please see \nameref{Methods} for details), are similar for each pair of distributions, both for the number of visits (see \cref{fig:national_scale}(d)) and the travelled distances (see \cref{fig:national_scale}(e)). 
In addition, both graphics in \cref{fig:national_scale}(d-e) show very small values for all the above-mentioned metrics. The values related to the comparison among distributions at different weeks, and indicating the number of visits (see \cref{fig:national_scale}(d)), vary from $0$ and $0.3$, while the values related to the distributions representing distances covered by visitors at different weeks, vary from $0$ and $0.1$. Values equal to $0$ would indicate identical distributions, while values equal to $1$ would represent completely dissimilar distributions. 
The graphical representation of the above-mentioned metrics showed in \cref{fig:national_scale}(d-e), would correspond to a matrix of vectors, $D=(d_{ijk})$, with $i,j=1,2,...,6$, indicating the estimated probability distributions $P_i$ and $P_j$ related to the $i-$week and $j-$week, respectively, and $k=1,2,...,6$ indicating the $k^{th}-$function of distance, or divergence, among the pair of distributions, that we can write as $f_k(P_i,P_j)$. Intuitively, the generic matrix element denotes how different $P_i$ and $P_j$ are, and it could be written as $d_{ijk}=f_k(P_i,P_j)$. In addition, the matrix elements are symmetric, i.e. $d_{ijk}=d_{jik}$ for each $k^{th}-$function, non-negative, i.e. $d_{ijk}\ge 0$, and they vanish on the diagonal, i.e. $d_{ii}=0$, making the matrix $D$ a dissimilarity matrix (of vectors) \cite{arabie1996clustering,carlsson2021topological,mathar2020fundamentals}.

\subsection*{Similarities between social events and a local flood at a local scale}
\label{similarities_local}

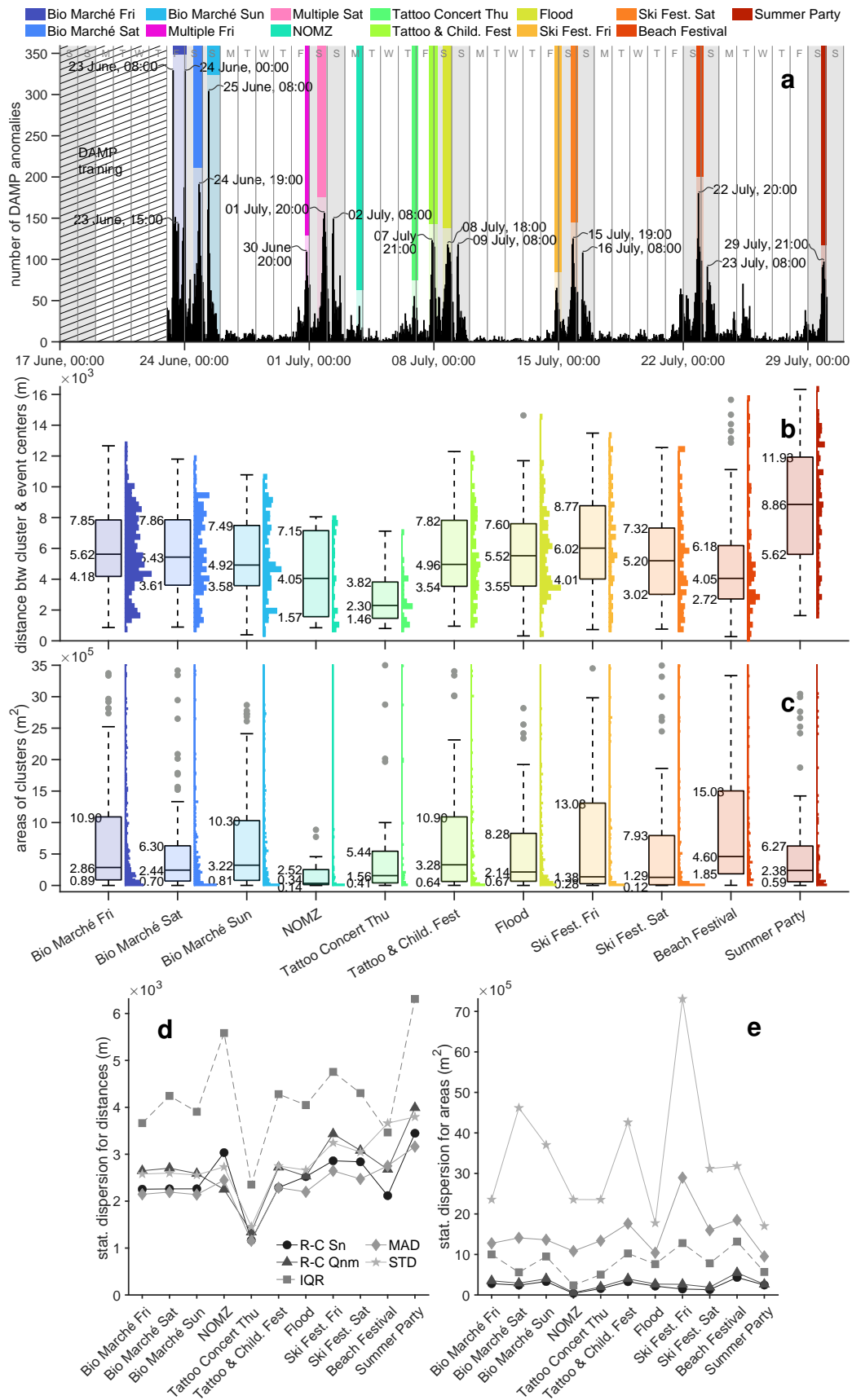
\begin{figure}[htp]
    \setlength{\lineskip}{0pt}
    \centering
    
    \begin{tikzpicture}
        \node[anchor=north west,inner sep=0pt] at (0,0)
        {\includegraphics[width=13.7cm]{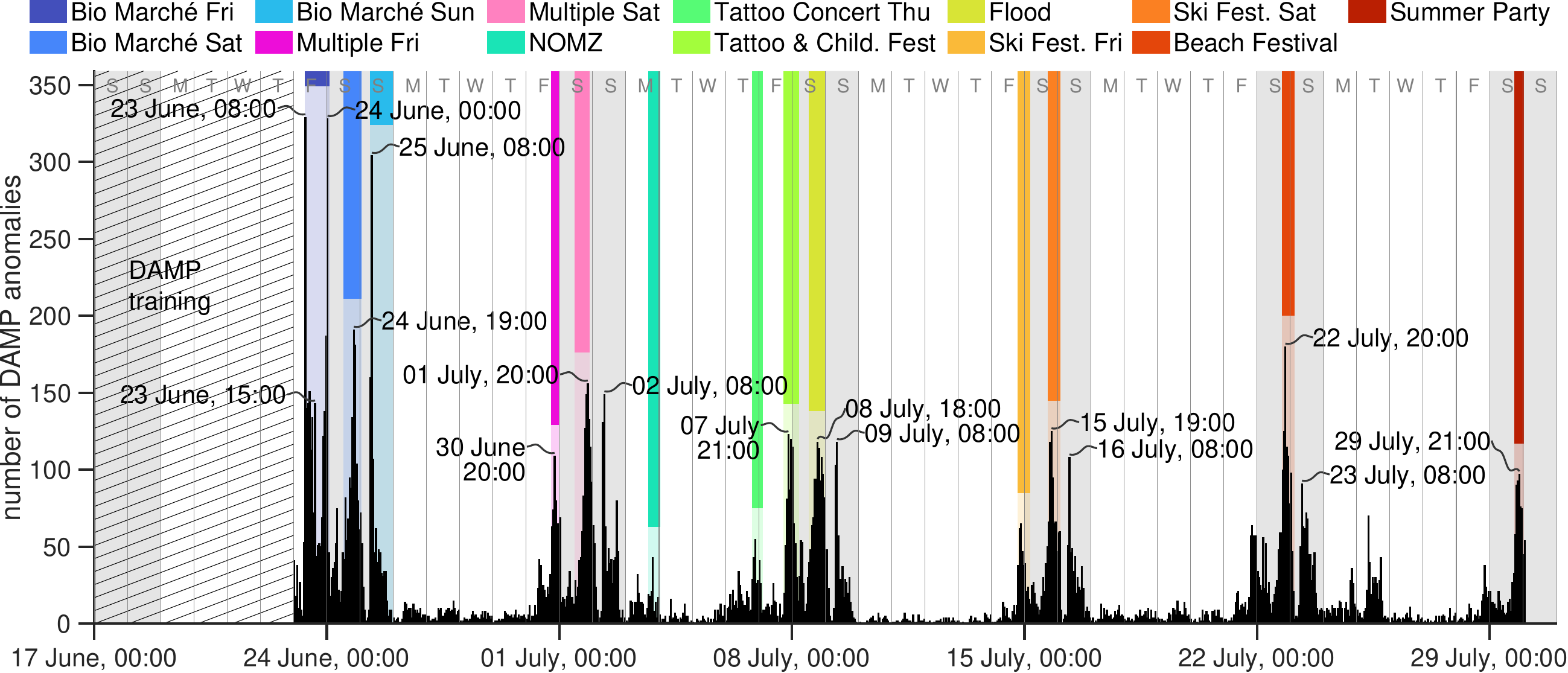}};
        \node[font=\sffamily\bfseries\large] at  (12.7cm,-1.2cm) {a};
    \end{tikzpicture}\\
    \begin{tikzpicture}
        \node[anchor=north west,inner sep=0pt] at (0,0)
        {\includegraphics[width=13.7cm]{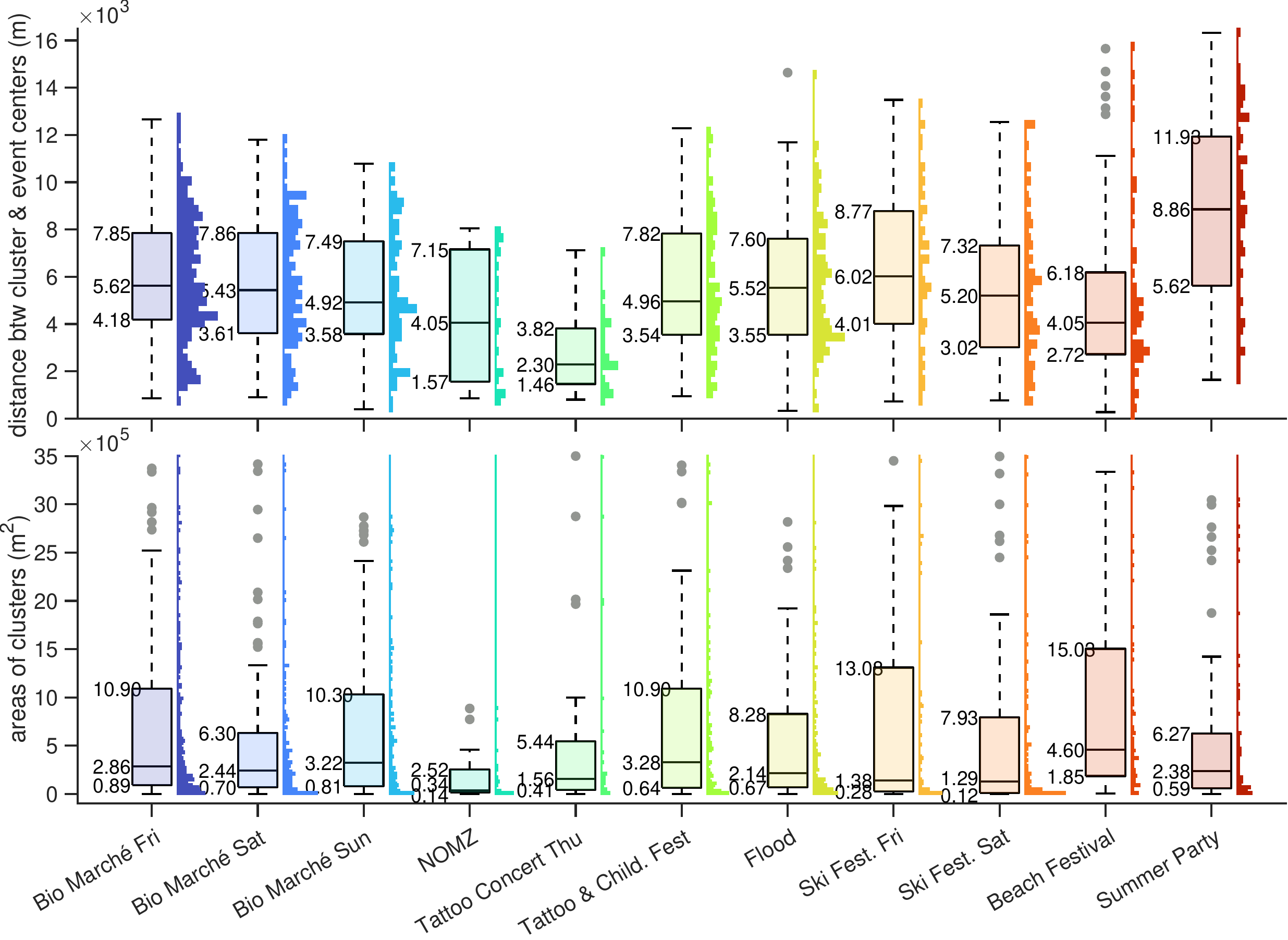}};
        \node[font=\sffamily\bfseries\large] at (12.7cm,-1cm) {b};
        \node[font=\sffamily\bfseries\large] at (12.7cm,-5.5cm) {c};
    \end{tikzpicture}\\
    \begin{tikzpicture}
        \node[anchor=north west,inner sep=0pt] at (0,0)
        {\includegraphics[width=11.4cm]{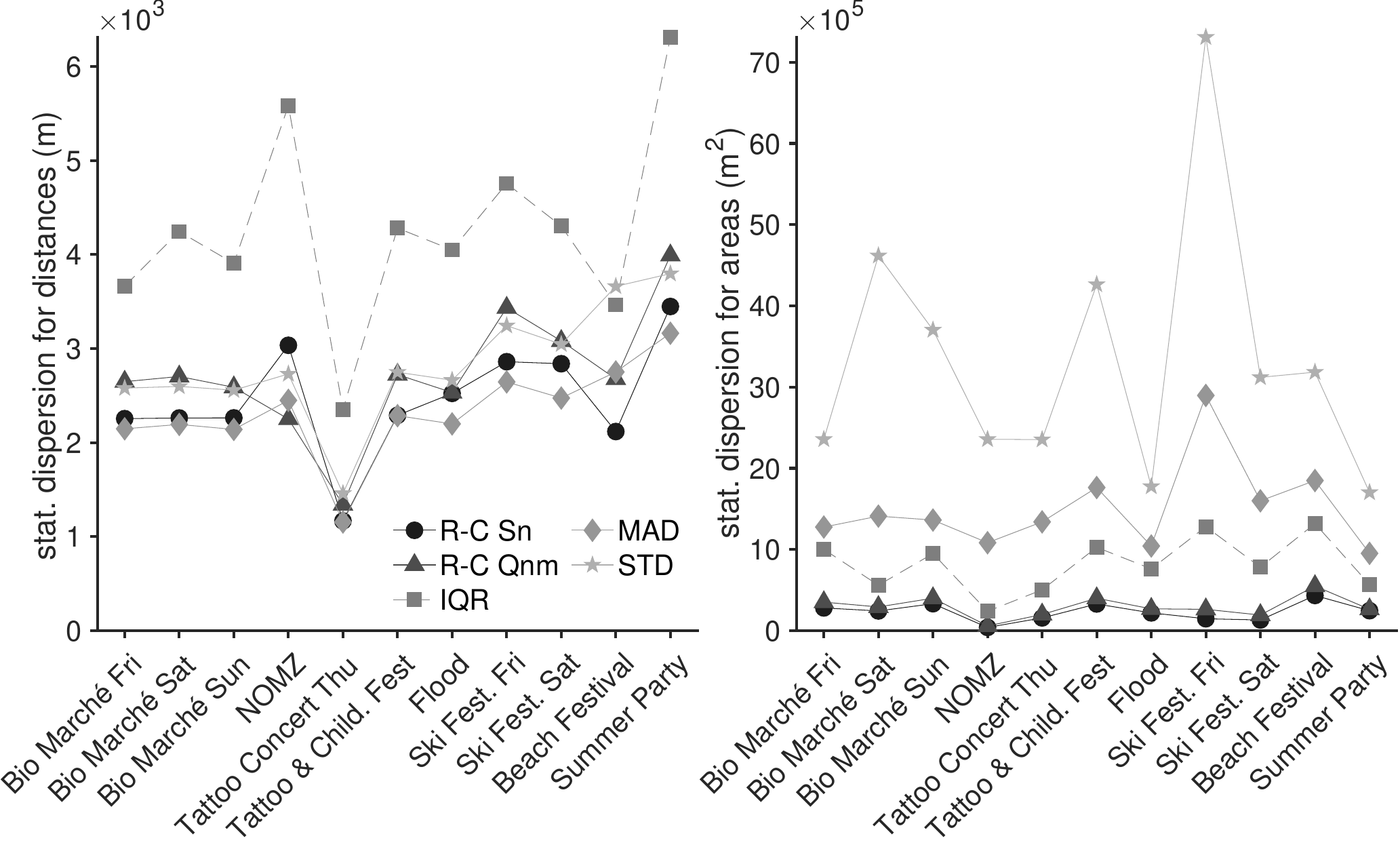}};
        \node[font=\sffamily\bfseries\large] at (1.4cm,-0.8cm) {d};
        \node[font=\sffamily\bfseries\large] at (11cm,-0.8cm) {e};
    \end{tikzpicture}
    
    \caption{\textbf{(a)} Temporal evolution of anomalies over 44 days. \textbf{(b)} Distribution of distances among clusters of anomalies and an event center, per each type of event. \textbf{(c)} Distribution of areas of clusters of anomalies, per each type of event. \textbf{(d)} Statistical dispersion of distances between the clusters of anomalies and an event center, per event type. \textbf{(e)} Statistical dispersion of cluster's areas, per event type.
    }
    \label{fig:local_scale}
    
\end{figure}

We now focus on the number of people who travelled each road or railway lines 
every hour, during the entire period of $44$ days, and within the local area of study (please see \nameref{Methods} for details about the datasets).
Those intra-town or intra-city travels typically follow the daily circadian rhythm \cite{Schneider}, leading to regular variation of the number of travellers on each road and railway track. When a disruptive event $-$ like a social or natural (loss) event
$-$ occurs, those variations might become irregular for a while, returning to the unperturbed state afterwards. If we consider the temporal evolution of the number of travellers as a \quotes{time series}, and the irregular variations of the number of travellers as \quotes{anomalies}, then, a disruptive event would cause
anomalies in a time series. 
We therefore use the anomalies discovered in travelers-related time series to measure the impact of social and flood events on human mobility, at a local scale. 
Clearly, an anomaly can represent either an excess of travelers (e.g. overflow of traffic) or a dearth of people (e.g. empty roads due to road closures or disruptions), but in this manuscript we do not deal with the two types of anomalies separately, i.e. we do not specify if an anomaly is related to a surplus or to a lack of travelers. Indeed, we are interested in the overall patterns of anomalies, without a distinction between the different types of anomalies.
Among the most relevant events that we found within the local area of study and throughout the $44$ days, seven of them are distinct social events, named Aargau Cantonal Shooting Festival (ACSF), Bio Marché, New Orleans Meets in Zofingen (NOMZ), Tattoo concert \& Children Festival, Ski Festival, Beach Festival and Summer Party, and only one of the largest events is a distinct natural event,
i.e. the flood of the $8^{th}$ of June.
With \quotes{distinct events} we mean events that occured singularly, without any temporal overlap with other significant events. A few of the most significant events were instead \quotes{multiple events}, i.e. different major events that occurred at the same time, or with some temporal overlap (details about all the social events are contained in the \textcolor{blue}{Supplementary Table 1}). 
Among the distinct social events, ACSF is the only one that occured in $7$ different municipalities. This means that the corresponding attendees did spread out over $7$ municipalities, instead of gathering in one municipality as in the other distinct social events. Therefore, due to the large dispersion of its attendees, we do not consider ACSF in the comparison among major events, throughout this work. 
For both distinct and multiple events, we detect anomalies with the Discord Aware Matrix Profile (DAMP) algorithm, a numerical approach based on the well-recognised concepts of \quotes{discord} $-$ that defines the notion of anomaly $-$, and \quotes{matrix profile} $-$ a structure for data storage. DAMP requires some training data and the period of our times series, called subsequence length, which is calculated automatically by the algorithm (please see \nameref{Methods} for details).
We set a training period equal to $6$ days, i.e. from Saturday $17^{th}$ of June to Thursday $22^{nd}$ of June 2017, both included. This means that the anomalies start to be detected by DAMP just after the midnight between the $22^{nd}$ of July
and the $23^{rd}$ of July 2017. In addition, for each road or railway track, DAMP seeks for the largest anomalies, up to $10$, if present. 
\\
\\
We observe that the largest amount of anomalies is detected both during the course of all the major recorded events, and on Sundays in which no significant event was reported (please see \cref{fig:local_scale}(a)). Since the majority of the most significant recorded events took place over the weekends (among them, only NOMZ and the Tattoo Concert that took place on weekdays),
we can therefore say that the anomalies are concentrated around weekends, rather than on weekdays. 
Please note that this statement would be still valid even including a build-up of anomalies occurring on Tuesday $25^{th}$ of July, which origin is unknown to us (and that we did not mention so far).
Interestingly,
we note that all Sundays are characterized by a peak of anomalies in the morning, at $08$ a.m.
At least partially, the peaks on Sunday $25^{th}$ of June and on Sunday $2^{nd}$ of July are due to, respectively,
Bio Marché in Zofingen, and the School \& City Festival in Olten (which started in the morning, around $07$ a.m.).
However, no relevant documented events took place on the other Sundays, early in the morning. 
We therefore associate
those peaks of anomalies, that regularly happen at $08$ a.m. every Sunday,
to recurrent leisure activities, typical of weekends \cite{Maeder,Mohammadi,Pappalardo}. 
Hence, if every Sunday features regular leisure activities, regardless the occurrence of any event, when an event then takes place on Sunday, it would contribute to increase the peak of anomalies. This would be quite evident for the peak of anomalies observed on Sunday $25^{th}$ of June at $08$ in the morning, during Bio Marché.
This observation would suggest that a relationship exists between the number of anomalies and the number of attendees, but it would be difficult to find an accurate one, since we just have rough estimates of the number of participants to the events (and not even for all the most significant events). However, if we consider the largest local maxima in \cref{fig:local_scale}(a) and the corresponding approximate numbers of attendees in \textcolor{blue}{Supplementary Table 1}, we see
that the largest local maxima in the distribution of anomalies are directly proportional to the estimated numbers of attendees. 
Still looking at the local maxima we can notice that the height of the flood peak ($8^{th}$ of July) is very similar to the peaks' height of both the Tattoo \& Children Festival on Friday $7^{th}$ of July (with $\sim1000$ attendees) and of the Ski Festival on Saturday $15^{th}$ of July (with $\sim1500$ attendees). Also, the height of the flood peak ($8^{th}$ of July) is larger than the peaks' height of both NOMZ ($3^{rd}$ of July) and Tattoo \& Children Festival on Thursday $6 ^{th}$ of July, and smaller than all the most prominent peaks' height of Bio March\'e ($23^{rd}$-to-$25^{th}$ of June, with $\sim12000$ attendees per day), Beach Festival ($22^{th}$ of July, with $\sim5000$ attendees), and of the multiple events that occurred the $1^{st}$ of July (with $\sim6500$ attendees). Therefore, such a visual comparison among the local maxima in \cref{fig:local_scale}(a)
allow us to identify
a similarity between the impact of social events, with an estimated number of attendees between $\sim1000$ and $\sim1500$ individuals, and the impact of a local flood, on human mobility.
\\
\\
Although the sole number of anomalies might give already an indication on the similarities between the social events and the flood, it does not convey
any spatial information, except the name and border of the municipality where an event took place. 
As a first spatial information about the roads and railways' anomalies, we observe $-$ for example from the \textcolor{blue}{Supplementary Figures $1-4$} $-$ that the anomalies are scattered all over the local area of study, with varying density, and showing  different patterns every hour.
Given such variations of density, we hypothesize that a sufficiently significant event would be responsible for the occurrence of anomalies mainly in the surroundings of the same event. In other words, we expect
that anomalies group around the location where an event takes place (for example, please see in
\textcolor{blue}{Supplementary Figure 2}, in the time range between $6$ p.m. of Thursday $6^{th}$ of July and 1 a.m. of Friday $7^{th}$ of July, and in
\textcolor{blue}{Supplementary Figure 3}, between $7$ and $8$ p.m. of Saturday $22^{nd}$)
Since in data mining and machine learning the groups of data-points in (high-dimensional) data spaces are called \quotes{clusters} $-$ which are detected with techniques of data clustering $-$ \cite{Aggarwal2013}, we define the collections of nodes that represent the groups of roads and railway tracks with anomalies as \quotes{clusters}.
At the same time, roads and railways tracks represent the edges of the corresponding transport networks, and 
the word \quotes{community} could be used here to indicate a group of edges
that are \quotes{densely connected} within the community and \quotes{sparsely connected} to the other communities \cite{Newman2018}.
This would be a common terminology in network science, where
a group of nodes is called community if those nodes are more
densely connected within the community than with the rest of the network \cite{Newman2018}.
But then, can we call
those groups of roads or railway tracks with anomalies as communities, if densely connected? Still from \textcolor{blue}{Supplementary Figures $1-4$}, we notice that some edges with anomalies are relatively far from the rest of the edges with anomalies, and they could be then considered as \quotes{outliers}. Indeed, those isolated anomalies could be caused by a single road disruption, or by a single technical issue on the railway line, and not related to any major event. Also, we can notice that relatively close edges with anomalies, might not have common nodes, but the corresponding anomalies could have been originated by the same event. The latter observation would mean that those edges with anomalies might belong to the same community (i.e. resulting from the same event), but being internally disconnected within it, i.e. a part of the community could be reached only through a path outside the community. 
Although there is no universally accepted
definition of community in network science \cite{Fortunato2010,Fortunato2022}, the connectedness among nodes is considered as a basic requirement to define a community \cite{Fortunato2010}, and internally disconnected communities are generally deemed to be bad partitions of a network \cite{Traag,Wolf}. 
Therefore, we could use the noun community if we extend its definition \cite{Fortunato2022} to those ones that might be internally disconnected, or if
we just consider definitions of community based on node similarity \cite{Fortunato2010}.
In this  way, the terms \quotes{cluster} and \quotes{community} might be used interchangeably.
\\
\\
By following our observations on the detected anomalies positions, we use RNN-DBSCAN \cite{Bryant} $-$ a data clustering algorithm $-$ as a density-based community detection technique \cite{bedi,Rostami}.
Indeed, RNN-DBSCAN can detect disconnected communities of arbitrary shapes and various densities, and filter noise, as those isolated edges with anomalies that would unlikely result from major events \cite{Wang2018_denoise,Jia2017_noise}. 
Once the clusters are detected, we can assess our hypothesis (that anomalies occur around the events) for every event, by measuring the distances between the clusters centers and the event center.
A cluster center might be calculated by simply averaging the coordinates of the cluster's data points. However, the resulting center would be influenced by the number of data points, and would be driven towards denser areas of data points. Thus, we draw a tight 2-D boundary \cite{MATLABboundary} that surrounds the cluster's data points, and we calculate the centroid, or center of mass, of the resulting polygon \cite{Waller_Gotway}. 
A centroid is
indeed less influenced by different densities of the cluster's data points \cite{Waller_Gotway}. For each event, the resulting distribution of the distances among the clusters centers and the event center is shown in \cref{fig:local_scale}(b), while the distribution of the clusters areas (i.e. the polygon areas) is illustrated in \cref{fig:local_scale}(c). 
Please note that some of the events in \cref{fig:local_scale}(b-c) took place over two or three days, and, therefore, the number of distributions is larger than the number of distinct events.
Alongside the distributions we show the corresponding boxplots,
which provide descriptive statistics and facilitate a visual comparison among (the distributions of) different events.
Probably, the first thing that strikes the reader about the boxplots in \cref{fig:local_scale}(b) (related to the distributions of the distances) is that 
$7$ out of $11$ boxplots are remarkably similar to each other, i.e. those ones related to Bio March\'e (all days), Tattoo concert \& Children Festival (on Friday the $7^{th}$ of July), the flood, and to the Ski Festival (both days).
We used the expression \quotes{remarkably similar} since the boxplots look very similar to each other, despite the great difference in the estimated number of attendees who participated in those events. Indeed,  $\sim 12000$ people were estimated for each day of Bio Marché, $\sim 1000$ people for the Tattoo concert \& Children Festival (on Friday the $7^{th}$ of July) and $\sim 2000$ people for the Ski Festival. But, how could events of different magnitude (i.e. expressed through the number of attendees) lead to similar spatial dispersions of anomalies (i.e. similar boxplots)?
We can exclude
any reason related to the process of clustering the edges with anomalies through RNN-DBSCAN, 
since the boxplots in \cref{fig:local_scale}(b) are very similar to the boxplots in
\textcolor{blue}{Supplementary Figure 5} (related to the distributions of distances among the edges with anomalies and the event centers, i.e. before any usage of RNN-DBSCAN).
We then suppose
that the spatial dispersion of anomalies might depend on the network topology, since it is the same for all the events, rather than on the number of attendees. In particular, all the events with similar boxplots, took place in the same town (Zofingen), except the Ski Festival which was held in a town around $5$ km far, in a beeline (Rothrist).
\\
\\
Still comparing the boxplots in \cref{fig:local_scale}(b) we notice that the boxes representing the interquartile ranges of NOMZ and the beach festival,  are slightly lower than the previous ones. However, we might still include those two events in the previous list of similar events (i.e. Bio Marché, Tattoo concert \& Children Festival on Friday the $7^{th}$ of July, flood, and Ski Festival).
Indeed, if we use the \quotes{average median} calculated over the initially mentioned $7$ boxplots, i.e. $\left \langle Q_2 \right \rangle \approx 5.38$ km, as one of the reference measures to uniquely characterize all the similar events, the medians related to NOMZ and the beach festival, i.e. $Q_2 \approx 4.05$, can be still considered reasonably close to the average median.
The boxplots related to the Tattoo concert (on Thursday the $6^{th}$ of July) and to the Summer Party, instead, look quite dissimilar to the previously mentioned boxplots, and their medians, which are $Q_2 \approx 2.30$ km and $Q_2 \approx 8.86$ km, respectively, are farther from the average median, than the NOMZ and beach festival's medians.
About the dissimilarity of the boxplot related to the Summer Party, we notice that a large number of anomalies detected during that event are quite far from the site where that event took place. The large distances between the edges (or clusters) of anomalies and the event location is then reflected in the values of the boxplot which are larger than all the other boxplots values.
The dissimilarity of the Summer Party's boxplot might be due to the event size ($\sim 500$ attendees) which was not sufficiently large for producing a considerable number of anomalies in the surroundings of the event, and at the same time, another event might have occurred, about which, however, we did not find any record. 
Overall, the boxplot associated to the Tattoo concert (on Thursday the $6^{th}$ of July) is the \quotes{lowest} one in \cref{fig:local_scale}(b), i.e. it has the smallest boxplot's summary statistics (except the minimum) than the other boxplots. This case would support our hypothesis about the occurrence of
anomalies mainly in the surroundings of an event. However, why aren't the other boxplots closer to distance zero as the boxplot related to the Tattoo concert? Or, other way around, if we take the $7$ similar boxplots as reference, why does the Tattoo concert's boxplot is considerably lower than them?
We know that the number of trips (i.e. trip frequency) is usually higher during weekdays than during weekends \cite{Agarwal2004,Charlton2002,Lockwood2005,OFallon2003} (with some exception \cite{Raux2016}), and that weekday trips are generally shorter \cite{OFallon2003,kagho2022,Lockwood2005,Yuan2012} and more regular \cite{schlich2001,Schlich2003,Buliung2008,Susilo2005,McInerney2012,VanAcker2010} than weekend trips. Despite there is some discordant results on trips regularities \cite{Yuan2012,Song2010}, we think that the relative stability and the short distances of weekday trips limit the occurrence of 
irregularities, i.e. the anomalies.
Therefore, given a less occurrence of anomalies during the weekdays, when a significant weekdays event takes place, most of the anomalies would be originated by that event. This means that, generally, a weekdays event would be more distinguishable $-$ i.e. showing a clear(er) grouping of anomalies around the event $-$ than other events occurring during weekends.
This is the case of the Tattoo concert (on Thursday the $6^{th}$ of July), which is the only major event that took place on a weekday not belonging to \quotes{long weekends} (i.e. Fridays and Mondays). Here, we exclude
Fridays and Mondays, since the Monday travel behaviour might be similar to the weekend one \cite{Buliung2008}, and the patterns of activity-travel behaviour on Fridays could be dissimilar to the behaviour of other weekdays \cite{McInerney2012,Jiang2012,kumar1996}.
Now, if an event that takes place on weekdays shows clearer patterns of anomalies around and close to the event due to the stability of weekdays trips, once an event occurs during the weekend or in the run-up to the weekend (or either side of the weekend), the clusters of anomalies originated by the major event might be partially overlapping or mixed with 
the clusters of anomalies caused by irregular and long-distance trips, not necessarily related to the event. 
Therefore, the spatial distribution of the (clusters of) anomalies would include the irregular travel behaviour, occurring potentially anywhere. Consequently, once we show the spatial distribution of anomalies related to a major weekend event, this would be larger than the distribution of anomalies related to a weekday major event. However, despite a possible mix of anomalies resulting from major events and anomalies given by unpredictable weekend trips, in \textcolor{blue}{Supplementary Table 2} we notice that
almost all the spatial distributions of anomalies' distances
(in \cref{fig:local_scale}(b)) 
show a general tendency to skew towards the event centers (i.e. positive skewness).
This is another commonality between the flood and social events.
Indeed, the Bowley’s coefficient of skewness (BS), and the Fisher-Pearson coefficient of skewness (FPS) in \textcolor{blue}{Supplementary Table 2} are positive for all the events, with the only exceptions of the Ski Festival
on Saturday the $15^{th}$ of July, where BS and FPS have discordant sign, and the Summer Party, where both BS and FPS are negative. Perhaps, such negative signs for both the BS and FPS might be interpreted as a lack of connection between the detected anomalies and the Summer Party. 
The reason why we use both BS and FPS is that, often, (strongly) skewed distributions have skewed boxes. Indeed, the Bowley’s coefficient assesses the skewness of the central box, while the Fisher-Pearson coefficient quantifies the skewness of the entire distribution.
However, the skewness should generally refer to the entire distribution, which would bring
our attention primarily to the Fisher-Pearson coefficient (even though the Bowley's skewness is preferable
in presence of outliers \cite{Kim2004}).
As possible qualitative descriptions, some authors  \cite{Bulmer1979,Hair2017} suggest that distributions need to have a $\left| \text{FPS}\right| \geq 1$ 
to be considered significantly skewed. Therefore, we can say that only the Tattoo concert (on Thursday the $6^{th}$ of July) and the Beach Festival are fairly
skewed (with FPS $\approx 0.97$ and FPS $\approx 1.25$ respectively), while the flood and the Ski Festival on Saturday $15^{th}$ of July are instead moderately skewed (with FPS $\approx 0.54$ and FPS $\approx 0.56$, respectively). 
All the other distributions show a light positive skewness, with $0.05 \leq \text{FPS} \leq 0.39$ (excluding the Summer Party). However, values of FPS less than $0.5$ are often associated to approximately symmetric distributions.
\\
\\
While the distributions of the distances exhibit a light-to-moderate departure from symmetry (or they are approximately symmetric), 
all the distributions of the clusters areas $-$ including the flood-related one $-$ show a very high positive skewness, with a large concentration of values below the median \cref{fig:local_scale}(c) (please see \textcolor{blue}{Supplementary Table 2} for the exact values of BS and FPS). 
This means that (at least) half of the detected clusters are small or very small, in comparison to the entire range of values included in the boxplots.
If we look at the medians of all the events, we can see that the approximate average median and the largest one would be, respectively, $\approx 0.21$ km$^{2}$ and $\approx 0.33$ km$^{2}$, if we do not include the Beach Festival (otherwise they would be, respectively, $0.23$ km$^{2}$ and $0.46$ km$^{2}$). Indeed, although the difference between the second quartile (i.e. median) and the first quartile in the Beach Festival is similar to other events, the first quartile of the Beach festival is the largest one among all the events ($Q_1 \approx 0.18$ km$^{2}$), and it is around double of the second largest first quartile, which belongs to the Bio Marché (on Friday $23^{th}$ of June). Therefore, the clusters of anomalies in the lowest $25\%$ of the Beach festival's distribution, reach at least twice the size of clusters in the lowest $25\%$ of other events' distributions. In other words, 
the lowest $25\%$ of the Beach festival's distribution shows
a larger range of cluster sizes than in the lowest $25\%$ of other events' distributions, where the clusters have
instead a similar size among each other. Another feature that stands out in \cref{fig:local_scale}(c) $-$ and that demans some attention $-$ is the large quantity of boxplots outliers. Indeed, although some outliers are associated to proper and correct assessments of the polygon sizes, other ones might be related to overestimates of the polygon sizes, and should therefore be interpreted with caution. The latter case occurs when long network's edges, i.e. railways or motorways, have anomalies, and they are included into polygonal shapes, i.e. the clusters of anomalies (from which we calculate the clusters areas). Basically, the long edges stretch the polygons towards themselves, leading to an increase of the polygon area. 
\\
\\
Although the boxplots in \cref{fig:local_scale}(b) and \cref{fig:local_scale}(c) provide a synthetic summary of the distributions of the clusters distances and areas, they do not directly represent variance or other measures of data variability. It is indeed a common practice to report mean and standard deviation for symmetric or approximate symmetric distributions, as in \cref{fig:local_scale}(b), and median and interquartile range for skewed distributions, as in \cref{fig:local_scale}(c) \cite{Cooksey2020,Lock2021}. In addition, so far, our evidence of similarity among the boxplots in \cref{fig:local_scale}(b) was mainly based on a visual inspection, which could be less robust than a numeric comparison. We therefore calculate and compare a set of measures of dispersion, both based on the mean, i.e. standard deviation (STD) and mean absolute deviation (MeanAD), and based on the median, i.e. interquartile range (IQR) and Rousseeuw-Croux scale estimators ($S_n$ and $Q_n$).
If we focus on \cref{fig:local_scale}(d), we notice that our previous observation related to the similarity among $7$ out of $11$ boxplots $-$ i.e. those ones associated to Bio March\'e (all days), Tattoo concert \& Children Festival (on Friday the $7^{th}$ of July), the flood, and to the Ski Festival (both days) $-$, is still fairly valid regardless the measure of dispersion used, and despite the slightly larger values of the Ski Festival (both days). 
Also, we mentioned that the boxplots related to NOMZ and the Beach Festival might still be considered similar to the previous $7$ boxplots, in terms of dispersion. This is quite true for NOMZ, if we consider all the measures of dispersion, except the interquartile range. About the Beach Festival, the answer depends on the measure of dispersion. If we do not consider both the interquartile range and the standard deviation, the Beach Festival shows a spatial dispersion similar to the previously mentioned $7$ events. In addition, the standard deviation is here large due to presence of outliers on the right tail (please see \cref{fig:local_scale}(b)).
About the remaining two events, i.e. the Tattoo concert (on Thursday the $6^{th}$ of July) and the Summer Party, they still look dissimilar to the previous events for any measure of dispersion (and not just for the interquartile range), showing, respectively, significantly lower and rather higher values of dispersion. Therefore, if we use any of the proposed measures of dispersion, except the interquartile range (that yields the largest excursion of values among all the measures of dispersion),
we can still fairly assert that $9$ out of $11$ events are approximately similar to each other, in terms of spatial variability of anomalies.
But then, should we avoid any usage of the interquartile range in \cref{fig:local_scale}(b)?
Although there is nothing stopping us from employing the interquartile range, this measure of dispersion is not generally applied to approximate symmetric or light-to-moderate skewed distributions as in \cref{fig:local_scale}(b), since it does not describe the entire distribution, as the standard deviation or the mean absolute deviation would do, but only the central half of the distribution.
The usage of the interquartile range is instead advised in cases of skewed distributions, as in \cref{fig:local_scale}(c). In particular, with the presence of outliers, which significantly affect the mean and all the mean-based measures of dispersion (in this case, the standard deviation and the mean absolute deviation). Indeed, it is the standard deviation that yields the largest (and the broadest range of) value(s) in \cref{fig:local_scale}(e), followed by the mean absolute deviation. On the contrary, the interquartile range, and in particular the Rousseeuw-Croux scale estimators, exhibit more robustness against the outliers, and thus less variations among the events. 
The fact that limited variations 
occur across the different events, might indicate a certain degree of similarity among the events (as between the flood and the social events), in terms of variability of the clusters areas (in particular the values provided by the Rousseeuw-Croux scale estimators). 
Interestingly enough, the Rousseeuw-Croux scale estimators yield values of dispersion which are very similar to the standard deviation and the mean absolute deviation ones in \cref{fig:local_scale}(d), and, as we have just mentioned, they provide some of the most robust values of dispersion in \cref{fig:local_scale}(e). This means that Rousseeuw-Croux scale estimators might be used as common tool for both types of dispersion, i.e. the dispersion of distances and the dispersion of the clusters areas.

\subsection*{Empirical relationships among the clusters of anomalies}
\label{empirical_relationships}

\begin{figure}[htp]
    \setlength{\lineskip}{0pt}
    \centering
    
    \begin{tikzpicture}
        \node[anchor=north west,inner sep=0pt] at (0,0){\includegraphics[width=12cm]{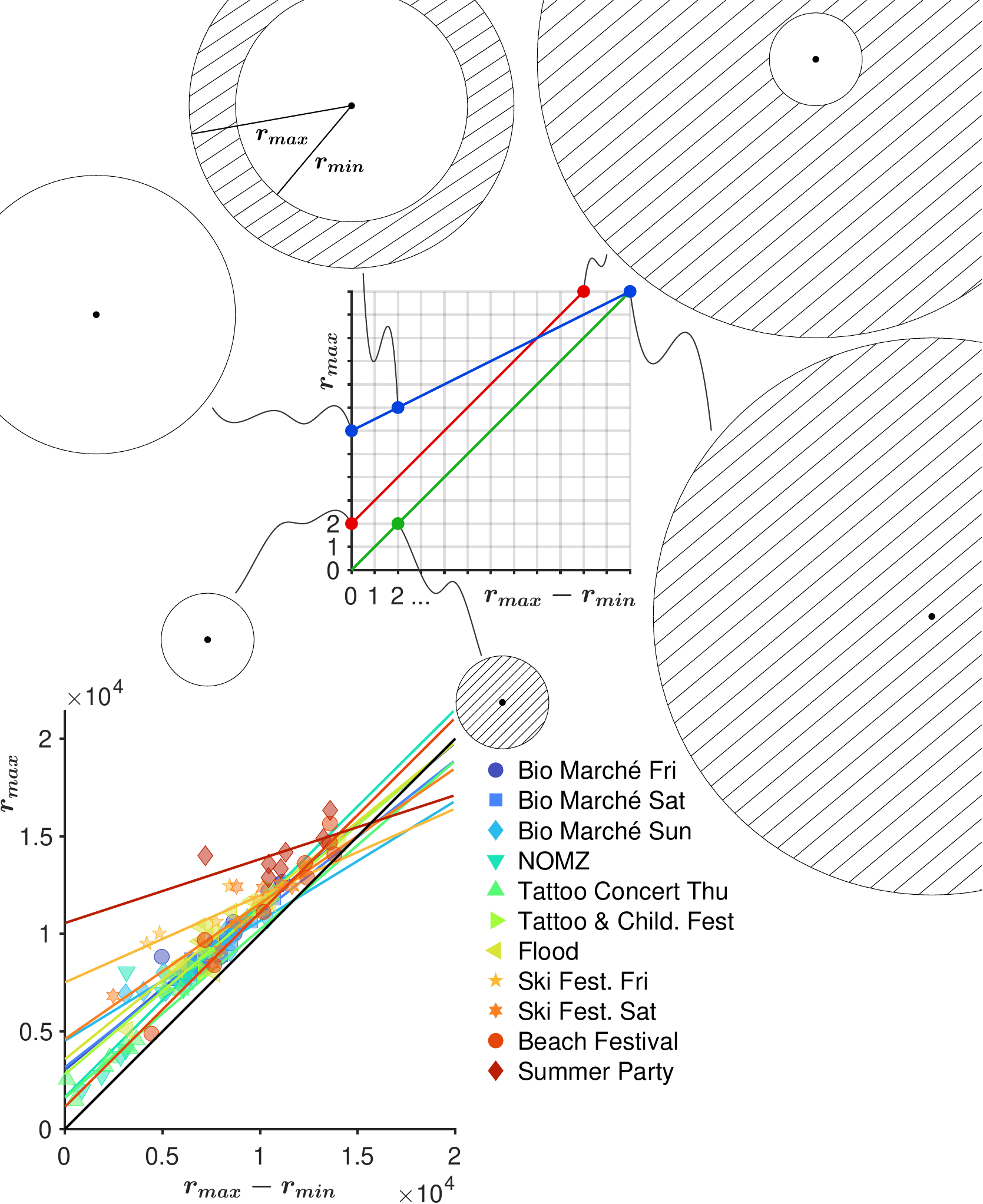}};
        \node[font=\sffamily\bfseries\large] at (1.4cm,-0.8cm) {a};
        \node[font=\sffamily\bfseries\large] at (1.4cm,-8.9cm) {b};
    \end{tikzpicture}\\
    \begin{tikzpicture}
        \node[anchor=north west,inner sep=0pt] at (0,0){\includegraphics[width=12cm]{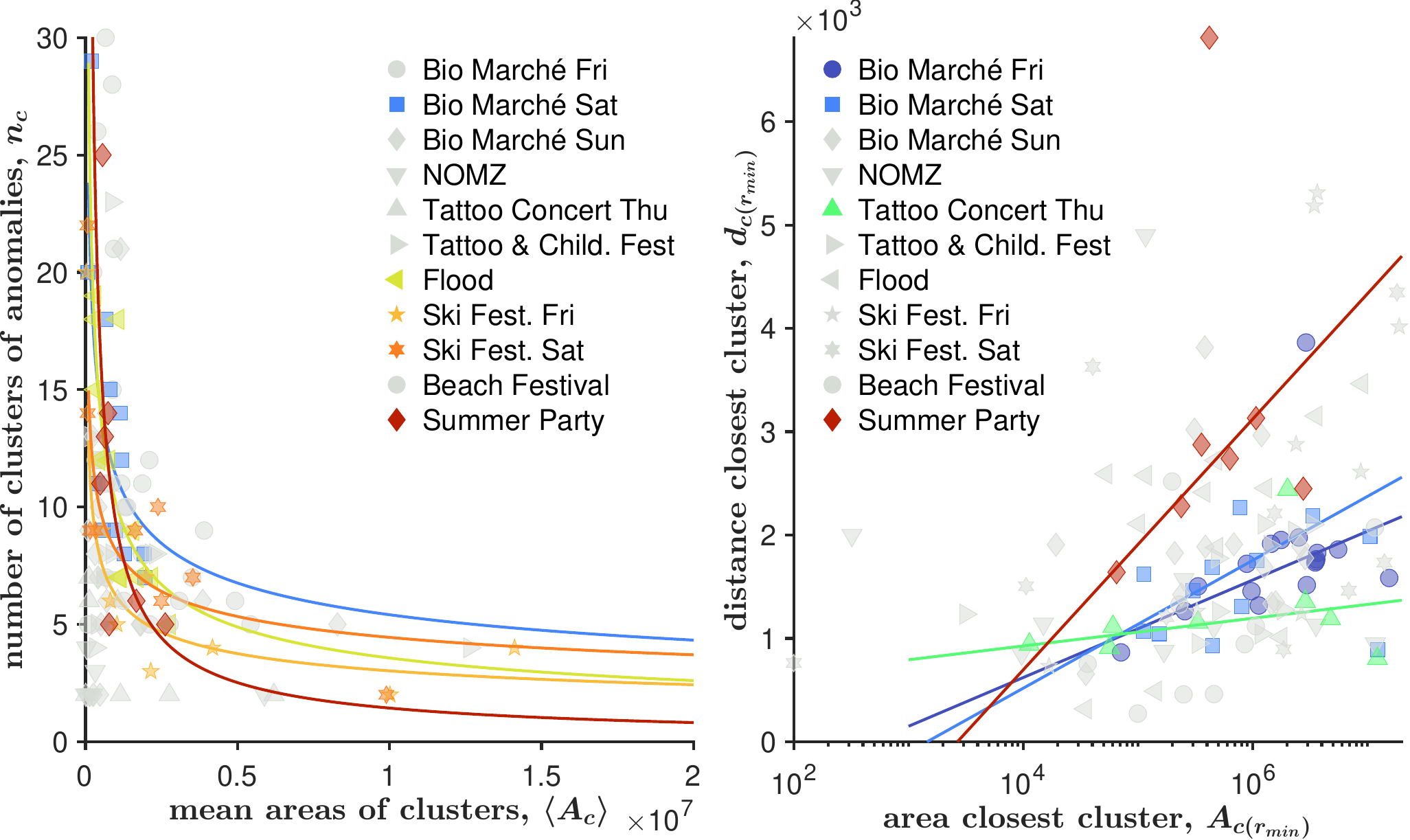}};
        \node[font=\sffamily\bfseries\large] at (1.4cm,-0.8cm) {c};
        \node[font=\sffamily\bfseries\large] at (11cm,-0.8cm) {d};
    \end{tikzpicture}
    
    \caption{\textbf{(a)} Graphical explanation of the relationship between the largest and the smallest distances among clusters of anomalies and an event center. \textbf{(b)} Largest distance between a cluster of anomalies and an event center, against the difference between the largest and the smallest distances among clusters of anomalies and an event center.  \textbf{(c)} Number of clusters of anomalies versus the mean value of the areas of the clusters of anomalies. \textbf{(d)} Distance of the closest cluster of anomalies to an event center against the area of the same cluster of anomalies.
    }
    \label{fig:empirical_relation}

\end{figure}
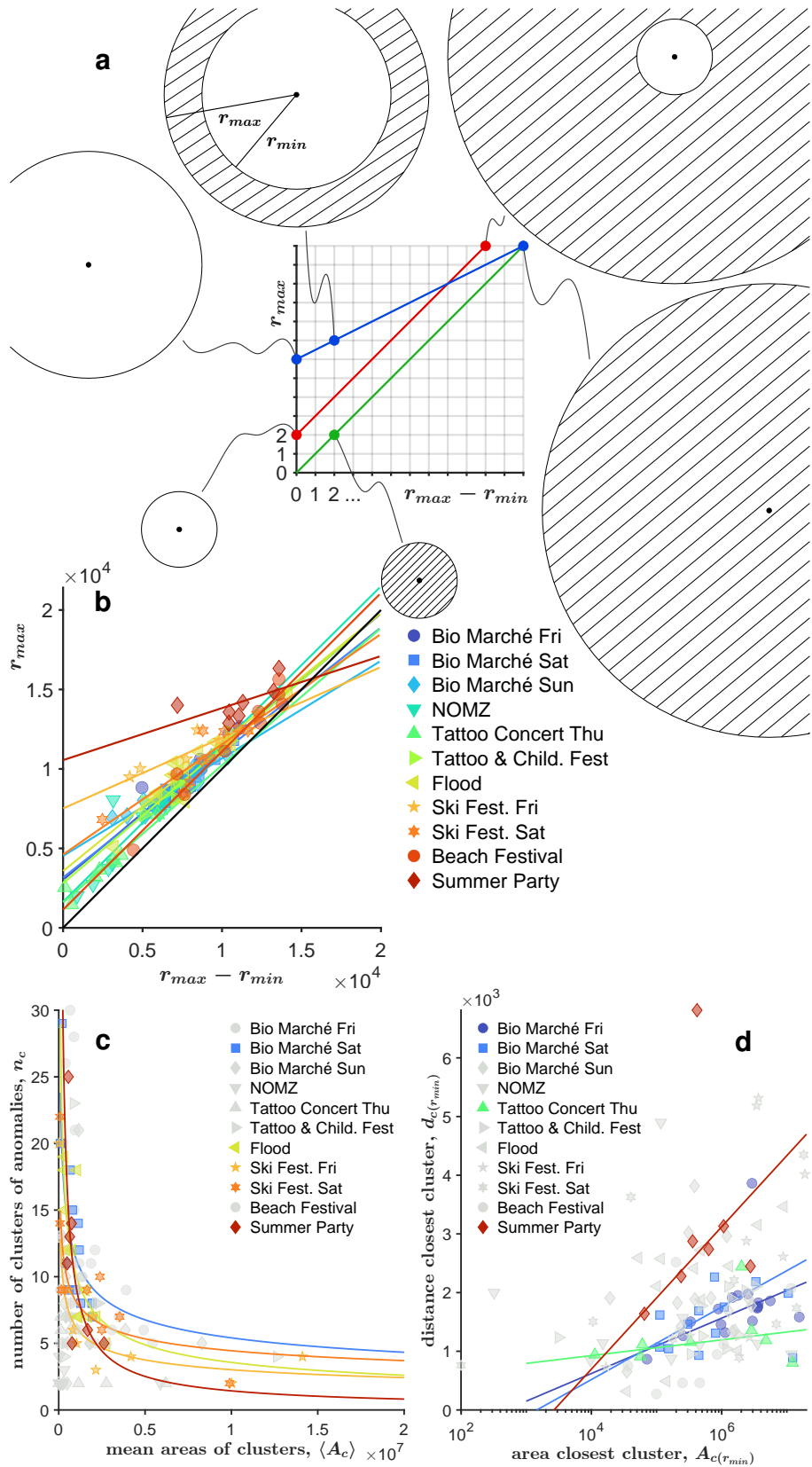

The distributions in \cref{fig:local_scale}(b) and \cref{fig:local_scale}(c) provide us a coarse-grained description of the events, since they include, respectively, all the distances and all the areas measured during the entire events, i.e. from the starting hour to the ending hour of the event. With this approach we therefore loose the details related to every single hour (in any event), that might be useful to identify possible relationships among distances, areas, and number of clusters. Indeed, for (every event and) every hour, we know the number of clusters detected by RNN-DBSCAN, as well as their distances from the event center, and their areas.
We can then extend our local scale-related analysis by using these hourly-based data as fine-grained information. 
If we consider for instance the distances between the event center and the farthest and closest clusters, i.e. $r_{\text{max}}$ and $r_{\text{min}}$, respectively, we observe that $r_{\text{max}}$ and $\Delta r=r_{\text{max}}-r_{\text{min}}$ are connected by a a linear relationship (please see \cref{fig:empirical_relation}(a-b)):
\begin{equation}
r_{\text{max}} = a_1 \Delta r + b_1
\label{eq:r_max_delta_r}
\end{equation}
where $a_1$ and $b_1$ are listed in \textcolor{blue}{Supplementary Table 3}, for each event. 
Although \cref{eq:r_max_delta_r} does not show a direct relationship between $r_{\text{max}}$ and $r_{\text{min}}$ (since we prefer to use $\Delta r$ instead of $r_{\text{min}}$, for reasons of clarity), we can always transform it into $r_{\text{max}} = \left(\frac{a_1}{a_1-1}\right) r_{\text{min}} - \left(\frac{b_1}{a_1-1}\right)$, but bearing in mind that it would be undefined for $a_1=1$ (i.e. we can use it only when $a_1\neq1$). 
By consulting \textcolor{blue}{Supplementary Table 3}, we notice that both slopes corresponding to NOMZ and the Beach Festival are $a_1\approx 1$. This means that $-$ for both social gatherings and their entire duration $-$ the closest cluster has a constant distance from the event center, $r_{\text{min}} = b_1$ (see the red line in \cref{fig:empirical_relation}(a)). 
All the other slopes, instead, are
smaller than the NOMZ and the Beach Festival's ones, being $0.45 \lessapprox a_1 \lessapprox 1$ (here, we exclude that one related to the Summer Party, due to the low coefficient of determination in \cref{eq:r_max_delta_r}, i.e. $R^2\approx0.37$). In this case, if we read from left to right the \cref{fig:empirical_relation}(a-b), not just the farthest cluster goes farther, but the closest cluster gets closer to the event center as well (see the blue line in \cref{fig:empirical_relation}(a)). 
In other words, the space occupied by all the clusters, and identified 
by $\Delta r$, increases radially towards both directions, i.e. inwards and outwards.
And, this behaviour is common to the vast majority of the events here presented, flood included.
Obviously, the upper bound for any cluster's distance is given by the perimeter of the local area of study, while the lower bound is zero, and it is reached when the cluster center overlaps the event center (see the black line in \cref{fig:empirical_relation}(a-b), i.e. any point on the $r_{\text{max}}=\Delta r$ line, also called the  $r_{\text{min}}=0$ line).
\\
\\
About the number of clusters, $n_c$, we observe that it decreases with the increase of the mean area of the clusters (averaged every hour), $\langle A_c \rangle$:
\begin{equation}
    n_c \sim \frac{1}{\langle A_c \rangle}.
\end{equation}
which might be somehow obvious. However, the way the number of clusters decreases might not be so trivial.
We see that some events exhibit a power law, $n_c=a_2\langle A_c \rangle^{b_2}$, where the exponent ranges around $b_2\approx -0.3$ (more precisely between $-0.45 \lessapprox b_2 \lessapprox -0.26$) for those events that have a relatively high coefficient of determination (i.e. Bio Marché, on Saturday $24^{th}$ of June,
with $R^2\approx 0.6$, Flood with $R^2\approx 0.65$, and Ski Festival, both days, with $R^2\approx 0.94$ and $R^2\approx 0.68$, respectively).
Still focusing on the areas of the clusters, we also observe a mild direct proportionality among the distance of the closest cluster, $d_{c\left(r_{\text{min}}\right)}$, and its area, $A_{c\left(r_{\text{min}}\right)}$:
\begin{equation}
    d_{c(r_{min})} \sim log(A_{c(r_{min})})
\end{equation}
In particular for Bio Marché (both Friday $23^{rd}$ and Saturday $24^{th}$ of June), the Tattoo \& Children Festival (on Thursday $6 ^{th}$ of July) and the Summer Party (on Saturday $29^{th}$ of July), we can write a logarithmic relationship, i.e. $d_{c(r_{min})} = a_3 log(A_{c(r_{min})}) + b_3$, where the coefficients are shown in the \textcolor{blue}{Supplementary Table 3}.

\subsection*{Anomalies in the framework of multi-layer transportation networks}
\label{anomalies_multi_layer} 

\begin{figure}[htp]
    \setlength{\lineskip}{0pt}
    \centering
    
    \begin{tikzpicture}
        \node[anchor=north west,inner sep=0pt] at (0,0){\includegraphics[width=14cm]{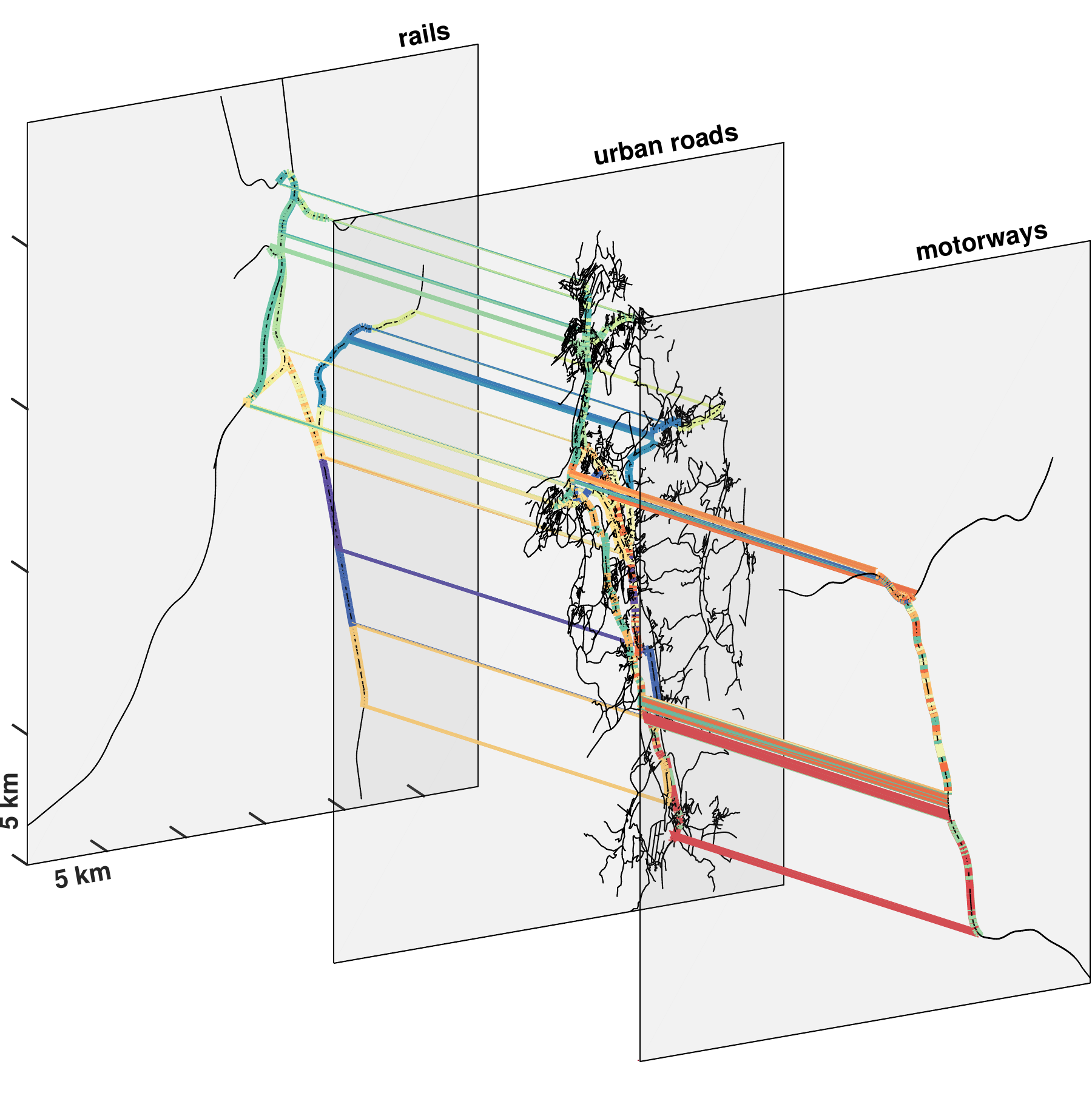}};
        \node[font=\sffamily\bfseries\large] at (12.4cm,-0.8cm) {a};
    \end{tikzpicture}\\
    \begin{tikzpicture}
        \node[anchor=north west,inner sep=0pt] at (0,0){\includegraphics[width=10cm]{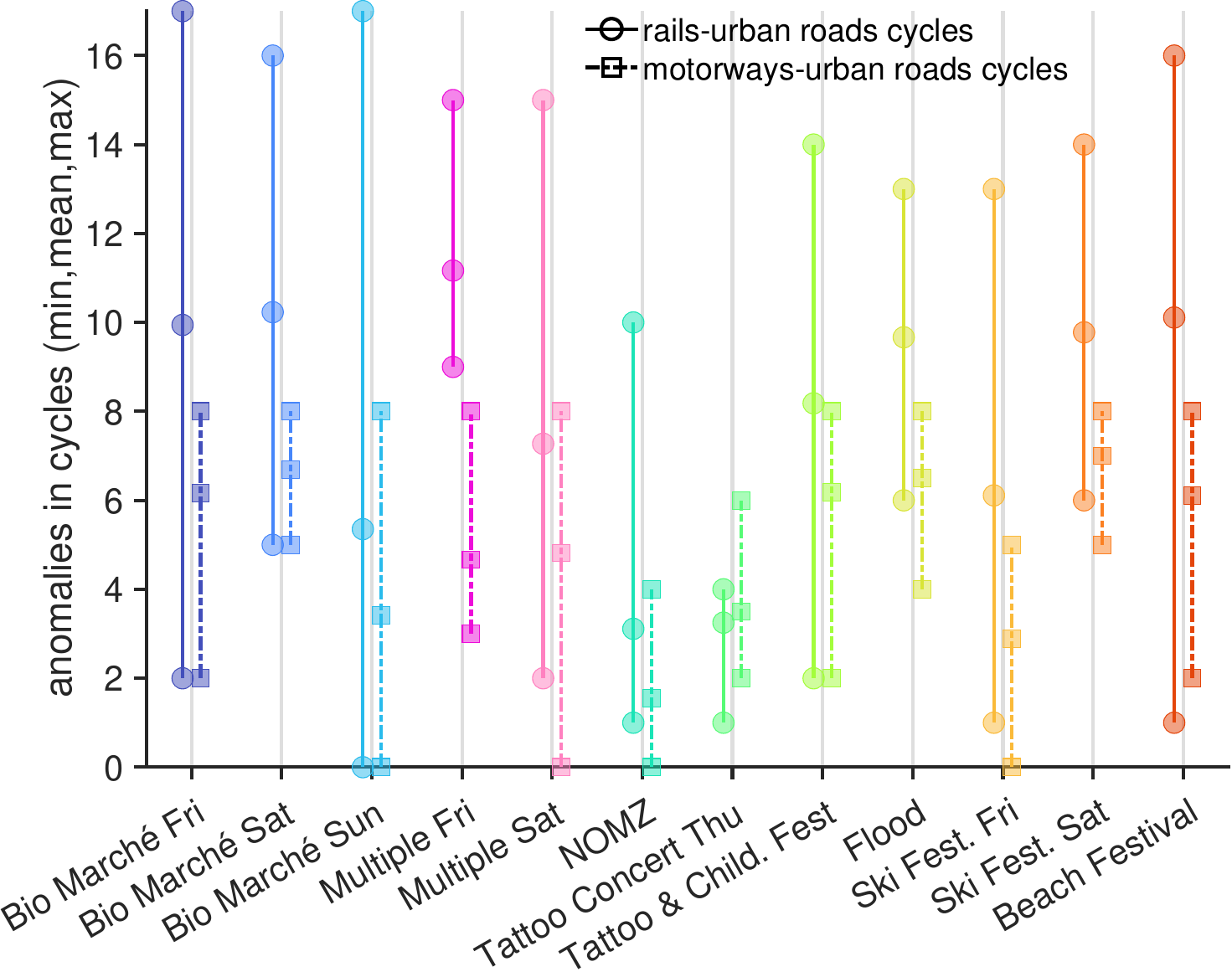}};
        \node[font=\sffamily\bfseries\large] at (11cm,-0.8cm) {b};
    \end{tikzpicture}
    
    \caption{\textbf{(a)} Multi-layer network representing the local area of study. The network includes three layers: the urban roads layer, the motorways layer and the rails layer. The coloured edges represent the unique cycles of the multi-layer network. Please notice that a number of edges are multicoloured due to the overlap of some cycles. \textbf{(b)} Min, mean and max number of anomalies detected within the cycles of the multi-layer network, for each social and flood event (the colors in \textbf{(b)} refer to \cref{fig:local_scale}(a)).
    }
    \label{fig:anomalies_multilayer}

\end{figure}

A multi-layer network \cite{Boccaletti2014,Kivela2014} is a framework consisting of coupled networks that are connected (to each other) through inter-links, and it allows to approach the complex reality of interconnections and interdependencies among the diverse networks, typical of real-world systems. 
Although the literature on coupled multi-layer networks is extensive \cite{Parshani2010,Shekhtman2018,Rinaldi2001,Havlin2015,Radicchi2015,Buldyrev2010,Gao2012,Gao2013,Li2019_cascading_failure}, there is a scarce evidence about the usage of multi-layer approaches for investigating the impact of floods on transportation networks \cite{Danziger2022,Yadav}. However, this framework was employed to model multimodal transportation 
\cite{Alessandretti2016,Aleta2017,Bassolas2020,Gallotti2021}, where each mode of transportation (e.g. bus, tram, metro) was represented by a distinct network (layer).
A typical analysis in multi-layer networks is related to their robustness, i.e. one investigates if the network functionality is preserved during random failures or targeted attacks (e.g. starting from nodes with highest degree or betweenness), up to the shattering of the coupled networks into small clusters. Once fragmented, a multi-layer network might still maintain a certain degree of functionality within the \quotes{mutually-connected components}, i.e. when there is an intra-layer path between two nodes in all of the intra-layer networks \cite{Kivela2014,Buldyrev2010}. Here, we do not perform any analysis of robustness (i.e. percolation) among the coupled networks, but we just focus on the final stage of the process of networks fragmentation, i.e. on all the possible smallest mutually connected components, where the change of transport mode and the return to the initial starting place are guaranteed. As shown in \cref{fig:anomalies_multilayer}(a), in our case we have three layers, i.e. rails, urban roads, and motorways, and the mutually connected components are represented by the differently coloured \quotes{cycles}. In this framework, interlinks would be \quotes{functionally interdependent} \cite{Ferrari2023}, and would start both at train stations (for the railway network) and at the motorways entrances or exits (for the motorways layer). What is surprising here is the presence of anomalies (detected with DAMP) in the smallest mutually connected components, and for all the events (see \cref{fig:anomalies_multilayer}(b)). Indeed, the mutually connected components should be the smallest elements in networks where the functionality should be preserved. Although anomalies could represent both a congestion or a smaller amount of people, in case of congestion, the functionality would be compromised.
Also, it is interesting to notice the similarity among the flood and other social events, in terms of mean number of anomalies, for both rail-urban roads cycles and motorways-urban roads cycles.
Interestingly,
all the events except the Tattoo \& Children Festival (on Thursday $6 ^{th}$ of July), have averagely more anomalies in the rails-urban roads cycles than in motorways-urban roads cycles. This might be interpreted as a larger impact of the events on the railway network, than on the motorways layer. This aspect would also make the flooding event look similar to the social events. The Tattoo \& Children Festival (on Thursday $6 ^{th}$ of July) is not just the only event that impacts slightly more the motorway
network than the railway one, but it also exhibits the lowest amount of anomalies in both rails-urban roads cycles and motorways-urban roads cycles among all the events in this study. This feature means that the event has a fairly small effect on both motorways and railways. In this case, the multi-layer approach for assessing the robustness of coupled networks might not be necessary (since the amount of anomalies in both types of cycles is quite small), while the traditional measures of robustness for single layers (monoplex) might be more appropriate, or at least sufficient.

\section*{Conclusions}
\label{Conclusions}

In this work, we presented a comparison between a severe flood event,
i.e. a flood that occurred in 2017 in Switzerland, and major social events, which took place in the weeks before and after the flood, but in the same geographical area, that we called as the \quotes{local area of study}. 
We showed that, both at a national scale and at a local scale, a local but highly damaging flood and social events have a similar impact on human mobility.
In particular, at the national scale, we compared the weekly-based distributions of visits and the weekly-based distributions of distances travelled by visitors, who reached the local area of study, and we found minimal differences among them (see \cref{fig:national_scale}(b-e)). At the local scale, we instead focused on the irregularities $-$ i.e. the anomalies $-$ of the number of people who travelled each road or railway line, within the local area of study, as well as on the spatial clusters of those irregularities.
Surprisingly, we observed that the largest number of anomalies that occurred during the flood (i.e. the local maxima in the distributions of anomalies) is comparable to those ones of major social events (see \cref{fig:local_scale}(a)). In fact, one might expect that a flood would easily surpass other social events, in terms of number of anomalies (related to mobility).
About the clusters of anomalies, that we detected every hour and for each event, we calculated both the distances between the clusters centers and the event center, and the areas of those clusters. We then compared the flood with the social events through the resulting distributions of clusters-event distances (see \cref{fig:local_scale}(b)) and the distributions of clusters areas (see \cref{fig:local_scale}(c)), finding compelling similarities between the corresponding summary statistics, i.e. boxplots and measures of skewness and dispersion (see \cref{fig:local_scale}(b-e)). For example, we found that the flood-related distribution of distances (see \cref{fig:local_scale}(b)) is positively skewed as most of the social events-related ones (which exhibit a light-to-moderate skewness), meaning that the anomalies tend to group around the event center. Still at a local scale, we found that all the events, except the Summer Party, follow a linear empirical relationship between the distances of the farthest and the closest clusters of anomalies (see \cref{eq:r_max_delta_r}), allowing us to distinguish two types of spatial distributions of clusters. One, where the distance between the closest cluster and the event center is constant during the event (at NOMZ and at the Beach Festival), and a second type, where the distance of the closest cluster varies during the event. We observed the second behaviour for most of the events, meaning that the flood is similar, in this aspect, to the majority of the social events. A last comparison that we performed among the flood and social events was related to the framework of multi-layers, in network science. Indeed, we found that, during the flood, the average number of anomalies in the smallest mutually connected components
of the multi-layer networks (i.e. that we called \quotes{cycles}), is comparable to the average number of anomalies occurring during the social events (see \cref{fig:anomalies_multilayer}). We also found that the average number of anomalies is larger in the rails-urban roads cycles than in motorways-urban roads cycles, in all the events except the Tattoo \& Children Festival (on Thursday $6 ^{th}$ of July), denoting another commonality between the flood and the social events. In addition, regardless the comparison among different events, the presence of anomalies in the smallest mutually connected components
of a multi-layer network is a surprising fact by itself, since those components should preserve the functionality of the coupled network.
\\
\\
All our observations were supported by a robust methodology. 
At the national scale
we used six measures of (dis)similarity among distributions (please see \cref{fig:national_scale}), while at the local scale we used two state-of-the-art algorithms of data mining and machine learning (i.e. DAMP and RNN-DBSCAN) which require, overall, only one parameter. Indeed, to detect the anomalies in the number of people who travelled each road or railway lines (please see \cref{fig:local_scale,fig:empirical_relation,fig:anomalies_multilayer}), DAMP uses a set of training data, which size needs to be selected by the user before running the algorithm (and only once). RNN-DBSCAN, instead, employs a heuristics approach to automatically select the single required parameter, and to ultimately find the clusters of anomalies (please see \cref{fig:local_scale,fig:empirical_relation}).
Despite the solid methodology, this work has limitations, starting from the mobile phone data. Indeed, our mobile phone data represent only around $60\%$ of the entire set of Swiss mobile phone data, and this is due to the market share of Swisscom (at the end of 2017, Swisscom held $60\%$ of the entire Swiss mobile network \cite{OFCOM_2023}). Also, to preserve user anonymity, the mobile phone datasets provided by Swisscom were hourly-based and did not include the number of people travelling a certain road or a railway line, if less than $20$ travellers were observed. Arguably, a hourly time granularity might be considered a little coarse for social or natural events.
Another limitation of this study is related to the lack of documentation (mainly on the Web) about the minor events that took place in the local area of study (e.g. a road closure due to maintenance works). Although we were able to identify the origin of almost all the anomalies (see \cref{fig:local_scale}(a)), a few anomalies remain undocumented, and are likely related to minor events. Also, the presence of undocumented minor events during the course of a major event, could be the cause (at least partial) of the dissimilarity of the Summer Party’s boxplot in 
\cref{fig:local_scale}(b).
Indeed, the presence of (clusters of) anomalies, relatively far from a major event, and likely related to unknown (since undocumented) minor events, would affect both the distribution of distances between the clusters center and the event center, and the distribution of the clusters areas, as well as the corresponding summary statistics (\cref{fig:local_scale}(b-e)). And this could still happen despite RNN-DBSCAN is a noise-resistant algorithm. 
One way to avoid such a situation would be to prevent it with an adequate documentation about the minor events. In that case, the documents would help to identify the minor events, which will be then classified, together with the Summer Party, as \quotes{multiple events}, and therefore not treated with our proposed methodologies (that are designed for \quotes{distinct events}).
However, future efforts may develop further our methodologies for \quotes{multiple events}, to better understand the physics of anomalies during both natural and human-made events, and strengthen our empirical relationships. 
For example, the physics of anomalies could be  further investigated for \quotes{multiple} social events, being scheduled and frequent, and then applied to the assessment of rare flood events.
In conclusion, our work represents a bridge between the studies on human mobility during natural disasters \cite{Wang2016,Li2022} and human mobility during social events \cite{Calabrese2010,Pintér2018}, and can therefore have an impact on both (flood) risk management and city and event management, supporting urban planners, social scientists, city authorities, and traffic engineers. We hope that our findings could also inspire the development of more sophisticated data mining and machine learning algorithms for time series anomaly detection and clustering, as well as stimulate the advancement of the physics of anomalies $-$ related to human mobility $-$, likely in the context of network science.

\section*{Methods}
\label{Methods}

\subsection*{Nomenclature}
\label{Nomenclature}

Please refer to \textcolor{blue}{Supplementary Table 4} for a complete list of symbols used in this work, and their description.

\subsection*{Datasets}
\label{Datasets}

To perform our analysis at the national scale, we employ both the estimated (by Swisscom) number of users who, coming from the entire Switzerland were detected in the local area of study during the period of $44$ days, and the postal code of their municipality of residence. 
In particular, for each of the $44$ days, our dataset is composed of (i) the postal code of the municipality of residence, (ii) the name of the municipality of residence, (iii) the hour of the day during which travellers were detected, and (iv) the estimated count of detected people. We therefore derive the frequency of visits by using the estimated number of travellers, and, both time and day in which they were detected in the local area of study. About the way to derive the frequency of distances travelled by the users, we just employ the postal code of their municipality of residence. We then calculate those distances by selecting the shortest path, on the Swiss road network, between the centroid of every municipality of residence of the users, and the centroid of the local area of study.
At the local scale, we use both the estimated number of people who travelled each road or railway of the transport networks (which includes $\sim 5000$ edges, and $\sim 9000$ nodes), and the features of the network's edges. In particular, our dataset is composed of (i) the edge's start node, (ii) the edge's end node, (iii) the edge path, i.e. the nodes constituting the polyline edge, (iv) the edge type, i.e. if urban road, motorway or railway track, (v) the hour of the day during which travellers were detected, and (vi) the estimated count of detected people.

\subsection*{Measures of dissimilarity among distributions}
\label{Methods_dissimilarity_distributions}

Given a measurable space $\left( \X,\A\right)$, i.e. a pair consisting of a set $\X$ (called sample space) and a $\sigma-$algebra $\A$, we consider two probability measures, $P_1$ and $P_2$, on $\left( \X,\A\right)$, and their respective densities, $\rho_1$  and $\rho_2$. 
\\ 
\\
\textbf{Jensen–Shannon divergence.} Lin \cite{Lin} defines the Jensen–Shannon divergence through the Shannon entropy function, $f_{SE}(P_i)=-\sum_{x \in X} P_i(x) \log P_i(x)$, and the weights of $P_1$ and $P_2$, i.e. $\omega_1$ and $\omega_2$ \cite{Lin,Mendez}:
\begin{align}
      f_{JS}(P_1,P_2) \nonumber
      &\overset{\mathrm{def}}{=} f_{SE}(\omega_1 P_1 + \omega_2 P_2) - \omega_1 f_{SE}(P_1) - \omega_2 f_{SE}(P_2) 
      \\ 
      \intertext{We then rewrite the Jensen–Shannon 
      divergence in terms of the Kullback–Leibler divergence, $f_{KL}(P_1\mathrel{\Vert}P_2)$, using uniform weights, $\omega_1=\omega_2=\frac{1}{2}$:}
      &\overset{\mathrm{def}}{=} -\sum_{x \in X} \biggl(\frac{P_1(x)}{2}+\frac{P_2(x)}{2} \biggr)\log \biggl(\frac{P_1(x)}{2}+\frac{P_2(x)}{2} \biggr) \nonumber
      \\
      &\qquad + \frac{1}{2}\sum_{x \in X}P_1(x)\log P_1(x) + \frac{1}{2}\sum_{x \in X}P_2(x)\log P_2(x) \nonumber
      \\
      &= -\sum_{x \in X} \biggl(\frac{P_1(x)}{2}\biggr)\log \biggl(\frac{P_1(x)}{2}+\frac{P_2(x)}{2} \biggr) -
          \sum_{x \in X} \biggl(\frac{P_2(x)}{2} \biggr)\log \biggl(\frac{P_1(x)}{2}+\frac{P_2(x)}{2} \biggr) \nonumber
      \\
      &\qquad + \frac{1}{2}\sum_{x \in X}P_1(x)\log P_1(x) + \frac{1}{2}\sum_{x \in X}P_2(x)\log P_2(x)
      \nonumber
      \intertext{We use $H(x)=\frac{P_1(x)+P_2(x)}{2}$,}
      &= -\frac{1}{2} \sum_{x \in X} P_1(x)\log H(x) - \frac{1}{2} \sum_{x \in X} P_2(x) \log H(x) 
      + \frac{1}{2}\sum_{x \in X}P_1(x)\log P_1(x) + \frac{1}{2}\sum_{x \in X}P_2(x)\log P_2(x) \nonumber
      \\
      &= \frac{1}{2} \biggl(- \sum_{x \in X} P_1(x)\log H(x) - \sum_{x \in X} P_2(x) \log H(x) 
      + \sum_{x \in X}P_1(x)\log P_1(x) + \sum_{x \in X}P_2(x)\log P_2(x) \biggr) \nonumber
      \\
      &= \frac{1}{2} \biggl(\sum_{x \in X} P_1(x)\log \frac{P_1(x)}{H(x)} + \sum_{x \in X}P_2(x)\log \frac{P_2(x)}{H(x)} \biggr) 
      \label{KLexplicit}
      \\
      &\overset{\mathrm{def}}{=} \frac{1}{2}f_{KL}(P_1\mathrel{\Vert}H) + \frac{1}{2}f_{KL}(P_2\mathrel{\Vert}H) \nonumber
\end{align}
In this study, we use \cref{KLexplicit} to calculate the Jensen–Shannon divergence.
\\ \\
\textbf{Wasserstein distance.} 
We consider the Monge–Kantorovich minimization problem \cite{villani2009},
also called the Kantorovich's optimal transportation problem \cite{villani2021}:
\begin{align}
    \textit{minimize}\qquad 
    &\int_{\X \times \Y} \text{cost}(x,y) d\pi(x,y) \qquad\text{for}\; \pi \in \Pi(\mu,\nu) \nonumber
    \intertext{
    That would correspond to find the optimal transport cost between two probability measures $\mu$ and $\nu$ \cite{villani2009,villani2021}
    which is defined as:}
    \inf_{\pi \in \Pi(\mu,\nu)}
    &\int_{\X \times \Y} \text{cost}(x,y) d\pi(x,y) \nonumber
    \intertext{If both our two probability measures, $\mu$ and $\nu$, and the cost function are on $\mathbb{R}$, and the cost function is a convex non-negative symmetric function that takes the form $\text{cost}(x,y)=\text{cost}(x-y)$, then, the optimal transportation cost can be written as follows (details in \cite{villani2021}, from both Theorem 2.18 and Remarks 2.19):}
    \inf_{\pi \in \Pi(\mu,\nu)}
    &\int_{\X \times \Y} \text{cost}(x,y) d\pi(x,y)
    =\int_{0}^{1} c(F^{-1}(t)-G^{-1}(t)) \, dt \nonumber
    \intertext{where $F^{-1}$ and $G^{-1}$ are the quantile functions of the probability measures $\mu$ and $\nu$, respectively. $F$ and $G$ are instead the two cumulative distribution functions (CDFs), associated with $\mu$ and $\nu$, respectively. If the cost function takes the form $\text{cost}(x,y)=\left|x-y\right|$, by applying the Fubini's theorem, we would then get the following optimal transportation cost:}
    &\int_{0}^{1} \left| F^{-1}(t)-G^{-1}(t) \right| dt
    =\int_{\mathbb{R}} \left| F(x)-G(x) \right| dx \nonumber
    \intertext{We now consider a Polish metric space, $\X$, equipped with a metric $d(x,y)$, and we define the cost function as the $p-$th power of $d(x,y)$, i.e. $\text{cost}(x,y)=d(x,y)^{p}$, where $p\in [1,\infty)$. Then, we can establish a distance between $\mu$ and $\nu$, that we call the Wasserstein distance of order $p$ between $\mu$ and $\nu$ 
    \cite{villani2009,villani2021}:}
    W_{p}(\mu,\nu) = \Bigg( \inf_{\pi \in \Pi(\mu,\nu)} 
    &\int_{\X} d(x,y)^{p} d\pi(x,y) \Bigg) ^{\frac{1}{p}} \nonumber
    \intertext{The Wasserstein distance represents the metric side of optimal transportation. Therefore, when two probability measures, $\mu$ and $\nu$, are on the real line, i.e. $\mathbb{R}$, then the Wasserstein distance has a closed form, and reads as $W_{p}(\mu,\nu) = \left(\int_{0}^{1} \left| F^{-1}(t)-G^{-1}(t) \right|^{p} dt \right) ^{\frac{1}{p}}$. When both $\mu$ and $\nu$ are on $\mathbb{R}$, and the exponent is $p=1$, the Wasserstein distance takes the $W_1$ form, also called the Kantorovich–Rubinstein distance:}
    W_{1}(\mu,\nu) = 
    &\int_{\mathbb{R}} \left| F(x)-G(x) \right| dx
    \label{W1}
\end{align}
In this study, we use \cref{W1} to calculate the Wasserstein distance  $W_1$.
\\ \\
\textbf{Total Variation distance and Hellinger distance.} The total variation distance is defined as \cite{tsybakov2008}:
\begin{align}
      f_{TV}(P_1,P_2)
      &\overset{\mathrm{def}}{=} \sup_{A \in \mathcal{A}} \left| P_1(A)-P_2(A) \right| \nonumber
      \intertext{If $\X$ is finite, then, any $\sigma-$algebra $\A$ is finite (since the power set is finite). When $\A$ is finite, the $\operatorname{\max}_{A \in \A}$ exists, and it is equivalent to $\sup_{A \in \A}$. Then, if $\A$ is the power set, $\operatorname{\max}_{A \in \A}$ is equivalent to $\operatorname{\max}_{A \subseteq \X}$ leading to $f_{TV}(P_1,P_2) = \operatorname{\max}_{A \subseteq \X} \left| P_1(A)-P_2(A) \right|$. As an alternative characterization, the total variation distance can be expressed as half of the $L^{1}-$norm \cite{levin2017}. This can be shown by using subsets of $\X$,  i.e. $B={\{ x\, : \,P_1(x)>P_2(x) \}}$, and $B^{c}={\{ x\, : \,P_1(x)<P_2(x) \}}$ \cite{TVproof}:}
      f_{TV}(P_1,P_2)  \nonumber
      &= \operatorname{\max}_{A \subseteq \X} \left| P_1(A)-P_2(A) \right| \\ \nonumber
      &= P_1(B)-P_2(B) \\  \nonumber
      &= \frac{1}{2} \left( P_1(B)-P_2(B) +  P_1(B^{c})-P_2(B^{c})  \right) \\ \nonumber
      &= \frac{1}{2} \left( \sum_{x\in B}  \rho_1(x)-\rho_2(x)  +  \sum_{x\in B^{c}}  \rho_1(x)-\rho_2(x)  \right) \\ \nonumber
      &= \frac{1}{2} \sum_{x \in \X} \left| \rho_1(x)-\rho_2(x) \right| \\
      &\overset{\mathrm{def}}{=} \frac{1}{2} \left\lVert P_1 - P_2 \right\rVert_{1}
    \label{TVdist}
\end{align}
The Hellinger distance between two probability measures is defined as the $L^{2}-$norm of the difference between the square roots of the densities, $\sqrt{\rho_1}$ and $\sqrt{\rho_2}$ \cite{tsybakov2008,pollard2002}:
\begin{align}
      f_{H}(P_1,P_2) \nonumber
      &\overset{\mathrm{def}}{=} \frac{1}{\sqrt{2}}  {\left( \sum_{i = 1} {\left(  \sqrt{\rho_1(i)} - \sqrt{\rho_2(i)} \right)}^{2}  \right)}^{\frac{1}{2}} \\ 
      &\overset{\mathrm{def}}{=} \frac{1}{\sqrt{2}} \left\lVert P_1 - P_2 \right\rVert_{2}
    \label{Hdist}
\end{align}
In this study, we use \cref{TVdist} and \cref{Hdist} to calculate, respectively, the total variation distance and  the Hellinger distance.
\\
\\
\textbf{Complementary empirical distance correlation.} Proposed by Székely et al. in 2007 \cite{Székely2007}, the distance correlation $f_{DC}$ measures the degree of dependence between two multivariate random variables, or random vectors, i.e. $X$ in $\mathbb{R}^{u}$ and $Y$ in $\mathbb{R}^{v}$, with $u$ and $v$ positive integers. If we consider $X \sim P_1$ and $Y \sim P_2$, we can write the empirical distance correlation among two probability measures as:
\begin{align}
      f_{DC}(X,Y) &= 
      \begin{cases}
            \sqrt{\frac{\V^2(X,Y)}{\sqrt{ \V^2(X)\V^2(Y) }}}, 
            & \V^2(X)\V^2(Y) > 0\\
            0,              
            & \V^2(X)\V^2(Y) = 0
            \label{DCdist}
      \end{cases}
      \intertext{where}
      \V^2(X,Y) \nonumber
      &= \frac{1}{v^2} \sum_{i,j=1}^{v} A_{ij}B_{ij} && A_{ij} = a_{ij} + \frac{1}{v} \sum_{i=1}^{v}a_{ij} + \frac{1}{v} \sum_{j=1}^{v}a_{ij} + \frac{1}{v^{2}} \sum_{i,j=1}^{v}a_{ij}  \\ \nonumber
      \V^2(X) = \V^2(X,X)
      &= \frac{1}{v^2} \sum_{i,j=1}^{v} A_{ij} && B_{ij} = b_{ij} + \frac{1}{v} \sum_{i=1}^{v}b_{ij} + \frac{1}{v} \sum_{j=1}^{v}b_{ij} + \frac{1}{v^{2}} \sum_{i,j=1}^{v}b_{ij}  \\ \nonumber
      \V^2(Y) = \V^2(Y,Y)
      &= \frac{1}{v^2} \sum_{i,j=1}^{v} B_{ij} && a_{ij} = \left|X_i-X_j \right|_{u} \qquad b_{ij} = \left|Y_i-Y_j \right|_{v} \nonumber
\end{align}
and $n$ is the sample size. In \cref{DCdist}, $f_{DC}(X,Y) = 0$, when $\V^2(X)\V^2(Y) = 0$, means that the distance correlation is zero if $X$ and $Y$ are independent. In this work, we use \cref{DCdist} to calculate the complementary empirical distance correlation, defined as $f^{\complement}_{DC}(X,Y) \overset{\mathrm{def}}{=} 1 - f_{DC}(X,Y)$.
\\
\\
\textbf{Cosine distance} Among the measures of distributional (dis)similarity \cite{cha2007}, we also consider the cosine similarity, with $X \sim P_1$ and $Y \sim P_2$:
\begin{align}
      f_{CS}(X,Y) \nonumber
      &= \frac{\sum_{i=1}^{v}(X_i,Y_i)}{\sqrt{\sum_{i=1}^{v}X_{i}^{2} \sum_{i=1}^{v}Y_{i}^{2}}} \\ 
      &\overset{\mathrm{def}}{=} \frac{X \cdot  Y}{\left\lVert X \right\rVert_{2} \left\lVert Y \right\rVert_{2}}
      \label{CosSim}
\end{align}
In this study, we use \cref{CosSim} to compute the cosine distance $f_{CD}(X,Y) = 1 - f_{CS}(X,Y)$.

\subsection*{Discord Aware Matrix Profile (DAMP) algorithm}
\label{Methods_DAMP}

The DAMP algorithm \cite{Lu2022_DAMP,Lu2023_DAMP} is based on two state-of-the-art techniques for time series anomaly detection (TSAD), i.e. the \quotes{time series discord} \cite{Hehir2023,Nilsson2023,Zymbler2019,Zymbler2023}, 
proposed by Keogh et al. \cite{Keogh2005} in 2005, and the \quotes{matrix profile} \cite{DePaepe2020_article,DePaepe2020_book,Tata2022,Quislant2023}, introduced by Yeh et al. \cite{Yeh2016} in 2016. Intuitively, a time series discord defines the anomaly in a time series, while the matrix profile is a data structure used for time series analysis. 
Despite there being no single method that is superior in every case of anomaly detection
\cite{Paparrizos2022,Schmidl2022,Chandola2009_TR}, the matrix profile-based approaches $-$ of which DAMP is one $-$, performed  well in different benchmarks \cite{Paparrizos2022,He2023}.
On a large benchamark containing $13766$ univariate time series (i.e. TSB-UAD), Paparrizos et al. \cite{Paparrizos2022} compared $12$ methods for anomaly detection, which belonged to different method families, i.e. deep learning (CNN, AutoEncoder, LSTM), outlier detection (Isolation Forest, LOF), classic machine learning (OCSVM, PCA, HBOS) and data mining (matrix profile, NormA). They found that, on average, the best methods for anomaly detection were the matrix profile-based algorithms and NormA \cite{Boniol2021}.
Another comparison between anomaly detection algorithms was made by He et al. \cite{He2023}, on a collection of $250$ univariate time series, called the KDD CUP 2021 (KDD21) dataset \cite{KDD21}. They used algorithms based on deep learning (LSTM, USAD, TranAD), seasonal-trend decomposition (NSigma, OnlineSTL, OneShotSTL) and matrix profile (NormA, STOMPI, SAND, DAMP), and they showed the superiority of DAMP, in terms of accuracy.
\\ 
\\
\textbf{Basic concepts and definitions in DAMP.} We consider a time series $T$, defined as a sequence of temporal data, i.e. $T=(t_k)_{k=1}^{n}$, and a time subsequence $T_{i,m}=(t_k)_{k=i}^{i+m-1}$, where $k,n,i,m \in \N$. In particular, $n$ and $m$ are the length of the time series and the length of the time subsequence, respectively. A time subsequence $T_{i,m}$ is therefore characterised by two values: one is the starting position $i$ on the temporal line, while the second is its length $m$. We then consider the Euclidean distance among two subsequences, $T_{i,m}=(t_k)_{k=i}^{i+m-1}$ and $T_{j,m}=(t_{k})_{k=j}^{j+m-1}$, that is defined through the $L^2-$norm, i.e. 
$\left\lVert T_{i,m}-T_{j,m} \right\rVert_{2} \overset{\mathrm{def}}{=} 
\sqrt{\sum_{l=0}^{m-1}(t_{i+l}-t_{j+l})^{2}}$. 
Here, we used $t_{i}=(t_k)_{k=i}$ and $t_{j}=(t_k)_{k=j}$ to indicate the first elements of $T_{i,m}$ and $T_{j,m}$, respectively. More precisely, for this algorithm, we will use the $z-$normalised Euclidean distance, 
$\sqrt{\sum_{l=0}^{m-1}(\widetilde{t}_{i+l}-\widetilde{t}_{j+l})^{2}}$
, where 
$t_{i+l}\rightarrow \widetilde{t}_{i+l} = \left( \frac{t_{i+l} - \mu_i}{\sigma_i} \right)$ and $t_{j+l}\rightarrow \widetilde{t}_{j+l} = \left( \frac{t_{j+l} - \mu_j}{\sigma_j} \right)$, and, for ease of notation, we will denote it with the same $L^2-$norm notation of the non-normalized Euclidean distance, i.e. $\left\lVert T_{i,m}-T_{j,m} \right\rVert_{2}$.
Distances are calculated only for non-overlapping subsequences $T_{i,m}$ and $T_{j,m}$, also called as \quotes{non-self match} subsequences, i.e. when $\left| i-j \right| \geq m$. We can name $\left| i-j \right| \geq m$ as the \quotes{non-overlapping condition} as well. If we focus on one specific $T_{i,m}$, i.e. we fix the index $i$ to a certain number, and we only vary the index $j$ such that $\left| i-j \right| \geq m$, we can calculate all the distances among our chosen subsequence $T_{i,m}$ and all the other subsequences $T_{j,m}$ forming the same time series. We define the set of distances among the reference subsequence $T_{i,m}$ and all  the other subsequences $T_{j,m}$, of the same time series, as  \quotes{distance profile} $\delta_i$:
\begin{equation}
\delta_{i}
=\left\{  
\left\lVert T_{i,m}-T_{j,m} \right\rVert_{2} 
\, \colon \,
1 \leq j \leq (n-m+1)
\enspace \land \enspace 
\left| i-j \right| \geq m
\right\}
\end{equation}
Given the \quotes{distance profile} $\delta_{i}$ of a reference sebsequence $T_{i,m}$, we can then find the \quotes{nearest neighbour} of $T_{i,m}$, i.e. that $T_{j,m}$ subsequence that has the minimum Euclidean distance from $T_{i,m}$. In other words, the \quotes{nearest neighbour} is defined by $\operatorname{\operatorname{\min}} (\delta_{i})$. Obviously, we can find the nearest neighbour for each reference subsequence $T_{i,m}$, and the collection of all the minimum distances, $\operatorname{\min} (\delta_{i})$, is called as the \quotes{matrix profile}, $M$ \cite{Yeh2016}:
\begin{equation}
    M = 
    \{ \operatorname{\min} (\delta_{1}), \operatorname{\min} (\delta_{2}), \ldots, \operatorname{\min} (\delta_{n-m+1}) \}
\end{equation}
The subsequence corresponding to the maximum value of the matrix profile, i.e. the subsequence with the largest Euclidean distance to its non-overlapping nearest neighbour, is called as \quotes{time series discord}.
Indeed, a discord is a subsequence which is a maximally different from all the other subsequences, and can therefore be thought as an anomalous subsequence. 
Despite the discovery of time series discords is a well recognised and popular anomaly detection approach \cite{Hehir2023,Nilsson2023,Zymbler2019,Zymbler2023}, it fails when 
unusual subsequences (discords) occur more than once in the same time series. This is a widely known issue, called as the \quotes{twin freak problem} \cite{Wei2008,He2020}. However, the current algorithm is able to mitigate this problem, by computing the \quotes{left-discords} \cite{Lu2022_DAMP,Lu2023_DAMP}. Essentially, those discords are calculated as previously described, but by adding the $1 \leq j \leq (i-m)$ condition, i.e. by considering only subsequences $T_{j,m}$ on the left of the reference subsequence $T_{i,m}$. With this additional condition, the profile distance would become the \quotes{left distance profile}:
\begin{align}
\delta_{i}^{*}
&=\left\{  
\left\lVert T_{i,m}-T_{j,m} \right\rVert_{2} 
\, \colon \,
1 \leq j \leq (i-m) \leq (n-2m+1)
\enspace \land \enspace 
\left| i-j \right| \geq m
\right\} 
\label{left_DP}
\\
\intertext{and, consequently, the matrix profile would become the \quotes{left matrix profile}:}
M^{*} 
&= 
\{ \operatorname{\min} (\delta_{1}^{*}), \operatorname{\min} (\delta_{2}^{*}), \ldots, \operatorname{\min} (\delta_{n-m+1}^{*}) \}
\label{left_MP}
\end{align}
Therefore, the DAMP algorithm mainly computes \cref{left_DP} and \cref{left_MP} to find the time series left-discords. Then, those discords can be sorted to identify the \quotes{top left-discords}, i.e. the largest anomalies.
\\ 
\\
\textbf{Parameters in DAMP.} 
We use DAMP\textunderscore topK \cite{DAMP_topK}, a variant of DAMP that computes the top left-discords, and that splits the data into two intervals, i.e. the \quotes{training data} and the \quotes{test data} \cite{Lu2022_DAMP,Lu2023_DAMP}. With such a division of data, we compare the current subsequence $T_{i,m}$ with those subsequences $T_{j,m}$ that are located within the training interval. 
We set the split point $I$ between the training data and the test data $-$ called \quotes{CurrentIndex} $-$, equal to $I=6 \times 24$, i.e. $6$ days. During those $6$ days, ranging from Saturday $17^{th}$ of June to Thursday $22^{nd}$ of June, we did not find significant events that would affect the anomaly detection analysis. For details on the events that occurred during the entire examined temporal window of $44$ days, within the local area of study, please see \textcolor{blue}{Supplementary Table 1}. Once the CurrentIndex is given by the user, DAMP\textunderscore topK calculates automatically the subsequence length. In our case, it would result in $m=24$, corresponding to one period, i.e. $1$ day, of our time series. 
With these numbers, we would get a ratio between the CurrentIndex and the subsequence length equal to $6$, which is larger than the recommended ratio, $\frac{I}{m}>4$ \cite{LoretiLuKeoghBlind}. 
We then set the number of top left-discords we want to extract with DAMP\textunderscore topK equal to $10$. This means that for every road and railway track, DAMP would identify up to the $10$ largest left-discords (anomalies), if they exist.

\subsection*{Reverse Nearest Neighbor Density-Based Spatial Clustering of Applications with Noise (RNN-DBSCAN) algorithm}
\label{Methods_RNN-DBSCAN}

DBSCAN is one of the most popular algorithm for clustering detection, it is resistant to noise and can handle clusters of different shapes and sizes. However, it does not work well with varying densities and it is sensitive to its two parameters. RNN-DBSCAN \cite{Bryant} is an extension of DBSCAN, based on the construction of a directed network over the dataset of points. RNN-DBSCAN reduces the amount of input parameters from two to one, and has the ability to handle regions that can vary greatly in density, showing a great accuracy in cluster detection \cite{Liu2020_DDNFC}.
\\ 
\\
\textbf{Basic concepts and definitions in DBSCAN.} Let $\Upsilon$ be a set of points in a $2-$dimensional space $\R^{2}$, where the generic point $i$ and the generic point $j$ have, respectively, coordinates $(x_i,y_i)$ and $(x_j,y_j)$. Please note that, here, we use $i$, $j \in \N$ to label our points, while in DAMP, we use $i$ and $j$ as $1-$dimensional coordinates over the time line, i.e. to indicate the position of subsequences $T_{i,m}$ and $T_{j,m}$. Also, we use the same indices, $i$ and $j$, for two different contexts, to facilitate those readers who have a background in the respective scientific areas, and are therefore used to see $i$ and $j$ as common notations. 
After this brief digression, we continue with DBSCAN and we define the $\epsilon-$neighborhood of point $i$ as a subset of points, whose Euclidean distance from point $i$ is less than or equal to $\epsilon$, i.e., $\epsilon N_{i}=\{ j \, \colon \, d_{ij}\leq \epsilon \}$. Intuitively, the $\epsilon-$neighborhood would allow us to distinguish points inside a cluster, that we call \quotes{core points}, from points on the border of a cluster, that we name \quotes{border points}, and isolated points, simply identified as \quotes{noise}. Indeed, the $\epsilon-$neighborhood of a border point would contain less points than the $\epsilon-$neighborhood of a core point. 
But what is the exact number of points, within a $\epsilon-$neighborhood, to distinguish core points from border points (or noise points)? 
We can unambiguously identify core points from border points by introducing a minimum number of points, $\gamma$, that lie within a $\epsilon-$neighborhood, i.e. by assigning a minimum local density. Therefore, a core point needs to contain at least $\gamma$ points within a $\epsilon-$neighborhood, i.e. 
$\left| \epsilon N_{i} \right| \geq \gamma$, while a border point contains less than $\gamma$ points within a $\epsilon-$neighborhood, i.e. $\left| \epsilon N_{i} \right| < \gamma$. Any point $j$ that belongs to the neighborhood of a core point $i$, i.e. 
$\left\{ j \, \colon \, j \in \epsilon N_{i} \enspace \land \enspace \left| \epsilon N_{i} \right| \geq \gamma \right\}$, 
is called \quotes{directly density-reachable} from $i$. This definition can be then extended to pairs of points $i$ and $j$ that are not directly reachable. Indeed a point $j$ is  \quotes{density-reachable} from a point $i$, if we can form a chain of \quotes{directly density-reachable} points, starting at point $i$ and ending at point $j$. 
If both points $i$ and $j$ are core points, the property of density-reachability is symmetric, i.e. holds for both directions, either starting at point $i$ and moving towards point $j$, or starting at point $j$ and moving towards point $i$. However, if $i$ and $j$ are both border points, they are not generally density-reachable, since the core point condition (i.e. $\left| \epsilon N_{i} \right| \geq \gamma$) does not hold for border points. Nevertheless, two border points can be \quotes{density-connected} if they are both density-reachable from a common point between them. The notion of density-reachability allows us to define a cluster as a set of points that are density-reachable from a core point (and, consequently, density-connected among each other). Indeed, when DBSCAN starts to retrieve all points that are density-reachable from an arbitrary point, if it is a core point, DBSCAN detects the entire cluster, to which that core point belongs. Otherwise, if the arbitrary point is a border one, or a noise one, DBSCAN does not find any density-reachable point from it, and goes to visit the next point in the dataset.
\\ 
\\
\textbf{Basic concepts and definitions in RNN-DBSCAN.}
RNN-DBSCAN is a graph-based interpretation of DBSCAN, where the concept of $\epsilon-$neighborhood (with a fixed radius) of a point $i$, $\epsilon N_{i}$, is replaced by the graph's notion of reverse $k-$nearest neighborhood of node $i$, $RkNN_{i}$. 
The concept of reverse $k-$nearest neighborhood, $RkNN_{i}$, is
related to the construction of a $k-$nearest neighbor graph ($k-$NNG) over our set of points $\Upsilon$. Indeed, for any set of points $\Upsilon$, we can build up a directed graph $k-$NNG with nodes equal to $\Upsilon$, and directed edges that connect every node to each of its $k-$nearest neighbors. 
Once the $k-$NNG graph is built, we can introduce the $k-$nearest neighborhood of an arbitrary node $i$, i.e.
$ %
{kNN}_i = \left\{ j \, \colon \, d_{ij}<d_{ij'} \, ,\forall {j} \in \Upsilon{\setminus}\left( \{i\} \cup \{j'\} \right)
,\forall {j'} \in \Upsilon{\setminus}\left( \{i\} \cup \{j\} \right) \enspace \land \enspace 
\left| \{j\} \right| = k \right\}
$
and, likewise, ${kNN}_j$, i.e. the $k-$nearest neighborhood of node $j$, that we can use to define the \quotes{outgoing degree} of node $i$, $k^{\text{out}}_i = \left| \{ (i,j) \, \colon \, j \in {kNN}_i \} \right|$ and the \quotes{ingoing degree} of the same node $i$, $k^{\text{in}}_i = \left| \{ (j,i) \, \colon \, i \in {kNN}_j \} \right|$. Here, $(i,j)$ denotes the edge from node $i$ to node $j$, while $(j,i)$ symbolizes the edge from node $j$ to node $i$. The outgoing degree $k^{\text{out}}_i$, abbreviated as \quotes{out-degree}, represents the number of edges that point from node $i$ to other nodes, while the ingoing degree $k^{\text{in}}_i$, abbreviated as \quotes{in-degree}, represents the number of edges that point to node $i$. Therefore, the out-degree $k^{\text{out}}_i$ corresponds to the size of the $k-$nearest neighborhood of an arbitrary node $i$, i.e.
$k^{\text{out}}_i = \left| {kNN}_i \right|$, while
the in-degree $k^{\text{in}}_i$ corresponds to the size of the reverse $k-$nearest neighborhood of node $i$, i.e.
$k^{\text{in}}_i = \left| {RkNN}_{i} \right|$.
A meticulous reader would confute both $k^{\text{out}}_i = \left| {kNN}_i \right|$ and $k^{\text{in}}_i = \left| {RkNN}_{i} \right|$, 
since ${kNN}_i$ and ${RkNN}_i$ would be sets of nodes and not a sets of edges, while both $k^{\text{out}}_i$ and $k^{\text{in}}_i$ would refer to edges. However, every node $j$ would be part of an edge $(i,j)$ or $(j,i)$, by construction (i.e. definition) of the $k-$NNG graph, and we could then use a set of edges $(i,j)$ $-$ having the same starting node $i$, and $j \in {kNN}_i$ $-$, to indicate ${kNN}_i$, and a set of edges $(j,i)$ $-$ having the same ending node $i$, and $i \in {kNN}_j$ $-$, to indicate ${RkNN}_i$.
For example, in this perspective, $j \in {RkNN}_i$ means that node $j$ is the starting node of an edge $(j,i)$ that points towards node $i$, such that $i \in {kNN}_j$. In other words, $j \in {RkNN}_i$ means $j \in \{ (j,i) \, \colon \, i \in {kNN}_j \}$, or, more precisely, $j \in \{ j \, \colon \, j \in (j,i) \enspace \land \enspace i \in {kNN}_j \}$. 
Due to the definition of $k-$NNG graph, all nodes in $k-$NNG have the same out-degrees, $k^{\text{out}}_i = \left| {kNN}_i \right| = k$, but they have different in-degrees, $k^{\text{in}}_i$. We can now replace the DBSCAN core point condition, $\left| \epsilon N_{i} \right| \geq \gamma$, with the RNN-DBSCAN core point condition, i.e. $k^{\text{in}}_i = \left| {RkNN}_i \right| \geq k$. 
Similarly to what we did in DBSCAN with its core point condition, here, we can use the RNN-DBSCAN core point condition to re-define the core and border points (and noise points), as well as the notions of \quotes{directly density-reachable}, \quotes{density-reachable}, \quotes{density-connected}  and \quotes{cluster}. For example, the set of points $j$ that are \quotes{directly density-reachable} from $i$ would read as $\left\{ j \, \colon \, j \in {RkNN}_i \enspace \land \enspace k^{\text{in}}_i = \left| {RkNN}_i \right| \geq k \right\}$.
\\ 
\\
\textbf{Parameters in RNN-DBSCAN.}
We use a Matlab implementation of RNN-DBSCAN \cite{Vannoy}, which requires the sole input parameter $k$, i.e. the number of nearest neighbours in the $k-$NNG graph, and we use a heuristics approach proposed by Bryant et al. \cite{Bryant}, to automatically select an appropriate value for $k$. For each $k = 1,2,\cdots,100$, we run RNN-DBSCAN and detect the number of resulting clusters, $c_{k}$. We therefore obtain a sequence of $100$ elements, that we denote with $C=(c_{k})_{k=1}^{100}$. Since we commonly get the same number of clusters for different values of $k$, i.e. $c_{k} = c_{k'}$ for $k \neq k'$, the sequence $C$ will contain duplicate elements. The duplicate elements can be then gathered to form subsequences of $C$, where each subsequence contains the same element, repeated a number of times. We then use the image of $C$ and the preimage of $C$ to build up those subsequences. 
The image of $C$ is a set, i.e. an unordered collection of distinct objects, $\operatorname{img}(C)= \{c_{k}\}_{k=1}^{100}$, and it is usually denoted using curly braces. Through the image of $C$ we therefore get the unique elements of $C$, i.e. unique values of the number of clusters. Indeed, for example, if we had $\operatorname{img}(C)=\{3,1,3,7,7,3,1,1,3,2\}$, this would be equivalent to $\{1,2,3,7\}$. Then, for any element of the image we can introduce the preimage of that element, $C^{-1}(\{c_{k}\})$, i.e. the set of $k$ indices that indicate the position of that element's duplicates within $C$. Hence, the preimage of $C$ allows us to see if those unique elements are repeated, while the cardinality of the preimage of $C$, denoted by $\left| C^{-1}(\{c_{k}\}) \right|$, tells us about the number of times those unique elements are repeated. Basically, the cardinality of the preimage of $C$ is the frequency of occurrence of the number of clusters.
Obviously, the union of preimages of singletons $\{c_{k}\}$, with respect to $C$, will give us the original sequence, i.e. $C = \bigcup_{c_{k}\in\operatorname{img}(C)} C|_{C^{-1}(\{c_{k}\})}$, 
where $C|_{C^{-1}(\{c_{k}\})}$ is the restriction of the function $C$ (a sequence is a function) to the preimage of $\{c_{k}\}$, with respect to $C$.
Once we know the frequency of occurrence for each number of clusters, we essentially have a distribution of the number of clusters calculated over the range $1 \leq k \leq 100$.
The heuristics approach proposed by Bryant et al. \cite{Bryant} consists of considering the leftmost local maxima in the distribution, to find first the best value of $k$, and then get the best clustering in a given dataset. Indeed, when they tested RNN-DBSCAN over a range $1 \leq k \leq 100$, on a set of artificial datasets with given ground truth, they noticed that the Adjusted Rand Index (ARI) performance was maximum at the leftmost local maxima of the distribution. Also, they observed a positive correlation between the maximum ARI and the ARI performance at minimum $k$. This means that, among all the $k$ input parameters which contributed to the construction of the leftmost local maxima of the distribution, the minimum $k$ could be safely used as input parameter for generating the correct number of clusters in a dataset of points. Please note that here, and only for this distribution of the number of clusters, we use the same convention employed by Bryant et al. \cite{Bryant} about the direction of the $x-$axis, where positive numbers increase towards left. Obviously, if we had used the classical convention for the $x-$axis, where positive numbers increase towards right, we would have focused on the rightmost local maxima of the distribution. Our tests on the proposed heuristic approach, and on the same artificial datasets used by Bryant et al. \cite{Bryant}, are illustrated in \textcolor{blue}{Supplementary Figure 6}.
There, we show pretty much the same results to those ones of Bryant et al. \cite{Bryant}.

\newpage
\section*{Acknowledgements}

We thank Eamonn Keogh, Yue Lu, and Trevor Vannoy for useful discussions, and Swisscom for providing the data.

\section*{Author contributions statement}

S.L. designed research, analyzed data, performed research, created the collaboration with Swisscom, and wrote the paper. M.K. and A.Z. reviewed the paper, M.K. initiated the project, A.Z. performed geodata analysis and flood exposure analysis.

\section*{Additional information}

\textbf{Competing interests:} The authors declare no competing interest.
\\
\\
\textbf{Costs for data:} Mobile phone data were purchased by Swisscom in November 2019, for a total cost of $32400$ CHF (i.e. $\sim 36000$ USD, in October 2023). The costs included $1800$ CHF for \quotes{Scope \& Alignment} ($1-$man day of work), $9000$ CHF for \quotes{Data mining \& preparation} ($5-$man days of work), and $21600$ CHF for \quotes{Data processing \& analysis} ($12-$man days of work).
\\
\\
\textbf{Data availability:} 
The Swisscom mobile phone data will be available on Zenodo.

\FloatBarrier

\bibliography{main}

\end{document}


\maketitle

\def\LT@output{%
  \ifnum\outputpenalty <-\@Mi
    \ifnum\outputpenalty > -\LT@end@pen
      \LT@err{floats and marginpars not allowed in a longtable}\@ehc
    \else
      \setbox\z@\vbox{\unvbox\@cclv}%
      \ifdim \ht\LT@lastfoot>\ht\LT@foot
        \dimen@\pagegoal
        \advance\dimen@-\ht\LT@lastfoot
        \ifdim\dimen@<\ht\z@
          \setbox\@cclv\vbox to \textheight{\unvbox\z@\copy\LT@foot}%
          \@makecol
          \@outputpage
          \setbox\z@\vbox{\box\LT@head}%
        \fi
      \fi
      \global\@colroom\@colht
      \global\vsize\@colht
      \vbox to \textheight
        {\unvbox\z@\box\ifvoid\LT@lastfoot\LT@foot\else\LT@lastfoot\fi}%
    \fi
  \else
    \setbox\@cclv\vbox to \textheight{\unvbox\@cclv\copy\LT@foot}%
    \@makecol
    \@outputpage
      \global\vsize\@colroom
    \copy\LT@head\nobreak
  \fi}
\NewDocumentCommand{\hoursbar}{ O{5cm} O{4pt} O{red} m m }{%
   \begin{tikzpicture}[baseline]
        \fill[#3] ({(#4/24)*#1},0) rectangle ({(#5/24)*#1},{#2});
        \draw (0,0) rectangle ({0.25*#1},{#2});
        \draw ({0.25*#1},0) rectangle ({0.5*#1},{#2});
        \draw ({0.5*#1},0) rectangle ({0.75*#1},{#2});
        \draw ({0.75*#1},0) rectangle ({#1},{#2});
   \end{tikzpicture}%
}

\def\pcbACSF#1{%
   {\color[HTML]{000000}\rule{\fpeval{#1/\percentscale*\barwidth} cm}{\barheight}} #1
} %
\def\pcbBioM#1{%
   {\color[HTML]{000000}\rule{\fpeval{#1/\percentscale*\barwidth} cm}{\barheight}} #1
} %
\def\pcbSeniorExcursion#1{%
   {\color[HTML]{000000}\rule{\fpeval{#1/\percentscale*\barwidth} cm}{\barheight}} #1
} %

\newcommand{\barwidth}{5} %
\newcommand{\barheight}{4pt} %
\newcommand{\percentscale}{10000} %

\section*{Supplementary Text}

\subsection*{A German$-$to$-$English translation of notes about the flood, from the 
\quotes{WSL Swiss Flood and Landslide Damage
Database} \cite{WSL,Andres2019}}
\quotes{Several thunderstorms passed over the region in quick succession. This led to large amounts of rain. The Cantonal building insurance company of Aargau (AGV) received around $2500$ damage reports for the entire canton of Aargau (floods including backwater and slides
). In total, buildings suffered damage of around CHF $55$ million.
\\
\\
\textbf{Zofingen.} A massive thunderstorm cell with hail and a lot of rain (around $3$h, $85-90$ mm over the catchment areas of the Zofingen city streams) unloaded over the city. Hail and leaves clogged the drains. There was flooding from surface water and overloaded sewer lines with backflows in the buildings. The city area was flooded over a wide area within 30 minutes. Landslides occurred (recorded separately). The areas of Riedtal and Mühlethal as well as areas along the Wigger river were particularly
badly affected. Two people who slipped on the mud had to be taken to hospital. Because nine transformer stations were under water, the power failed in large parts of the city.
The emergency call center and the fire department received about $550$ damage reports. Around $250-450$ rescue workers were deployed. The fire department pumped water out from more than 200 cellars. The storm flooded hundreds of businesses, basements, underground garages and underpasses. The Zofingen fire department performed $1800$ person$-$hours,
the civil defense organization of the region another $4700$ hours. 
The AGV assumed $977$ damages in Zofingen (incl. backwater). Based on this information, we calculated $42$ million CHF (including furniture). Note: Taking this information into account, an initial estimate after the event of over CHF $100$ million in damages seems too high.
\\
\\
\textbf{Mühlehtal.} Among the worst hit were the textile finishing company Bethge AG in Mühlethal, which was devastated by water and mud masses, and a private property in the same location, which was hit by a slides
(recorded separately). Parts of the hospital in Zofingen were up to $40$ centimeters under water. In the meantime, the elevators failed. The water ran from the sewage system into the hospital's archives $-$ $85000$ patient files were destroyed. A data recovery team froze the files and then placed them in a vacuum tank. The work took several months. In the military accommodation Rosengarten there was a major fire department operation. The water damaged all the equipment and technology. Several walls and floor slabs were destroyed by the lateral pressure and buoyancy of the water. The accommodation had to be completely renovated; it was not expected to be rentable again until the beginning of $2018$.
\\
\\
\textbf{Industrial area (west of the railway tracks).} Water penetrated into the industrial facility of Swissprinters: there was considerable damage to the printing devices, the building, and the logistics
hall. The dirty water left a picture of destruction. Up to $2000$ tons of paper were rendered unusable. The loss amounted of several
millions Swiss Francs. At the pharmaceutical company Siegfried AG, the basements of production facilities, warehouses and administration buildings were under water. There was major damage. Production could be resumed in the night of the $9^{th}$ of June.
There was only minor damage at Muller Martini and the Delta company. Cellars were also flooded on Henzmannstrasse. A traffic circle on Strengelbacherstrasse was under water.
\\
\\
\textbf{Old town.} At the Blumenheim retirement center, the basements of all buildings were affected (in part, $1$ m high water and mud). Heating, elevator, washing machines, cooling systems and power distribution were damaged. The home's manager estimated the damage at over $1$ million CHF. The rooms in the basement of the Lloyd Café were unusable for months. Huge damage also occurred in the bar \quotes{Key69}.
The entire wooden furnishings had to be replaced. In addition to a flooded basement, various stores had to deal with soaked and broken floors on the first floor. In the town hall, all carpets had to be replaced. 
Bar
staff and guests were locked in for about 2h. Food and beverage storage, storage and tank rooms were flooded. The water in the basement was so high that the ladies' room was no longer visible. At the Xpert AG insurance company's store on Rathausgasse, the water penetrated through the toilets into the basement. The store was flooded. At the Leutwyler bakery, all the equipment was broken due to the flooding of the basement. The building of the regional administration was under water.
The water penetrated into the parking facility of the main station and flooded two basements with over $100$ parked vehicles (they suffered total damage). $7$ vehicles of the regional police Zofingen were under water in the $2^{nd}$ basement of the parking station and had to be replaced.
On the $21^{st}$ of July the parking garage could be reopened. The power supply and electrical controls, the ventilation, the sprinkler system as well as the fire doors and the ticket machines had to be repaired beforehand. From the Blumenheim retirement center, the water flowed onto General Guisan road and on to the Luzernerstrasse traffic circle. Pomernweg was under water. The Strengelbacher road underpass under the railway tracks was flooded several meters deep. A motorist who got stuck in the flooded Henzmann road underpass was able to rescue himself from the car. One vehicle got stuck in the Altachen road underpass. Due to the storm, the Bahnhofplatz and two railway tracks in Zofingen were also under water, which had an impact beyond the region: the line between Basel and Ticino was interrupted between Olten and Lucerne. 
The flood washed out and damaged several roads. As a result of the heavy hail, leaves, branches and ferns were cut off in the Zofingen forest, which was washed away by the rainwater and filled road ditches and blocked the passages. The water then sought a new path, sweeping away mud and gravel. Half of the road in the Brunngraben area was washed away. At the Stöckliacher junction, the tar pavement was damaged. The damage to roads was estimated at $0.2$ million CHF.}

\section*{Supplementary Tables and Figures}

\aftertwo{
\onecolumn
\begingroup
\setlist[itemize]{label={--},nosep, leftmargin=*,
before=\vspace*{-\baselineskip}}
\setlength{\extrarowheight}{1.5pt}
\begin{small}
\rowcolors{1}{light-gray}{light-blue}

    \begin{longtblr}[theme = supplementary,
         caption = {This is a summary of the major social events that occurred between Friday, the $16^{th}$ of July 2017, and Sunday, the $30^{th}$ of July 2017, in the Swiss region surrounding the flooded area.
         The inundation occurred on the $8^{th}$ of July 2017.},
        ]{  
            colspec={m{0.9cm} m{0.1cm} m{0.6cm} m{1cm} m{3cm} m{7.3cm} m{1.5cm}},
            column{6}={preto={\minipage{\linewidth}}, appto={\endminipage}},
            row{1}={font=\bfseries}, %
            row{2,5,7-8,11-13,20-26,30,33-34,36,39-40,42}={bg=light-blue},
            hline{1,Z}=0.8pt, hline{2}=0.4pt, 
            hline{4,8,10,12-13,15-19,21-26,28-29,32,34,38,40}=dotted, 
            hline{3,5-7,9,11,14,20,27,30-31,33,35,36-37,39,41,42-43}=solid,
            rowhead=1
        }
        Day
        &
        & Month
        & Hours
        & Event
        & Description
        & Municipality
        \\
        Friday
        & $16$
        & June
        & 08 $-$ 19
        & 30th Aargau Cantonal Shooting Festival (ACSF) \cite{ACSF,Zofingen_eventsBlind} 
        & 
        A target shooting competition that included both rifle ($50$ and $300$ m) and pistol ($25$ and $50$ m) disciplines. A total of around $6500$ shooters attended the competition of ten days, which was spread out over three consecutive weekends (i.e. $3$ Fridays, $3$ Saturdays and $3$ Sundays, plus $1$ Monday). The first long weekend included both Friday and Monday. Each long weekend, approximately $2300$ marksmen took part in the competition. The competition was held in 7 shooting ranges, in 7 different municipalities. Here following the exact positions of the shooting ranges:

        {\centering
        \begin{tblr}{ |c|c|c| }
            \hline
             Municipality & Latitude (N) & Longitude (E) \\  
             \hline
             Zofingen & $47\degree16'59.09''$ & $07\degree57'30.56''$ \\ 
             Kölliken & $47\degree19'15.96''$ & $08\degree01'12.40''$ \\         
             Aarburg & $47\degree19'30.11''$ & $07\degree54'33.44''$ \\
             Brittnau & $47\degree14'57.44''$ & $07\degree57'08.93''$ \\         
             Reitnau & $47\degree15'15.52''$ & $08\degree01'53.98''$ \\
             Staffelbach & $47\degree16'39.54''$ & $08\degree02'27.28''$ \\         
             Muhen & $47\degree20'41.96''$ & $08\degree02'48.12''$ \\
             \hline
        \end{tblr}
        \par}
        \pcbACSF{505}Attendees
        & Multiple 
        \\
        \SetCell[r=2]{l} Saturday 
        & \SetCell[r=2]{l} 17 
        & \SetCell[r=2]{l} June
        & 08 $-$ 19
        & ACSF \cite{ACSF,Zofingen_eventsBlind} 
        & On Saturdays, the shooting match took place until 18:00 only in Aarburg and in Zofingen. 
        \newline
        \pcbACSF{802}Attendees
        & Multiple 
        \\
        &
        &
        & 9 $-$ 12
        & Edi Schneider Cup \newline\cite{RothristBlind,Rothrist_Edi_Schneider_Cup} 
        & Hockey tournament organised by the Inline Hockey Club (IHC) of Rothrist.
        \newline
        \pcbACSF{150}Attendees
        & Rothrist
        \\
        %
        Sunday
        & $18$
        & June
        & 08 $-$ 17
        & ACSF \cite{ACSF,Zofingen_eventsBlind} 
        & As all Sundays, in Aarburg, the shooting was from 09:00 to 16:00. Only on Sunday $18^{th}$, the afternoon session of shooting did not take place in Zofingen.
        \newline
        \pcbACSF{444}Attendees
        & Multiple 
        \\
        %
        Thursday
        & $22$
        & June
        & 
        & Senior citizens \newline excursion \cite{OftringenBlind} 
        & \pcbSeniorExcursion{300}Attendees
        & Oftringen 
        \\
        %
        \SetCell[r=2]{l} Friday 
        & \SetCell[r=2]{l} 23 
        & \SetCell[r=2]{l} June
        & 08 $-$ 19
        & ACSF \cite{ACSF,Zofingen_eventsBlind} 
        & \pcbACSF{803}Attendees
        & Multiple 
        \\
        &
        &
        & 14 $-$ 21
        & Bio Marché \cite{BioMBlind,Zofingen_eventsBlind}
        & The largest Swiss organic trade fair with approximately between $35000$ and $40000$ visitors over a $3-$day weekend. The fair was held in  the old town of Zofingen. On Friday $23^{rd}$, gastronomy events and concerts took place until 24:00.
        \newline
        \pcbBioM{12000}A.
        & Zofingen
        \\
        %
        \SetCell[r=2]{l} Saturday 
        & \SetCell[r=2]{l} 24 
        & \SetCell[r=2]{l} June
        & 08 $-$ 19
        & ACSF \cite{ACSF,Zofingen_eventsBlind} 
        & On Saturdays, the shooting match took place until 18:00 only in Aarburg and in Zofingen. 
        \newline
        \pcbACSF{969}Attendees
        & Multiple 
        \\
        &
        &
        & 10 $-$ 21
        & Bio Marché \cite{BioMBlind,Zofingen_eventsBlind}
        & On Saturday $24^{th}$, gastronomy events and concerts took place until 24:00.
        \newline
        \pcbBioM{12000}A.
        & Zofingen 
        \\
        %
        \SetCell[r=3]{l} Sunday 
        & \SetCell[r=3]{l} 25
        & \SetCell[r=3]{l} June
        & 08 $-$ 17
        & ACSF \cite{ACSF,Zofingen_eventsBlind} 
        & On Sundays, the shooting match took place from 09:00 to 16:00 only in Aarburg.
        \newline
        \pcbACSF{593}Attendees
        & Multiple 
        \\
        & 
        & 
        & 10 $-$ 18
        & Bio Marché \cite{BioMBlind,Zofingen_eventsBlind}
        & \pcbBioM{12000}A.
        & Zofingen 
        \\
        &
        &
        & 
        & Day of People \cite{RothristBlind} 
        & Religious event of the Roman Catholic Parish of St. Paul.
        & Rothrist 
        \\
        %
        \pagebreak
        \SetCell[r=6]{l} Friday 
        & \SetCell[r=6]{l} 30
        & \SetCell[r=6]{l} June
        & 08 $-$ 19
        & ACSF \cite{ACSF,Zofingen_eventsBlind} 
        & \pcbACSF{948}Attendees
        & Multiple 
        \\
        & 
        & 
        & 20 $-$ 02
        & School Festival \& City Festival \cite{Olten}
        & Two Festivals combined together, taking place over a $3-$day weekend. The event was held among two fairgrounds (\quotes{Kirchgasse} and \quotes{Schützenmatte}) in Olten, approximately where the Dünnern river meets the Aare river. On Friday $30^{th}$,  City Lounge since 20:00. It was not possible to estimate the number of attendees for the entire event \cite{Olten_Miriam_SiegristBlind}.
        & Olten
        \\
        &
        &
        & 17 $-$ 23
        & Youth Festival \cite{Safenwil,Safenwil_SchelshornBlind}
        & Festival for young people, on Friday evening and Saturday. The schedule for Friday evening included disco music (17:30 $-$ 19 and 20:30 $-$ 23), a concert (19 $-$ 20), and amusement park-related activities (from 17).
        \newline
        \pcbACSF{600}Attendees
        & Safenwil \& Walterswil
        \\
        &
        &
        & 18$-$ 23
        & Children's Festival \cite{OftringenBlind}
        & 
        & Oftringen
        \\
        &
        &
        & $16^{30}-$ $20^{30}$
        & Music Chat Reiden Mitte \cite{Reiden}
        & One evening of short concerts performed by four different music groups of the Reiden's Music Society (the Brassinis, the Junior Brass Band, the Brass Band, and the Senior Music).
        & Reiden
        \\
        &
        &
        & $13^{30}-$ $24$
        & Children's Festival 
        \newline
        \cite{RothristBlind,Rothrist_School}
        & Festival for the pupils of the Rothrist School, over a $3-$day weekend. On Friday $30^{th}$, a school Open Day was organised for public. An amusement park was set up for the evening.
        \newline
        \pcbACSF{1150}Attendees
        & Rothrist
        \\
        %
        \SetCell[r=7]{l} Saturday 
        & \SetCell[r=7]{l} 1
        & \SetCell[r=7]{l} July
        & 08 $-$ 19
        & ACSF \cite{ACSF,Zofingen_eventsBlind} 
        & On Saturdays, the shooting match took place until 18:00 only in Aarburg and in Zofingen. 
        \newline
        \pcbACSF{971}Attendees
        & Multiple 
        \\
        & 
        & 
        & $09^{30}-$ 02
        & School Festival \& City Festival \cite{Olten}
        & Kids (09:30 $-$ 12:30) and adults (16 $-$ 18:30) races with office chairs, plus concerts and disco music (16 $-$ 24). Live musical performances (live-acts), In Kirchgasse, street food, bar service (09 $-$ 02), and live-acts, i.e. live musical performances (20 $-$ 02).
        & Olten
        \\
        & 
        & 
        & 10 $ - 22^{30}$
        & Youth Festival \cite{Safenwil,Safenwil_SchelshornBlind}
        & Several concerts, exhibitions, and amusement park-related activities took place over the whole day. Fireworks were set off at 22:30. Although a road section was closed to traffic, and a detour was arranged by the fire department, no traffic congestion was reported \cite{Safenwil_SchelshornBlind}.
        \newline
        \pcbACSF{2000}Attendees
        & Safenwil \& Walterswil
        \\
        &
        &
        & 10 $-$ 24
        & Children's Festival \cite{OftringenBlind}
        & \pcbACSF{1500}Attendees (btw. $1000$ and $2000$)
        & Oftringen
        \\
        & 
        &
        & 10 $-$ 24
        & Children's Festival 
        \newline
        \cite{RothristBlind,Rothrist_School}
        & On Saturday $1^{st}$, the major activities included a children's parade (10 $-$ 11:30), the speech of the President of the Canton of Aargau -  Benjamin Giezendanner - (11:45), several concerts (13:30 $-$ 17:30) and a recreational sport event (12 $-$ 16).  An amusement park was set up for the entire afternoon and evening (12 $-$ 24).
        \newline
        \pcbACSF{3000}Attendees
        & Rothrist
        \\
        &
        &
        & 17 $-$ 02
        & Midsummer Night \newline Festival \cite{Brittnau}
        & Concerts by the Music Society of Brittnau.
        & Brittnau
        \\
        & 
        & 
        & $16^{30}-$ 22
        & Village Festival \cite{Dagmersellen}
        & Concerts by the Brass Band Uffikon-Buchs and by the Bärgblueme Dagmersellen yodelling club.
        & Dagmersellen
        \\
        %
        \SetCell[r=3]{l} Sunday 
        & \SetCell[r=3]{l} 2
        & \SetCell[r=3]{l} July
        & 08 $-$ 17
        & ACSF \cite{ACSF,Zofingen_eventsBlind} 
        & On Sunday $2^{nd}$, the shooting range of Reitnau was closed. In Aarburg, the shooting match lasted from 09:00 to 16:00, as all Sundays.
        \newline
        \pcbACSF{358}Attendees
        & Multiple 
        \\
        & 
        &
        & 07 $-$ 21
        & School Festival \& City Festival \cite{Olten}
        & Drums concert in the city center (starting at 07:00), parade (starting at 09:45), Olten Music School's and Adrian
         Stern's concerts (12:30 $-$ 18:00), kids games (13 $-$ 17), and electronic music (up to 21).
        & Olten
        \\
        &
        &
        & $09^{30}-$ $19$
        & Children's Festival \cite{Rothrist_School}
        & Ecumenical worship (09:30 $-$ 10:30) and amusement park (11 $-$ 19).
        & Rothrist
        \\
        %
        Monday
        & 3
        & July
        & 17 $-$ 00:30
        & New Orleans Meets in Zofingen (NOMZ) \cite{Zofingen_NOMZ}
        & An evening-long jazz \& blues festival (17 $-$ 00:30). Eight bands performed over three stages in the old town of Zofingen (in Thutplatz, in Kirchplatz and in Chorplatz). Food and beverage service provided by around 20 stands.
        & Zofingen
        \\
        %
        \SetCell[r=2]{l} Thursday 
        & \SetCell[r=2]{l} 6
        & \SetCell[r=2]{l} July
        & 18 $-$ 03
        & Tattoo Concert \& \newline Children's Festival \cite{Zofingen_Children_FestivalBlind}
        & Music parade of marching drummers in the old town of Zofingen (19:30 $-$ 20), followed by \quotes{Tattoo} concerts (typically called \quotes{Zapfenstreich} in the German-speaking countries) still in the old town of Zofingen (20 $-$ 02). Music stopped at 02 in the Sundays's morning. The overall event started with some opening activities, promoted by the Reformed Church's community (18 $-$ 19:30).
        & Zofingen
        \\
        &
        &
        & 17 $-$ 20
        & Fashion Show by SMG Olten \cite{Olten_Fashion_Show} 
        & Two fashion shows, where the students of SMG Olten presented their own work to the public. Each show lasted around one hour. The shows took place at the main entrance staircases of the BBZ building, where approximately 150 seats were set, as well as an area were around 100 guests could stand.
        \newline
        \pcbACSF{250}Attendees
        & Olten 
        \\
        %
        \SetCell[r=2]{l} Friday 
        & \SetCell[r=2]{l} 7
        & \SetCell[r=2]{l} July
        & 07 $-$ 23
        & Tattoo Concert \& \newline Children's Festival \cite{Zofingen_Children_FestivalBlind}
        & The morning and the first afternoon were marked by military-like parades of the Zofingen's Cadets Corp in the old town of Zofingen (07 $-$ 10, 11 $-$ 11:45, and 13:45 $-$ 14:45), and by a solemn celebration in the Reformed \quotes{Town} Church of Zofingen (07 $-$ 11:30). The afternoon was instead characterized by a battle reenactment on a parade ground, called \quotes{Heiternplatz}, southeast of the Zofingen's old town (14:45 $-$ 16), and by a number of concerts (16 $-$ 19:30). The Festival ended with a torch and lantern procession (22 $-$ 23).
        \newline
        \pcbACSF{1000}Attendees
        & Zofingen
        \\
        &
        &
        & 18 $-$ 23
        & Dinner \cite{Staffelbach}
        & Fish dinner of the Music Society of Staffelbach 
        & Staffelbach
        \\
        Friday
        & 14
        & July
        & 19 $-$ 03
        & Ski Festival \cite{Rothrist_Skifest_1,Rothrist_Skifest_2Blind}
        & Food and drink, and music festival.
        \newline
        \pcbACSF{2000}Attendees
        & Rothrist
        \\
        Saturday
        & 15
        & July
        & 18 $-$ 03
        & Ski Festival \cite{Rothrist_Skifest_1,Rothrist_Skifest_2Blind}
        & Food and drink, and music festival.
        \newline
        \pcbACSF{1500}Attendees
        & Rothrist
        \\
        \SetCell[r=2]{l} Friday 
        & \SetCell[r=2]{l} 21
        & \SetCell[r=2]{l} July
        & $20^{30}-$ 22
        & Variété Pavé \cite{Olten_Variete}
        & Street theatre in the old town of Olten (Kirchgasse).
        & Olten
        \\
        & 
        & 
        & 18 $-$ 02
        & Beach Festival \cite{Aarburg_Beach_Festival_1,Aarburg_Beach_Festival_3Blind}
        & Musical entertainment, artisan food products and amusement park.
        \newline
        \pcbACSF{2000}Attendees
        & Aarburg
        \\
        \pagebreak
        \SetCell[r=2]{l} Saturday 
        & \SetCell[r=2]{l} 22
        & \SetCell[r=2]{l} July
        & $20^{30}-$ 22
        & Variété Pavé \cite{Olten_Variete}
        & Street theatre in the old town of Olten (Kirchgasse).
        & Olten
        \\
        &
        &
        & 18 $-$ 02
        & Beach Festival \cite{Aarburg_Beach_Festival_1,Aarburg_Beach_Festival_3Blind}
        & Concerts and artisan food products. On Saturday, the highlight of the Festival was a choreographed musical firework display, where the background music was synchronized with the fireworks. Shortly after darkening (sunset around 21:15), fireworks were set off from the banks of the Aare river, and they lasted around half an hour.
        \newline
        \pcbACSF{5000}Attendees
        & Aarburg
        \\
        Sunday
        & 23
        & July
        & $20^{30}-$ 22
        & Variété Pavé \cite{Olten_Variete}
        & Street theatre in the old town of Olten (Kirchgasse).
        & Olten
        \\
        Saturday
        & 29
        & July
        & 17 $-$ 04
        & Summer Party \cite{Reiden_Summer_Party_1,Reiden_Summer_Party_2Blind}
        & Music performances and street food. A fireworks display in the evening.
        \newline
        \pcbACSF{500}Attendees
        & Reiden
        \\
    \end{longtblr}

\begin{table}[H]
\centering
\caption{\label{tab:boxplot}Descriptive statistics about \textbf{(a)} the distributions of distances among the clusters centers and event centers, and \textbf{(b)} the distributions of clusters areas, for each major event. With $Q_1$, $Q_2$ and $Q_3$ we indicate, respectively, the lowest quartile (i.e. the $25^{th}$ percentile), the median (i.e. the $50^{th}$ percentile), and the  upper quartile (i.e. the $75^{th}$ percentile) of the distribution. Those three quartiles are necessary to calculate the Bowley’s coefficient of skewness (BS), i.e. $BS = \frac{(Q_3+Q_1-2\times Q_2)}{(Q_3-Q_1)}$.
The Fisher-Pearson coefficient of skewness (FPS) needs instead the mean of the distribution, i.e.
$
FPS = \frac{\frac{1}{n_s}\sum_{i=1}^{n_s}(x_i - \left \langle x \right \rangle)^{3}}
{\left(\sqrt{\frac{1}{n_s}\sum_{i=1}^{n_s}(x_i - \left \langle x \right \rangle)}\right)^{3}}
$, where $x_i$ represents either each distance between a cluster and the event, or each cluster area, $\left \langle x \right \rangle$ indicates the mean of the distribution, and $n_s$ is the sample size (i.e. $i=1,2,\ldots,n_s$). In FPS, we can recognise the standard deviation of the distribution, i.e. $\sigma = \sqrt{\frac{1}{n_s}\sum_{i=1}^{n_s}(x_i - \left \langle x \right \rangle)}$.
}
\subcaption{Summary statistics about the distributions of distances among clusters centers and events centers.} %
\begin{tblr}
{
   width = \textwidth, 
   colspec = {lllllllll}, 
   hlines,
   row{1} = {font=\bfseries}, 
   row{2,4,6,8,10,12}={bg=light-blue},
   hline{1,Z}=0.8pt, hline{2}=0.4pt, 
   hline{1-12}=solid,
   rowhead=1
}
Event $\quad \left(\times 10^{3} \; m \rightarrow \right)$ 
& Min & Q1 & Q2 & Q3 & Max & Mean & Bowley Skewness & FP Skewness \\ 
Bio Marché Fri & 0.86 & 4.18 & 5.62 & 7.85 & 12.66 & 5.92 & 0.21 & 0.16 \\ 
Bio Marché Sat & 0.89 & 3.61 & 5.43 & 7.86 & 11.8 & 5.65 & 0.14 & 0.05 \\ 
Bio Marché Sun & 0.39 & 3.58 & 4.92 & 7.49 & 10.78 & 5.39 & 0.32 & 0.17 \\ 
NOMZ & 0.85 & 1.57 & 4.05 & 7.15 & 8.05 & 4.28 & 0.11 & 0.07 \\ 
Tattoo Concert Thu & 0.8 & 1.46 & 2.3 & 3.82 & 7.12 & 2.68 & 0.29 & 0.97 \\ 
Tattoo \& Child. Fest & 0.95 & 3.54 & 4.96 & 7.82 & 12.29 & 5.5 & 0.34 & 0.39 \\ 
Flood & 0.32 & 3.55 & 5.52 & 7.6 & 11.69 & 5.75 & 0.03 & 0.54 \\ 
Ski Fest. Fri & 0.73 & 4.01 & 6.02 & 8.77 & 13.48 & 6.36 & 0.16 & 0.18 \\ 
Ski Fest. Sat & 0.76 & 3.02 & 5.2 & 7.32 & 12.55 & 5.34 & $-$0.01 & 0.56 \\ 
Beach Festival & 0.27 & 2.72 & 4.05 & 6.18 & 11.12 & 5.03 & 0.23 & 1.25 \\ 
Summer Party & 1.64 & 5.62 & 8.86 & 11.93 & 16.33 & 8.72 & $-$0.03 & $-$0.1 \\ 
\end{tblr}
\bigskip
\subcaption{Summary statistics about the distributions of clusters areas.} %
\begin{tblr}
{
   width = \textwidth, 
   colspec = {lllllllll}, 
   hlines,
   row{1} = {font=\bfseries}, 
   row{2,4,6,8,10,12}={bg=light-blue},
   hline{1,Z}=0.8pt, hline{2}=0.4pt, 
   hline{1-12}=solid,
   rowhead=1
}
Event $\quad \left(\times 10^{5} \; m^{2} \rightarrow \right)$ 
& Min & Q1 & Q2 & Q3 & Max & Mean & Bowley Skewness & FP Skewness \\ 
Bio Marché Fri & 0 & 0.89 & 2.86 & 10.9 & 25.23 & 11.19 & 0.61 & 4.68 \\ 
Bio Marché Sat & 0 & 0.7 & 2.44 & 6.3 & 13.32 & 11.01 & 0.38 & 10.58 \\ 
Bio Marché Sun & 0.01 & 0.81 & 3.22 & 10.3 & 24.12 & 11.7 & 0.49 & 8.69 \\ 
NOMZ & 0 & 0.14 & 0.34 & 2.52 & 4.58 & 7.23 & 0.83 & 4.22 \\ 
Tattoo Concert Thu & 0 & 0.41 & 1.56 & 5.44 & 10 & 9.92 & 0.54 & 3.74 \\ 
Tattoo \& Child. Fest & 0.01 & 0.64 & 3.28 & 10.9 & 23.13 & 14.31 & 0.48 & 7.07 \\ 
Flood & 0 & 0.67 & 2.14 & 8.28 & 19.22 & 8.93 & 0.61 & 3.66 \\ 
Ski Fest. Fri & 0 & 0.28 & 1.38 & 13.08 & 29.83 & 21.5 & 0.83 & 6.12 \\ 
Ski Fest. Sat & 0 & 0.12 & 1.29 & 7.93 & 18.59 & 11.91 & 0.7 & 4.14 \\ 
Beach Festival & 0.02 & 1.85 & 4.6 & 15.03 & 33.36 & 16.32 & 0.58 & 3.88 \\ 
Summer Party & 0 & 0.59 & 2.38 & 6.27 & 14.22 & 8.23 & 0.37 & 4.42 \\ 
\end{tblr}
\end{table}

\begin{table}[H]
\centering
\caption{Coefficients of the empirical relationships among distances (between the clusters centers and the event center), clusters areas, and number of clusters, for each event. The goodness of fit of the three relationships is assessed through the coefficient of determination, $R^2$.}
\subfloat[Coefficients of $r_{\text{max}} = a_1 \Delta r + b_1$, where $r_{\text{max}}$ is the distance of the farthest cluster (from the event center), and $\Delta r=r_{\text{max}}-r_{\text{min}}$ is the difference between the farthest and the clostest cluster distances.]{\label{a}\begin{tblr}
{
   width = \textwidth, 
   colspec = {llll}, 
   hlines,
   row{1} = {font=\bfseries}, 
   row{2,4,6,8,10,12}={bg=light-blue},
   hline{1,Z}=0.8pt, hline{2}=0.4pt, 
   hline{1-12}=solid,
   rowhead=1
}
Event & $\mathbf{a_1}$ & $\mathbf{b_1}$ & $\mathbf{R^2}$ \\ 
Bio Marché Fri & 0.84 & 3020.22 & 0.87 \\ 
Bio Marché Sat & 0.79 & 3163.22 & 0.91 \\ 
Bio Marché Sun & 0.61 & 4525.32 & 0.91 \\ 
NOMZ & 0.99 & 1641.08 & 0.71 \\ 
Tattoo Concert Thu & 0.86 & 1596.33 & 0.93 \\ 
Tattoo \& Child. Fest & 0.85 & 2802.61 & 0.92 \\ 
Flood & 0.81 & 3573.71 & 0.87 \\ 
Ski Fest. Fri & 0.45 & 7512.03 & 0.79 \\ 
Ski Fest. Sat & 0.69 & 4611.82 & 0.8 \\ 
Beach Festival & 1 & 1138.07 & 0.95 \\ 
Summer Party & 0.33 & 10554.52 & 0.37 \\ 
\end{tblr}}
\hspace{0.05\textwidth} %
\subfloat[Coefficients of $n_c=a_2\langle A_c \rangle^{b_2}$, where $n_c$ is the number of clusters and $\langle A_c \rangle$ is the average size of the clusters areas.]
{\label{b}\begin{tblr}
{
   width = \textwidth, 
   colspec = {llll}, 
   hlines,
   row{1} = {font=\bfseries}, 
   row{2,4,6,8,10,12}={bg=light-blue},
   hline{1,Z}=0.8pt, hline{2}=0.4pt, 
   hline{1-12}=solid,
   rowhead=1
}
Event & $\mathbf{a_2}$ & $\mathbf{b_2}$ & $\mathbf{R^2}$ \\ 
Bio Marché Fri & 1078.56 & $-$0.32 & 0.28 \\ 
Bio Marché Sat & 968.44 & $-$0.32 & 0.6 \\ 
Bio Marché Sun & 16.7 & $-$0.06 & 0.01 \\ 
NOMZ & 1.59 & 0.05 & 0.03 \\ 
Tattoo Concert Thu & 6.07 & $-$0.03 & 0.03 \\ 
Tattoo \& Child. Fest & 31.04 & $-$0.09 & 0.03 \\ 
Flood & 5417.26 & $-$0.45 & 0.65 \\ 
Ski Fest. Fri & 478.91 & $-$0.31 & 0.94 \\ 
Ski Fest. Sat & 311.49 & $-$0.26 & 0.68 \\ 
Beach Festival & 5.86 & 0.02 & 0.01 \\ 
Summer Party & 725459.48 & $-$0.82 & 0.41 \\ 
\end{tblr}}\par 
\bigskip
\subfloat[Coefficients of $d_{c(r_{min})} = a_3 log(A_{c(r_{min})}) + b_3$, where $d_{c(r_{min})}$ is the distance closest cluster, and $A_{c(r_{min})}$ is the area closest cluster.]
{\label{c}\begin{tblr}
{
   width = \textwidth, 
   colspec = {llll}, 
   hlines,
   row{1} = {font=\bfseries}, 
   row{2,4,6,8,10,12}={bg=light-blue},
   hline{1,Z}=0.8pt, hline{2}=0.4pt, 
   hline{1-12}=solid,
   rowhead=1
}
Event & $\mathbf{a_3}$ & $\mathbf{b_3}$ & $\mathbf{R^2}$ \\ 
Bio Marché Fri & 204.84 & $-$1265.51 & 0.65 \\ 
Bio Marché Sat & 268.55 & $-$1955.41 & 0.4 \\ 
Bio Marché Sun & 82.13 & 743.56 & 0.04 \\ 
NOMZ & $-$86.25 & 2149.05 & 0.28 \\ 
Tattoo Concert Thu & 58.14 & 391.11 & 0.7 \\ 
Tattoo \& Child. Fest & 102.77 & 198.7 & 0.2 \\ 
Flood & $-$111.98 & 3438.23 & 0.03 \\ 
Ski Fest. Fri & 533.33 & $-$4724.38 & 0.37 \\ 
Ski Fest. Sat & 33.38 & 1280.12 & 0.02 \\ 
Beach Festival & 63.06 & 196.62 & 0.01 \\ 
Summer Party & 526.67 & $-$4154.87 & 0.92 \\ 
\end{tblr}}
\end{table}

\pagebreak

\begin{longtblr}[theme = supplementary,
    caption = {List of symbols in the order in which they appear.},]
    {  
        colspec={ll},
        row{1}={font=\bfseries},
        row{2,4,6,8,10,12,14,16,18,20,22,24,26,28,30,32,34,36,38,40,42,44,46,48,50,52,54,56,58,60,62,64,66,68,70}={bg=light-blue},
        hline{1,Z}=0.8pt, hline{2}=0.4pt, 
        hline{1-100}=solid,
        rowhead=1
   }
    Symbol & Description \\ 
    $i,j,k$ & Indices \\
    $P_i$ and $P_j$ & Estimated probability distributions \\ 
    $D=(d_{ijk})$ & Dissimilarity matrix (of vectors), with element $d_{ijk}$ \\ 
    $f_k(P_i,P_j)$ & $k^{th}-$function of distance or divergence \\ 
    $Q$ & Quartiles of a distribution \\
    $\left \langle Q_2 \right \rangle$ & Average median (boxplots) \\
    $r_{\text{max}}$ and $r_{\text{min}}$ & Maximum and minimum distances of clusters from the event center \\
    $\Delta r$ & $r_{\text{max}}-r_{\text{min}}$ \\
    $n_c$ & Number of clusters \\
    $A_c$ & Area of clusters \\
    $\langle A_c \rangle$ & Mean area of the clusters \\
    $d_{c(r_{min})}$ & Distance of the closest cluster \\
    $A_{c(r_{min})}$ & Area of the closest cluster \\
    $a_1$, $a_2$, $a_3$, $b_1$, $b_2$, $b_3$ & Coefficients of the empirical relationships \\
    $\X$ and $\Y$ & Sample spaces (Polish metric spaces in the Wasserstein distance) \\ 
    $\A$ & $\sigma-$algebra \\
    $\left( \X,\A\right)$ & Measurable space \\ 
    $\rho_1$  and $\rho_2$ & Densities of $P_1$ and $P_2$ \\
    $\omega_1$ and $\omega_2$ & Weights of $P_1$ and $P_2$ \\
    $H(x)$ & Mixture distribution of $P_1$ and $P_2$.\\
    $\text{cost}(x,y)$ & Cost function \\
    $\mu$ and $\nu$ & Probability measures \\
    $\pi$ & Joint probability measure \\
    $\Pi(\mu,\nu)$ & Set of all joint probability measures on $\X \times \Y$ whose marginals are $\mu$ and $\nu$ \\
    $F$ and $G$ & Cumulative distribution functions (CDFs) associated with $\mu$ and $\nu$ \\
    $F^{-1}$ and $G^{-1}$ & Quantile functions of $\mu$ and $\nu$ \\ 
    $\left( \X,d\right)$ & Polish metric space equipped with a metric $d(x,y)$ \\ 
    $p$ & $p-$th power of the metric $d(x,y)$ \\
    $f_{SE}(P_i)$ & Shannon entropy function \\
    $f_{JS}(P_1,P_2)$ & Jensen–Shannon divergence \\
    $f_{KL}(P_1\mathrel{\Vert}P_2)$ & Kullback–Leibler divergence \\
    $f_{TV}(P_1,P_2)$ & Total variation distance \\
    $f_{H}(P_1,P_2)$ & Hellinger distance \\
    $W_{p}(\mu,\nu)$ & Wasserstein distance of order $p$ between $\mu$ and $\nu$ \\
    $W_{1}(\mu,\nu)$ & $W_1$ form of the Wasserstein distance (Kantorovich–Rubinstein distance) \\
    $X$ and $Y$ & Random variables \\ 
    $f_{DC}(X,Y)$ & Empirical distance correlation \\
    $f^{\complement}_{DC}(X,Y)$ & Complementary empirical distance correlation \\
    $\V(X)$, $\V(Y)$, $\V(X,Y)$ & Coefficients to calculate the empirical distance correlation \\
    $a_{ij}$, $b_{ij}$, $A_{ij}$, $B_{ij}$ & Coefficients to calculate the empirical distance correlation \\
    $u$ and $v$ & Dimensions of spaces $\mathbb{R}^{u}$ and $\mathbb{R}^{v}$ (in the empirical distance correlation)\\
    $f_{CS}(X,Y)$ & Cosine similarity \\
    $f_{CD}(X,Y)$ & Cosine distance \\
    $n$ & Length of the time series \\
    $m$ & Length of the time subsequence \\
    $T=(t_k)_{k=1}^{n}$ & Time series, with element $t_k$ \\
    $T_{i,m}$ & Time subsequence of length $m$, starting at $i$ \\
    $\delta_{i}$ & Distance profile \\
    $M$ & Matrix profile \\
    $\delta_{i}^{*}$ & Left distance profile \\
    $M^{*}$ & Left matrix profile \\
    $I$ & CurrentIndex (split point between the training data and the test data) \\
    $\Upsilon$ & Set of points in $\R^{2}$ \\
    $\epsilon$ & Radius defining the neighborhood of point $i$ \\
    $\epsilon N_{i}$ & $\epsilon-$neighborhood of point $i$ \\
    $\gamma$ & Minimum number of points that lie within a $\epsilon-$neighborhood \\
    $k-$NNG & $k-$nearest neighbor graph \\
    ${kNN}_i$ & $k-$nearest neighborhood of an arbitrary node $i$ \\
    $RkNN_{i}$ & Reverse $k-$nearest neighborhood of node $i$ \\ 
    $k^{\text{out}}_i$ & \quotes{Outgoing degree} of node $i$ \\
    $k^{\text{in}}_i$ & \quotes{Ingoing degree} of node $i$ \\
    $(i,j)$ & Edge from node $i$ to node $j$ \\
    $c_{k}$ & Number of clusters for different values of $k$ (from RNN-DBSCAN) \\
    $C=(c_{k})_{k=1}^{100}$ & Sequence of ($100$ dfferent) number of clusters, when $k$ varies  \\
    $\operatorname{img}(C)= \{c_{k}\}_{k=1}^{100}$ & Image of $C$ \\
    $C^{-1}(\{c_{k}\})$ & Preimage of element $c_{k}$ \\
   \end{longtblr}

\end{small}
\endgroup
\twocolumn
}


\aftertwo{
\onecolumn
\begingroup
\setlist[itemize]{label={--},nosep, leftmargin=*,
before=\vspace*{-\baselineskip}}
\setlength{\extrarowheight}{1.5pt}
\begin{small}

\begin{figure*}
        \centering
        \includegraphics[width=\textwidth]{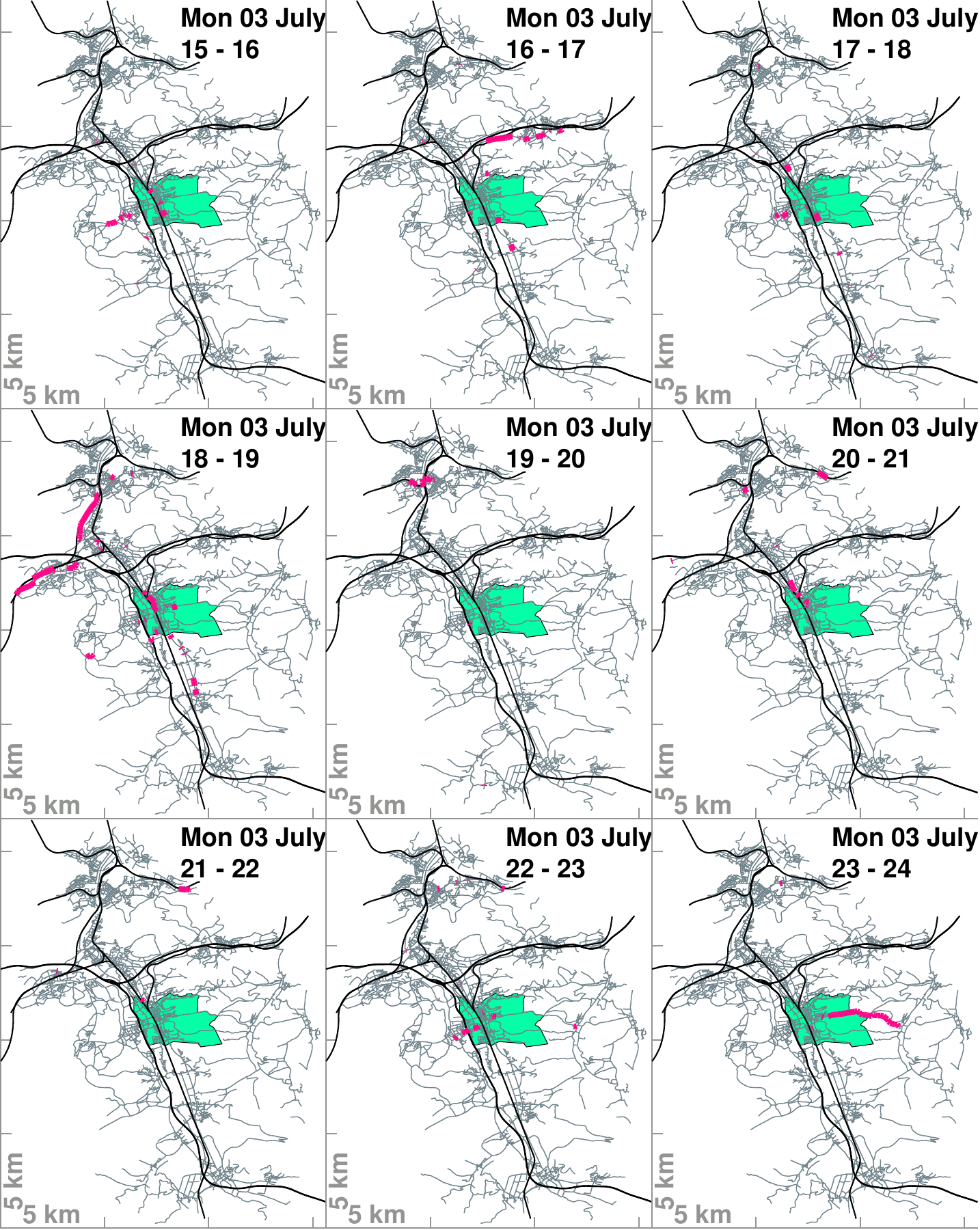}
        \caption{Spatio-temporal dynamics of anomalies (pink edges) during NOMZ in the municipality of Zofingen (green patch).}
        \label{fig:mylabel_2}
\end{figure*}

\begin{figure*}
        \centering
        \includegraphics[width=\textwidth]{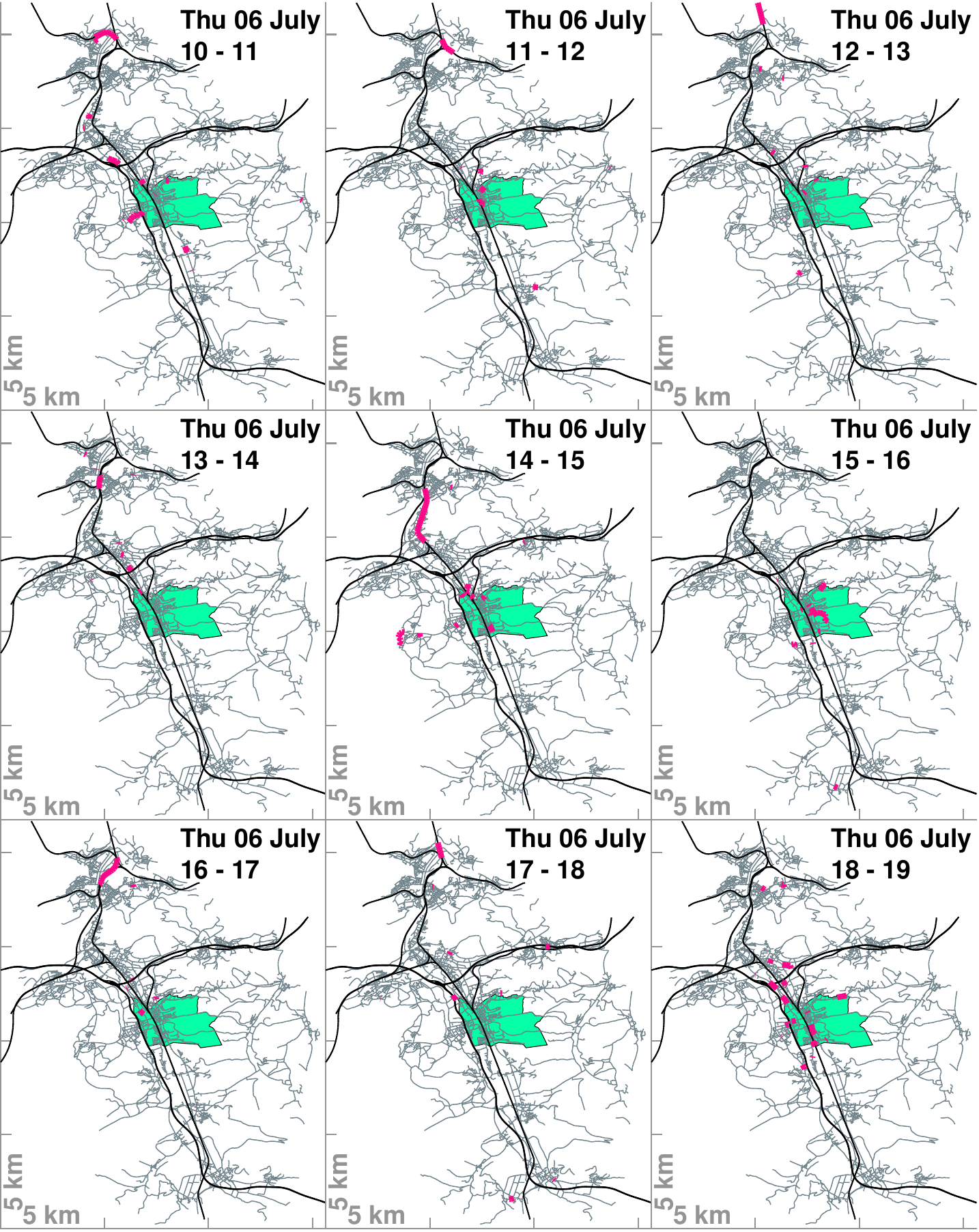}
        \label{sup_fig_anomalies_flood}
\end{figure*}
\begin{figure*}
        \centering
        \includegraphics[width=\textwidth]{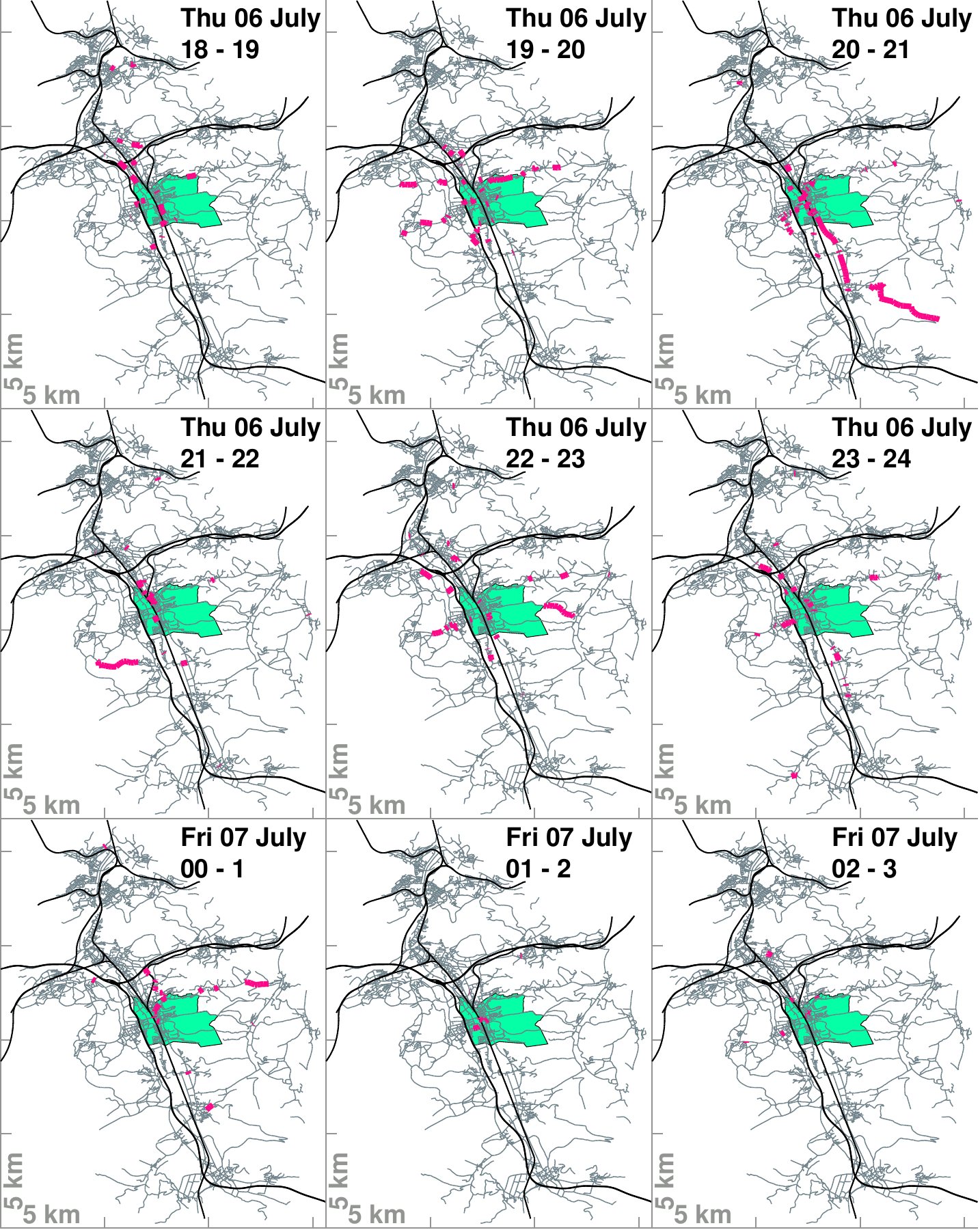}
        \caption{Spatio-temporal dynamics of anomalies (pink edges) during the Tattoo Concert \&
Children’s Festival on Thursday the $7^{th}$ of July that occurred in the municipality of Zofingen (green patch).}
\end{figure*}

\begin{figure*}
        \centering
        \includegraphics[width=\textwidth]{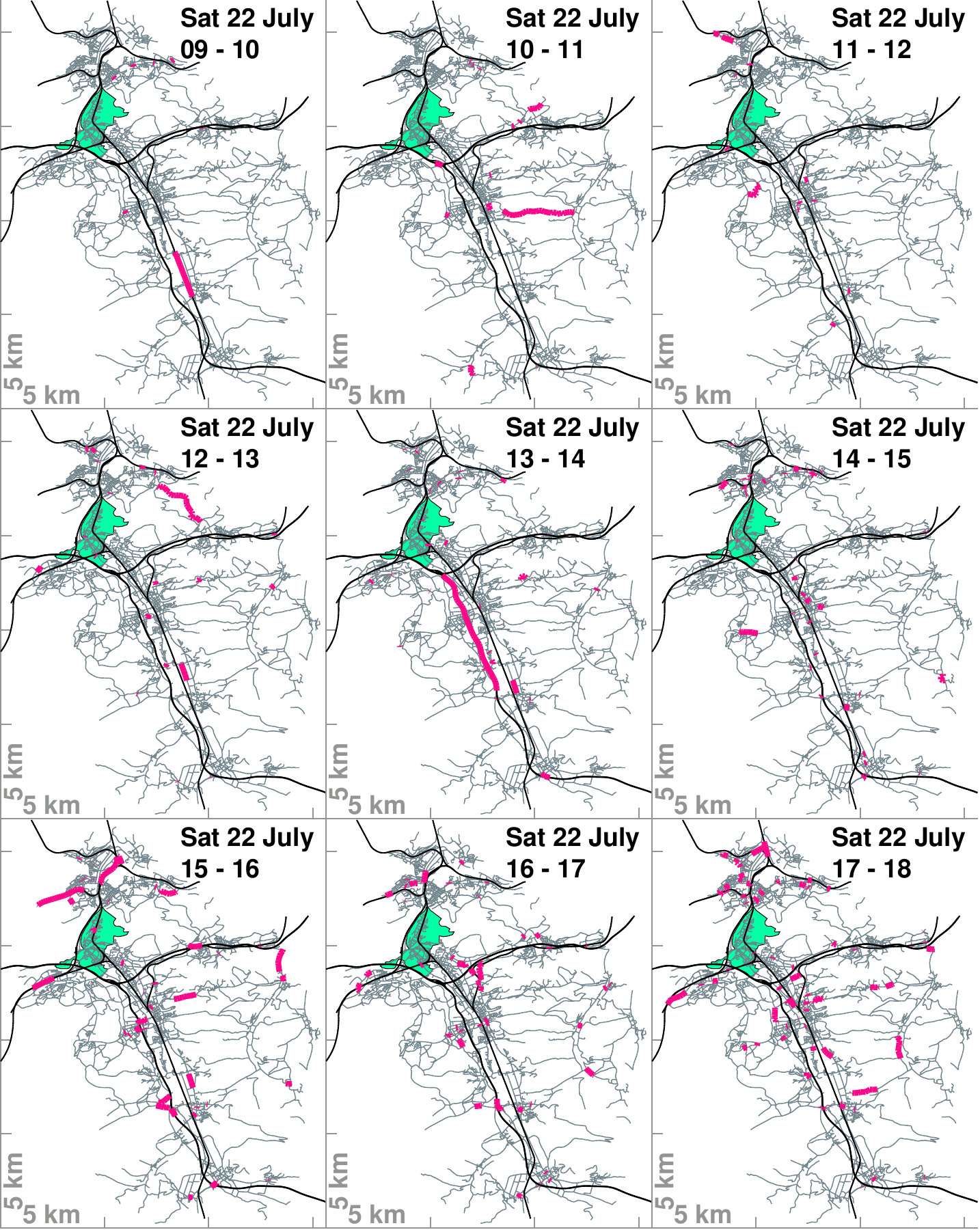}
        \label{sup_fig_anomalies_flood}
\end{figure*}
\begin{figure*}%
        \centering
        \includegraphics[width=\textwidth]{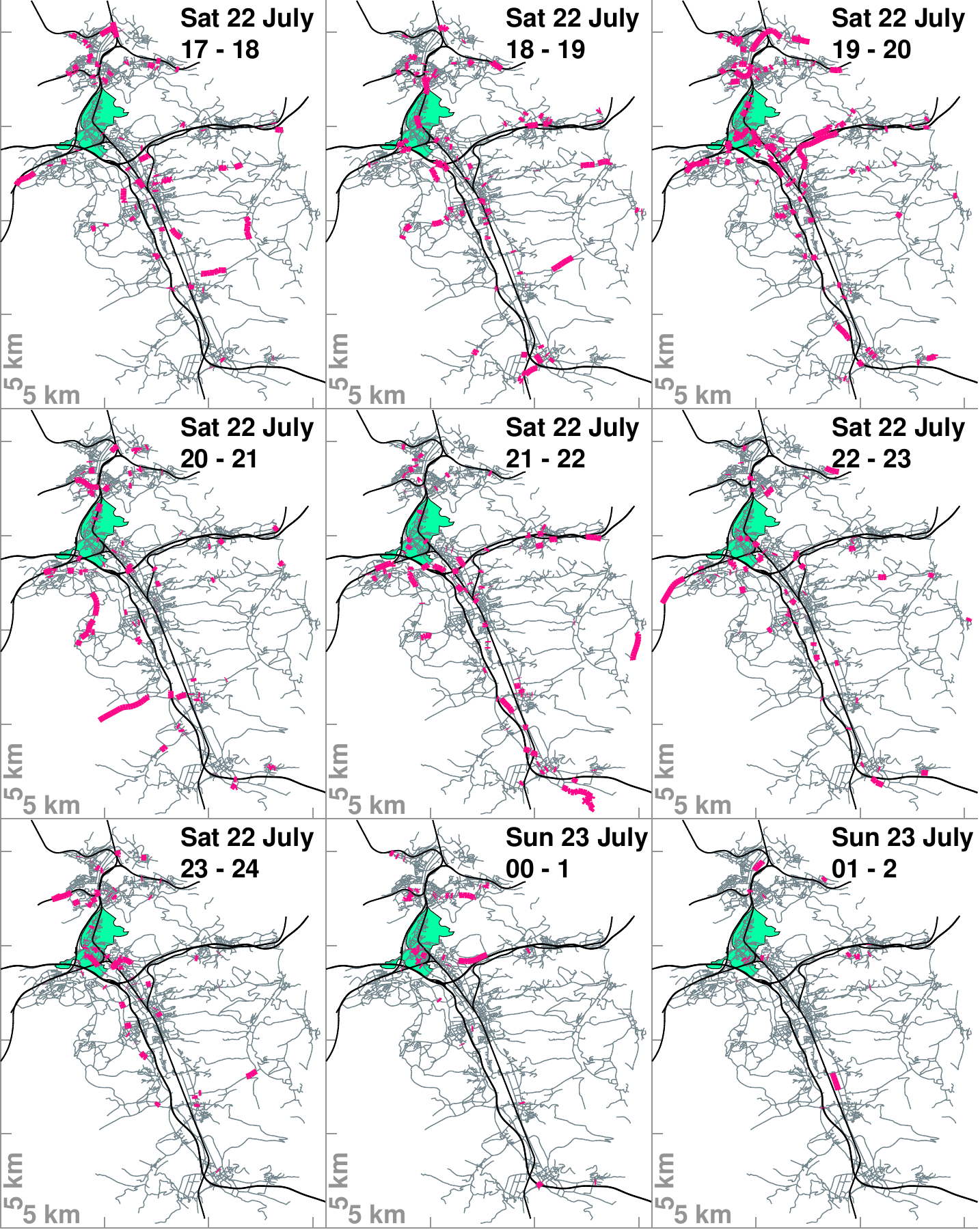}
        \caption{Spatio-temporal dynamics of anomalies (pink edges) during the Beach Festival that occurred in the municipality of Aarburg (green patch).}
\end{figure*}

\begin{figure*}
        \centering
        \includegraphics[width=\textwidth]{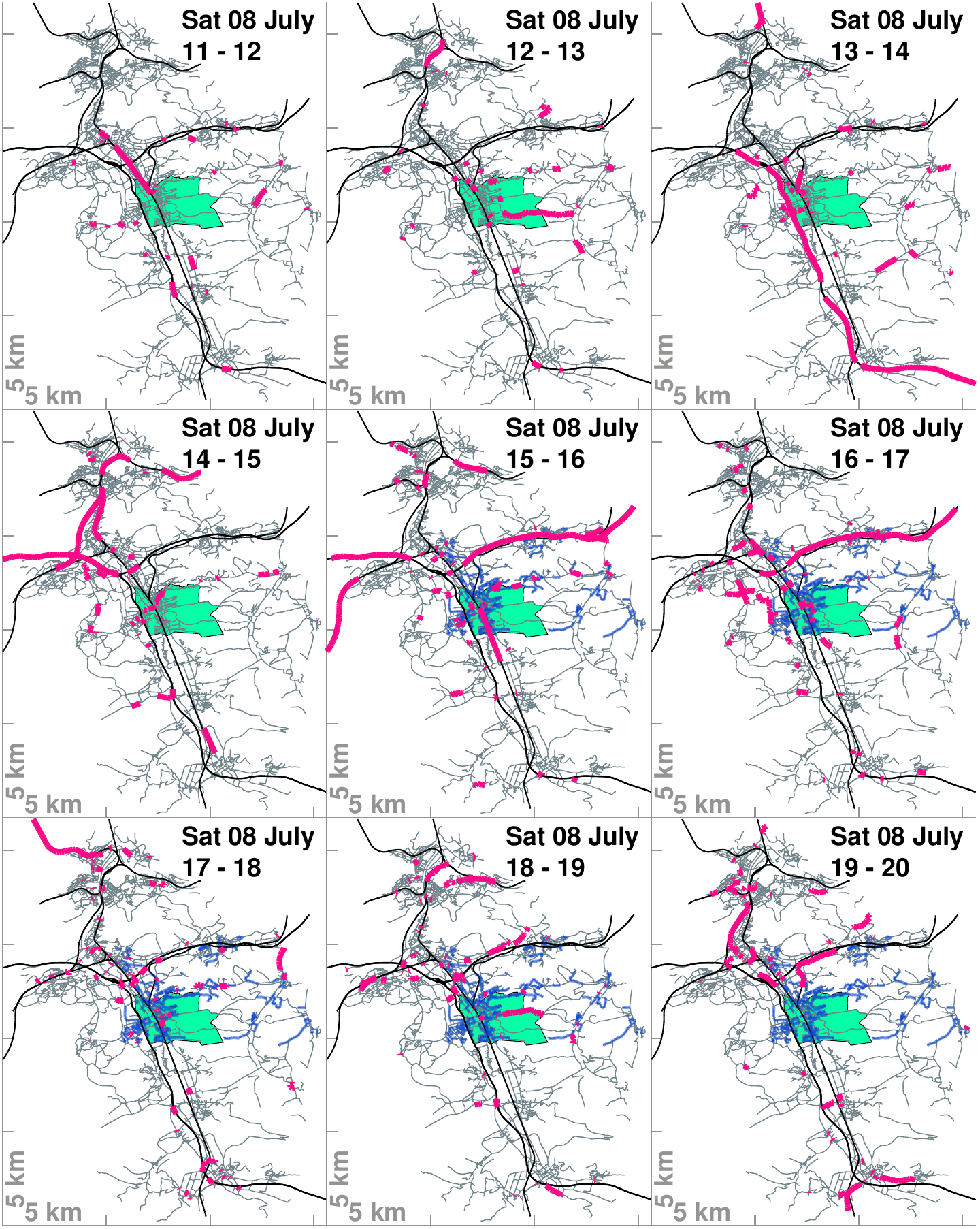}
        \label{sup_fig_anomalies_flood}
\end{figure*}
\begin{figure*}%
        \centering
        \includegraphics[width=\textwidth]{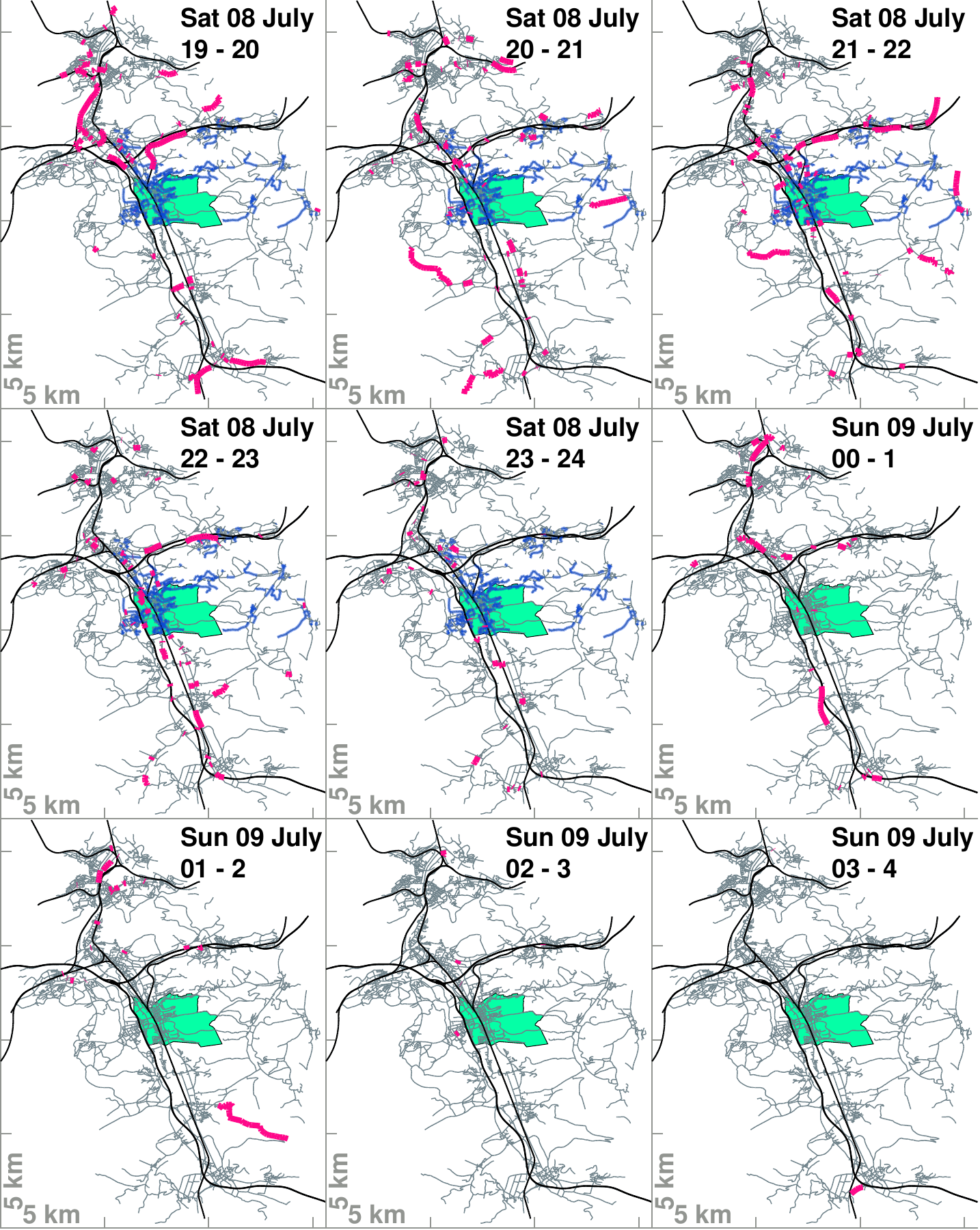}
        \caption{Spatio-temporal dynamics of anomalies (pink edges) during the flooding event that occurred in the municipality of Zofingen (green patch) and its surroundings. Roads with flooded buildings are highlighted in blue.}
\end{figure*}

\begin{figure*}[ht]
        \centering
        \includegraphics[width=0.75\textwidth]{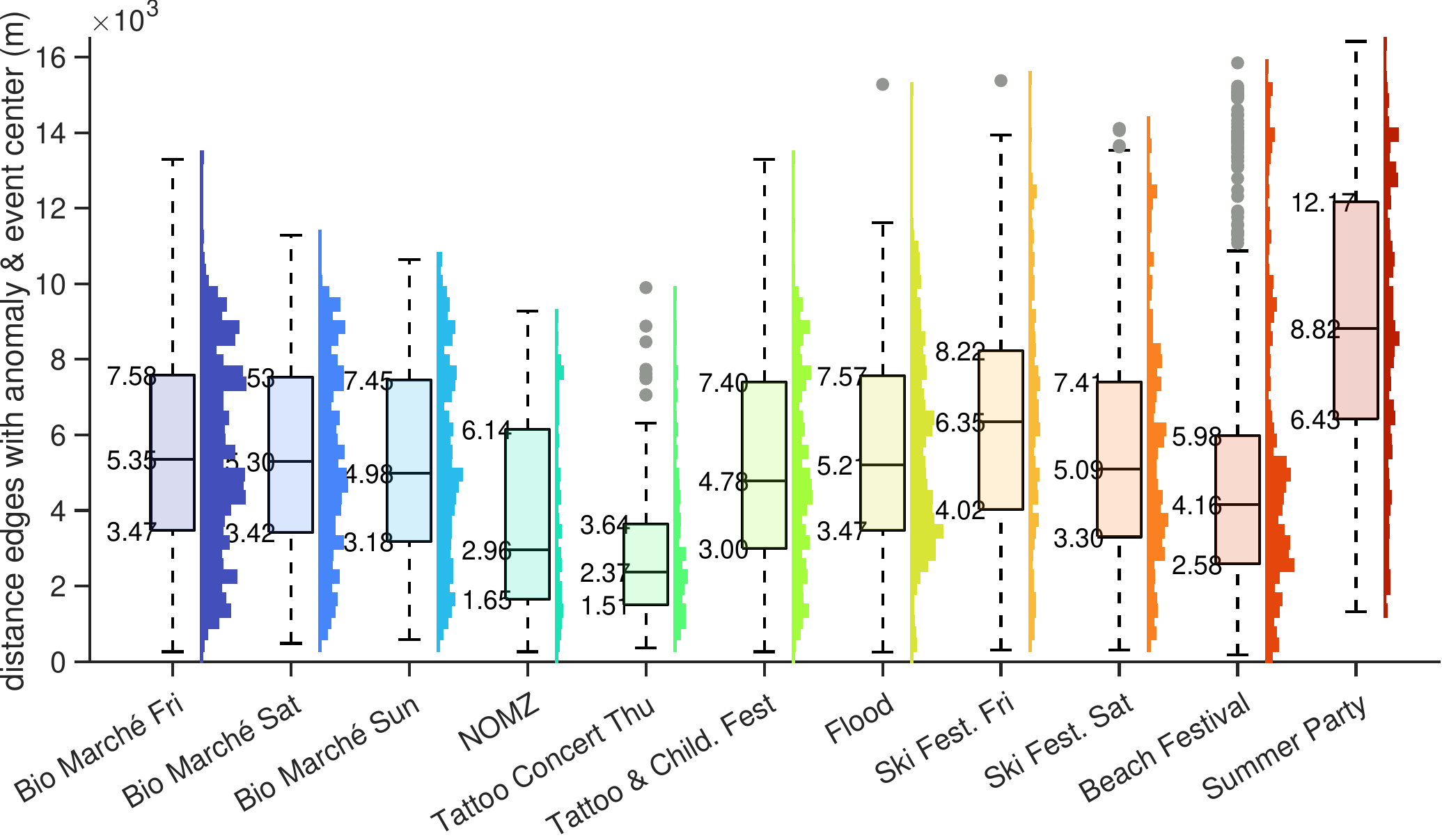}
        \label{sup_fig_distance_all_edges}
        \caption{Distribution of distances among the edges with anomalies and the municipality center where an event took place, for every major documented event.
        }     
\end{figure*}

\begin{figure*}
        \centering
        \includegraphics[width=0.75\textwidth]{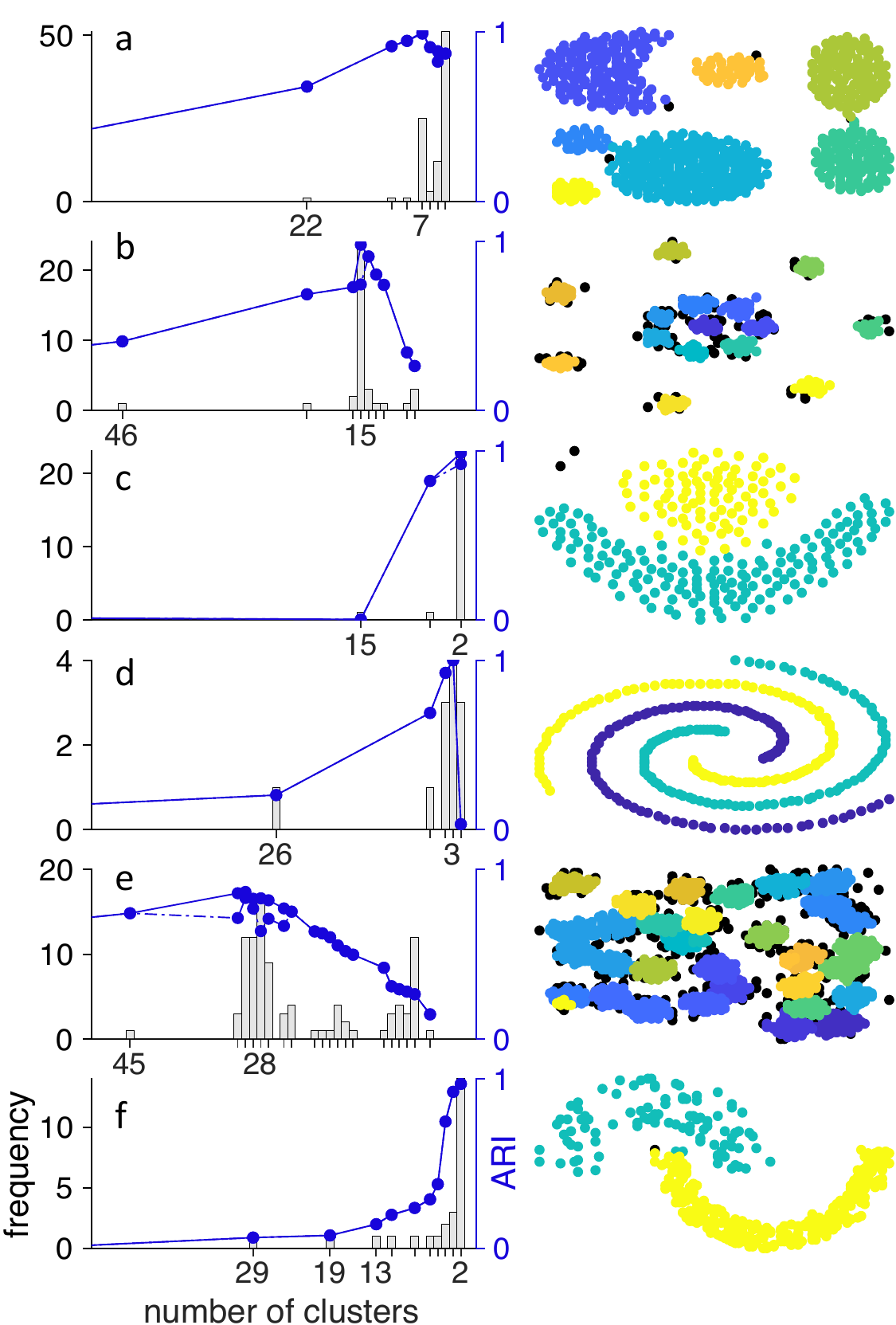}
        \label{sup_fig_RNN_DBSCAN}
        \caption{Test of RNN-DBSCAN clustering over the range $1 \leq k \leq 100$, on a number of artificial datasets \cite{ClusteringDatasets}:
        \textbf{(a)} aggregation, 
        \textbf{(b)} r15,
        \textbf{(c)} flame,
        \textbf{(d)} spiral, 
        \textbf{(e)} d31, and
        \textbf{(f)} jain. 
        On the left side of the figure, the number of clusters is shown against its frequency of occurrence (bars). This is depicted together with the maximum Adjusted Rand Index (ARI) performance (solid line) and the ARI performance at the minimum $k$ (dash-dotted line). On the right side of the figure, a graphical representation of the resulting clusters is depicted, where the number of resulting clusters corresponds to the local maxima on the furthest left of the adjacent histogram (this local maxima is called as the \quotes{first frequency spike} by Bryant et al. \cite{Bryant}). Since different $k$ values contributed to the construction of the bar corresponding to the leftmost local maxima, and since there is a positive correlation between the maximum Adjusted Rand Index (ARI) performance and the ARI performance at the minimum $k$ \cite{Bryant}, we use the minimum $k$ that produced the leftmost local maxima, as input parameter for producing the clusters on the right side of the figure. This test shows similar results to those ones presented by Bryant et al. \cite{Bryant} (please see Figure 4 of their paper).
        }     
\end{figure*}

\end{small}
\endgroup
\twocolumn
}

\clearpage

\bibliography{supp_info}